\documentclass{gGAF2e}
\usepackage{color}
\usepackage[pdf]{pstricks}
\definecolor{midgreen}{rgb}{0.2,0.5,0.1}
\definecolor{midblue}{rgb}{0.0,0.4,0.7}
\definecolor{midpurple}{rgb}{0.7,0.1,0.7}
\newcommand{\vect}[1]{{{\mbox{\boldmath $#1$}}}} 

\newcommand{\DR}{{\mathrm D}}

\def\SN{{_{\rm SN}}} 
\def\kms{{\,{\rm km\,s}^{-1}}} 
\def\cmcube{{\,{\rm cm}^{-3}}} 
\def\erg{{\,{\rm erg}}} 
\def\Myr{{\,{\rm Myr}}} 

\begin{document}

\jvol{00} \jnum{00} \jyear{2018} 

\markboth{GENT et al}{GEOPHYSICAL \& ASTROPHYSICAL FLUID DYNAMICS}


\title{Modelling supernova driven turbulence}

\author{
F. A. Gent$^{{\mathrm a},{\mathrm d}}$$^{\ast}$\thanks{
$^\ast$Corresponding author. Email: frederick.gent@aalto.fi\vspace{6pt}},
M-M. Mac Low$^{\mathrm b}$,
M. J. K\"apyl\"a$^{{\mathrm c},{\mathrm a}}$,
G. R. Sarson$^{\mathrm d}$ and
J. F. Hollins$^{\mathrm d}$\\
\vspace{6pt}  $^{\mathrm a}$ ReSoLVE Centre of Excellence, Dept. Computer
Science, Aalto University, FI-00076, Aalto, Finland\\ 
$^{\mathrm b}$ Dept.\ of Astrophysics, American Museum of Natural History, New York, NY 10024-5192, USA\\
{Center for Computational Astrophysics, Flatiron Institute, New York, NY 10010, USA} \\
$^{\mathrm c}$Max Planck Institute for Solar System Research, D-37707, G\"ottingen, Germany\\
$^{\mathrm d}$School of Mathematics, Statistics and Physics, Newcastle University, Newcastle upon Tyne, NE1 7RU, UK\\
\vspace{6pt}\received{submitted May 2018} }

\maketitle

\begin{abstract}
  High Mach number shocks are ubiquitous in interstellar turbulence. The Pencil Code is particularly well suited to the study of magnetohydrodynamics in weakly compressible turbulence and the numerical investigation of dynamos because of its high-order advection and time evolution algorithms. However, the high-order algorithms and lack of Riemann solver to follow shocks make it less well suited to handling high Mach number shocks, such as those produced by supernovae (SNe). Here, we outline methods required to enable the code to efficiently and accurately model SNe, using parameters that allow stable simulation of SN-driven turbulence, in order to construct a physically realistic galactic dynamo model. These include the resolution of shocks with artificial viscosity, thermal conductivity, and mass diffusion; the correction of the mass diffusion terms; and {a novel} generalization of the Courant condition to include all source terms in the momentum and energy equations. We test our methods with the numerical solution of the one-dimensional (1D) Riemann shock tube (Sod, {\itshape J. Comput. Phys.} 1978, {\bf 27}), also extended to a 1D adiabatic shock with parameters and Mach number relevant to SN shock evolution, including shocks with radiative losses. We extend our test with the three-dimensional (3D) numerical simulation of individual SN remnant evolution for a range of ambient gas densities typical of the interstellar medium and compare these to the analytical solutions of Sedov-Taylor (adiabatic) and the snowplough and Cioffi, McKee  and Bertschinger ({\itshape Astrophys. J.} 1988, {\bf 334})  results incorporating cooling and heating processes. We show that our new timestep algorithm leads to linear rather than quadratic resolution dependence as the strength of the artificial viscosity varies, because of the corresponding change in the strength of interzone gradients.
\end{abstract}


\section{Introduction}

Astrophysical turbulence often occurs in highly compressible flows  such as the interstellar medium (ISM), where turbulence is driven by repeated supernova (SN) explosions \citep{Elmegreen:2004,Scalo:2004}. The Pencil Code\footnote{https://github.com/pencil-code} \citep{Brandenburg:2002} has been extensively applied to weakly compressible flows, such as occur in stellar turbulence \citep{Haugen:2004a}, stellar magnetoconvection {\citep{KKB08,KKB09,KMB12,KKOBWKP16,BKMBFGK18}}, stellar and planetary dynamos \citep{DSB06,Mcmillan:2005}, and accretion disks \citep[][]{dVB06,KK11}. The Pencil Code is well suited to the investigation of dynamos, both small scale (fluctuation or random) modes and large scale (mean or system-wide) modes. It uses sixth-order in space and third-order in time advection algorithms to capture the flow with near-spectral accuracy, and is optimized for excellent performance on clusters of superscalar processors. 

This code has been applied to the study of SN driven ISM turbulence and the galactic dynamo \citep{Gent:2013a,Gent:2013b,KGVS17}{, building on more idealised Pencil Code experiments with high Mach numbers \citep{Haugen:2004a,Haugen:2004b}}. From these results a large scale dynamo (LSD) was obtained for a system resembling the solar neighbourhood of the Milky Way, but the parameters chosen resulted in magnetic Prandtl numbers varying strongly by phase, with the result that a small-scale dynamo (SSD) was only present in the hot phase. Here we shall report on improvements to the {hydrodynamic part of the} Pencil Code model --- which include using only the minimum artificial viscosity needed to permit the resolution of strong shocks with Mach numbers of order 100, improvement of the mass diffusion algorithm, and force-dependent time constraints to improve the stability of the code --- and test the results for various shocks.

Spectral methods are effective for accurately solving initial value problems without discontinuities, and well suited for elliptic equations. Reframing the problem as a superposition of basis equations can, however, be computationally intensive. Handling shocks still must occur in real space, applying similar tools as finite difference and volume element schemes. In this space either artificial diffusion or a Riemann solver is normally required.  Regardless of the order of accuracy of the various codes, artificial viscosity effectively reduces to a first order method in the vicinity of the shock. An alternative is to use Godunov methods to solve for the fluxes at zone boundaries.  These rely on exact or approximate solutions to the Riemann problem at each zone boundary.  Although accurate, they are computationally expensive and sensitive to the addition of new physics that can change the signal propagation characteristics.
 
In this article we shall explain the methods and parameter choices required for the Pencil Code to handle the 1D Riemann shock tube test (section\,\ref{sect:method}) and report its performance for various levels of shock reaching above Mach 100 (see section\,\ref{sect:sod}). We then describe some additional steps required to handle highly compressible SN driven turbulence, including radiative cooling, and in section\,\ref{sect:snowplough} present the results of Sedov-Taylor and snowplough tests for SN remnant evolution across a range of ambient gas density and model resolution. In section\,\ref{sect:timestep}, we describe some additions to the Pencil Code timestepping control, to maintain numerical stability for these challenging simulations and the interplay between the different timestep criteria in realistic models. Finally, we summarise our work in section\,\ref{sect:res}.
  

\section{Method}\label{sect:method}

Our strategy to model strong shocks is to use upwind differencing (effectively a form of hyperdiffusion) when solving each partial differential equation (PDE), to ensure the system is resolved at the grid scale; and to apply artificial viscosity and thermal diffusivity at the shock fronts, following \citet{SN92}, to avoid discontinuities in the derivatives by smoothing the shock profile. The fifth order implemetation of the upwind differencing applied here is detailed in the Pencil Code manual\footnote{{http://pencil-code.nordita.org/doc/manual}} section\,H.2. The current implementation of artificial diffusivities is adapted from the earlier treatments of \citet{Haugen:2004b} and \citet{Mee:2007}.

\subsection{Artificial shock viscosity}

In the momentum equation the shock capturing viscosity is applied as 
\begin{equation}\label{eq:viscosity}
\rho\frac{\DR \bm{u}}{\DR t}\,=\,\cdots\, + \bm\nabla\bigl(\rho\,\zeta_\nu\bm{\nabla\cdot u}\bigr)\,,
\end{equation}
where $\bm u$ denotes velocity, $\rho$ gas density and 
\begin{equation}\label{eq:matD}\frac{\DR}{\DR t}\,=\,\frac{\upartial }{\upartial t} + \bm{u \cdot \nabla}
\end{equation}
is the material derivative. The viscous coefficient takes the form 
\begin{equation} \zeta_\nu \,= \, \nu_{\rm shock}\,f_{\rm shock},
\end{equation} 
where 
\begin{equation}\label{eq:fshock}
f_{\rm shock}\,=\,\Bigl\langle\underset{5}{\rm max}\bigl[ \left(-{\bm {\nabla\cdot u}}\right)_{+}\bigr]\Bigr\rangle
     \bigl({\rm min}\left(\delta x,\delta y,\delta z\right)\bigr)^2.
\end{equation}
Taking only positive values of $-\bm{\nabla\cdot u}$ and otherwise zero, at any point the maximum value within two zones in any direction is applied\footnote{At sixth order accuracy, we may apply a maximum from within one, two or three zones, with three yielding more stability at the expense of increased smoothing. Empirical trials of SN-driven turbulence as discussed in section\,\ref{sect:timestep} indicate two zones to be sufficient.}. This field is then smoothed using a seven-point smoothing polynomial with gaussian weights [1,\,9,\,45,\,70,\,45,\,9,\,1]/180 to obtain $f_{\rm shock}$. Hence, the artificial viscosity is applied only locally at the shocks, and has quadratic dependence on the divergence. The dimensionless constant $\nu_{\rm shock} \simeq 1$.

An additional source term in the equation of energy arises from the viscous heating produced by the artificial viscosity. We solve the energy equation in the form of the specific entropy $s$, so we have
\begin{equation}\label{eq:vischeat}
\rho T\frac{\DR s}{\DR t}\,=\,\cdots\, + \rho\,\zeta_\nu\left(\bm{\nabla\cdot u}\right)^2\,,
\end{equation}
where $T$ denotes temperature.

\subsection{Thermal diffusion}

Including a similar artificial thermal diffusion $\zeta_\chi$ to the energy equation significantly damps numerical oscillations arising behind the shock front with negligible effect on the overall structure of the shock solutions. In the nonadiabatic system, particularly where cooling produces thermally unstable regimes, thermal diffusion can confine thermal instabilities to the limits of the numerical resolution. The thermal diffusion takes the form
\begin{equation}\label{eq:heat}
\rho T\frac{\DR s}{\DR t}\,=\,\cdots\, + \bm{\nabla\,\cdot}\,\bigl(c_{\rm{p}}\rho\,\zeta_\chi\bm\nabla T\bigr)\,,
\end{equation}
where $c_{\rm{p}}$ denotes the gas specific heat at constant pressure and the thermal diffusivity coefficient takes the form 
\begin{equation}
\zeta_\chi\, =\, \chi_{\rm shock}{f_{\rm shock}}\,.
\end{equation}
This coefficient is calculated using local maxima and smoothing ${f_{\rm shock}}$, as for the artificial viscosity, above. A modest value $\chi_{\rm shock}=0.5$ is adequate for weak to moderate adiabatic shocks.

\subsection{Mass diffusion} \label{sec:mass-diff}

Finally, we consider the inclusion of mass diffusion. Of course, there is no physical mass diffusion term in the continuity equation, so this is a purely artificial numerical device. Mass diffusion is not necessary to model even strong shock solutions with the Pencil Code (as considered in detail in Section~\ref{sect:sod}), although its use does tend to damp the oscillations in the wake of the shock. However, experiments with SN-driven turbulence show that {interacting} shocks in that context are prone to local numerical instabilities (effectively wall heating), where the density drops and the temperature rises without limit, producing a hot zone. The application of mass diffusion suppresses this problem.

With mass diffusion, in the absence of sinks or sources, the continuity equation becomes
\begin{equation}\label{eq:massdiff}
  \frac{\DR \rho}{\DR t}\,=\,-\,\rho \bm{\nabla\cdot u} + \zeta_D\nabla^2\rho + \bm\nabla\zeta_D\,\bm{\cdot\, \nabla}\rho\,,
\end{equation}
where 
\begin{equation}
  \zeta_D\, =\, D_{\rm shock}\,{f_{\rm shock}}\,,
\end{equation}
with $f_{\rm shock}$ as defined in (\ref{eq:fshock})  and $D_{\rm shock}\simeq1$. Adding a non-physical diffusion to the equation has consequences for the conservation of momentum and energy. Hence, corrections to each equation are required. If we consider the momentum and energy equations absent the artificial diffusion, we have
\begin{align*}
\rho\frac{\DR \bm u}{\DR t} \,\,=&\, {\rm RHS}\,, & \rho\frac{\DR e}{\DR t}\, =\,& {\rm RHS}\,,\\
\frac{\DR }{\DR t}(\rho\bm u)- \bm u\frac{\DR \rho}{\DR t}\,=\,&\, {\rm RHS}\,, &
\frac{\DR }{\DR t}(\rho e)- e\frac{\DR \rho}{\DR t}\,=\,&\, {\rm RHS}\,,\\
\frac{\DR }{\DR t}(\rho\bm u)+ \bm u\rho\bm{\nabla\cdot u}\,=\,&\, {\rm RHS}\,, &
\frac{\DR }{\DR t}(\rho e)+ c_{\rm{v}}\,T\rho\bm{\nabla\cdot u}\, =\,&\, {\rm RHS}\,,
\end{align*}
%
where $e=c_{\rm{v}}\,T$ denotes the internal energy and $c_{\rm{v}}$ is the specific heat at constant volume. When we include the mass diffusion in the continuity equation we obtain
\begin{align}
\frac{\DR }{\DR t}(\rho\bm u)
+ \bm u \bigl(\rho\bm{\nabla\cdot u}-\zeta_D\nabla^2\rho - \bm\nabla\zeta_D\,\bm{\cdot \,\nabla}\rho\bigr)\, \neq\,&\, {\rm RHS}\,,\\
\frac{\DR }{\DR t}(\rho e)
+ c_{\rm{v}}\,T\bigl(\rho\bm{\nabla\cdot u}-\zeta_D\nabla^2\rho - \bm\nabla\zeta_D\,\bm{\cdot\, \nabla}\rho\bigr)\, \neq\,&\, {\rm RHS}\,.
\end{align}
Hence, to conserve the properties of momentum and energy we must also subtract these extra terms from the respective RHS. For momentum we have
\begin{equation}\label{eq:fixu}
\rho\frac{\DR \bm u}{\DR t}\,=\,\cdots\, -\bm u\bigl(\zeta_D\nabla^2\rho + \bm\nabla\,\zeta_D\,\bm{\cdot\,\nabla}\rho\bigr)
\end{equation}
and for energy we have
\begin{equation}\label{eq:fixe}
\rho T \frac{\DR s}{\DR t}\, = \,\cdots\, - c_{\rm{v}}\,T\bigl(\zeta_D\nabla^2\rho + \bm\nabla\,\zeta_D\,\bm{\cdot\,\nabla}\rho\bigr)\,.
\end{equation}
We define an array of artificial diffusion coefficients as $c_{\rm shock} \equiv [D_{\rm shock},\nu_{\rm shock},\chi_{\rm shock}]$.

\subsection{Timestep determination\label{sect:timestep}}

The Pencil Code uses an explicit finite difference scheme, which can be faster than implicit schemes, but is not unconditionally stable. A necessary, although not necessarily sufficient, stability condition is to satisfy various Courant conditions on the time step, such as for advection and diffusion. For advection, this condition reads
\begin{equation}\label{eq:advect}
  \frac{1}{\delta t}\, \geq \,{\rm max}\left(\frac{U_{\rm max}}{c_{\delta t}\,\delta x_{\rm min}} \right),
\end{equation}
where $c_{\delta t}<1$ is the Courant number dedicated to the control of the advective timestep, $\delta x_{\rm min}={\rm min}(\delta x,\delta y,\delta z)$ is the minimum grid spacing at each location, and for the MHD case
\begin{equation}
U_{\rm max}\,=\,{\rm max}\left(|\bm u|+\sqrt{c_s^2+v_A^2}\,\right).
\end{equation}
$c_s$ and $v_A$ denote the sound speed and Alfv\'en speed, respectively. With maximal speeds of order $10^3\kms$ typical of such turbulence, however, this is rarely significant in determining the maximum timestep. We also must account for the artificial diffusion terms we have introduced to resolve shocks. The diffusive time step is controlled by the Courant condition 
\begin{equation}\label{eq:difft}
\frac{1}{\delta t}\, \geq\, {\rm max} \left(\frac{\zeta_{\rm max}}{c_{\delta t,v}\,\delta x^2_{\rm min}} \right),
\end{equation}
where $c_{\delta t,v}<1$ is a Courant coefficient for the diffusive timestep, $\zeta_{\rm max}={\rm max}(\zeta_\nu,\gamma\zeta_\chi,\zeta_D )$ is the maximum diffusive coefficient acting at each point in the grid. In the MHD case magnetic diffusivity $\zeta_\eta$ may also be included in $\zeta_{\rm max}$.

It has previously been observed that strong heat sources or sinks can cause stresses on the numerical solvers that the conventional timestep control described above does not address. In the context of ISM simulations and SN driven turbulence the effect of cooling and heating can cause numerical instability if not also accounted for. Minimum cooling times are typically around 100\,yr for the temperature and density ranges commonly considered in ISM simulations, but can be as low as 10\,yr. This is not usually a problem as other processes often require shorter time resolution. Heating by SN is instant and therefore presents a challenge primarily through heat diffusion timescales. The main source of heating driving the timestep due to the supersonic flows is viscous heating. To ensure the heating/cooling time is resolved, the net heating $H$ is summed and the timestep, $\delta t$ constrained by its absolute maximum throughout the grid as 
\begin{equation}\label{eq:Hmax}
\frac{1}{\delta t}\, \geq\, {\rm max} \left(\frac{|H|_{\rm max}}{c_{\delta t,s}\, e} \right),
\end{equation}
where $c_{\delta t,s}<1$ controls the fractional change of energy permitted in any cell.

In weakly compressible flows the largest values on the right hand side of the entropy equation can be adequately resolved in time, but for the highly compressible flows of SN driven turbulence the sum of all changes to the energy can be many orders of magnitude greater than the evolving entropy of order unity. To address this the $H_{\rm max}$ in Equation\,\eqref{eq:Hmax} is instead replaced by the maximal sum of the right hand side of the entropy equation $df(s)$, so that
\begin{equation}\label{eq:dfsmax}
 \frac{1}{\delta t}\,  \geq \, {\rm max}\left(\frac{|df(s)|_{\rm max}}{c_{\delta t,s}\, c_{\rm{v}}} \right)
\end{equation}
is used to constrain the timestep, with the maximum fractional change in entropy given by $c_{\delta t,s}$. We find empirically that this time step constraint dominates immediately after SN explosions during SN driven turbulent runs.

The momentum equation must also be considered as a whole, similar to the treatment of the energy equation (\ref{eq:dfsmax}). The troublesome contribution to the code stability is to be found in viscous force, in general expressed as 
\begin{equation}
\label{eq:fvisc}
\frac{\DR \vect{u}}{\DR t}\,
=\,\cdots\,+ \nu \nabla^{2} \vect{u} + \tfrac{\nu}{3}\vect\nabla \big(\vect\nabla\, \bm{\cdot}\,
\vect{u}\bigr) + 2 \bm{\mathsf{S}}\, {\bm \cdot}\,\bigl( \nu\vect{\nabla} \textrm{ln} \rho+\vect\nabla\nu\bigr)
+\zeta_{\nu}\vect{\nabla}{\bigl( \vect{\nabla}\,{\bm \cdot}\, \vect{u} \bigr)}
+\vect{\nabla}\zeta_{\nu} \bigl( \vect{\nabla}\,{\bm \cdot}\, \vect{u}\bigr)\,,
\end{equation}
involving the rate of strain tensor $\bm{\mathsf{S}}$ of the form 
\begin{equation}
  \label{eq:str}
  2 \mathsf{S}_{ij}\,=\, \frac{\upartial u_{i}}{\upartial x_{j}}+
  \frac{\upartial u_{j}}{\upartial x_{i}}
  -\frac{2}{3}\delta_{ij}\vect{\nabla}\, {\bm \cdot}\,\vect{u}\,. 
\end{equation}
Note also, the viscous heat applying to the energy equation is proportional to $\bm{\mathsf{S}}^2\equiv \mathsf{S}_{ij}\mathsf{S}_{ij}$. In the diffusive timestep only the coefficients $\nu$ and $\zeta_\nu$ are considered. The gradients in the expression are ignored, and in the case of SN turbulence, these contributions can be of order $\pm10^7$--$10^9${\,km\,s$^{-1}$\,Gyr$^{-1}$}. Increasing viscosity to smooth the gradients can be counter productive, making the viscous forces even larger. Instead, we limit the total change in momentum $df(\bm u)$ with a newly developed time step limitation of
\begin{equation}\label{eq:dfumax}
  \frac{1}{\delta t} \geq {\rm max}\left(\frac{|df(\bm u)|_{\rm max}}{c_{\delta t,f}\, |\bm u|_{\rm nom}} \right),
\end{equation}
where $c_{\delta t,f}<1$ is a Courant number applying to control of the forcing timestep, and $|\bm u_{\rm nom}|$ is a sufficiently small nominal velocity. In practice $c_{\delta t,f}|\bm u|_{\rm nom}\simeq20${\,km\,s$^{-1}$} is stable. An alternative approach might be to reduce $c_{\delta t,{v}}\ll 1$, but this would reduce the the timestep under all circumstances, whereas the forcing timestep control is more flexible and limits the timestep only when needed. The impact of this timestep control on a turbulent ISM simulation is
considered in section\,\ref{sect:timestep}

\subsection{{Sound-speed dependent shear viscosity}\label{sect:cs_visc}} 

The numerical solutions to the experiments included in this article are adequately modelled without any prescription for viscosity other than the artificial shock-dependent viscosity already described. However, in the turbulent SN-driven system, the numerical stability has been found to benefit from viscosity proportional to the speed of sound $c_s$. The ISM is typified by huge variation in temperature and associated characteristic turbulent velocities. Together with the artificial viscosity this applies a numerically stable constraint on the mesh Reynolds number. Whilst little is understood about the behaviour of turbulent viscosity in the ISM, such temperature dependent behaviour for the ISM can be argued to more closely approach the molecular Spitzer viscosity, $\nu\propto\rho^{-1}T^{2.5}$, than the usual application of constant $\nu$.

Various applications of physical viscosity may be considered, such as  \emph{Laplacian} for $\nu\vect\nabla^2 \vect{u}$ or \emph{bulk} which applies to the trace, omitted from the rate of strain tensor $\bm{\mathsf{S}}$. Here we apply \emph{shear} viscosity arising from the divergence of the \emph{traceless} rate of strain tensor. In this form the contribution of $\vect\nabla\nu$ in (\ref{eq:fvisc}) is nonzero, and additional viscous heating $2\rho\nu\bm{\mathsf{S}}^2$ applies to the energy equation. The sound-speed dependent viscosity is not included in section\,\ref{sect:sod}. To demonstrate that these results hold for the prescription we intend for the modelling of SN turbulence, we include this viscosity in the form $\nu\simeq c_s\,\delta x$ throughout section\,\ref{sect:snowplough}.

\section{Riemann shock tube test}\label{sect:sod}

\begin{figure}
\begin{center}
\begin{minipage}{150mm}
{\resizebox*{7.25cm}{!}{\includegraphics[trim=0.25cm 1.7cm 0.5cm 0.4cm,clip=true]{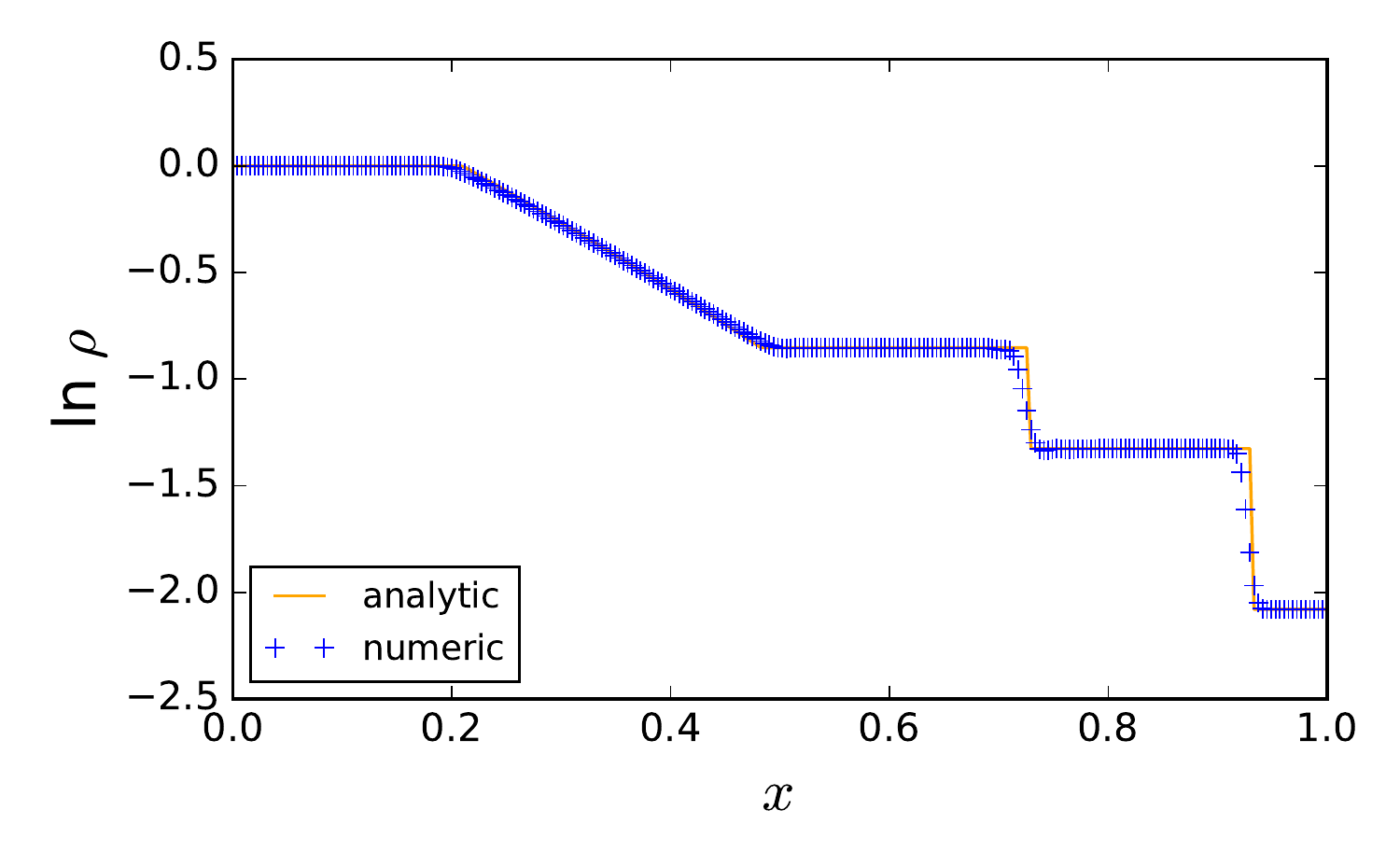}}}%
{\resizebox*{7.25cm}{!}{\includegraphics[trim=0.10cm 1.7cm 0.5cm 0.4cm,clip=true]{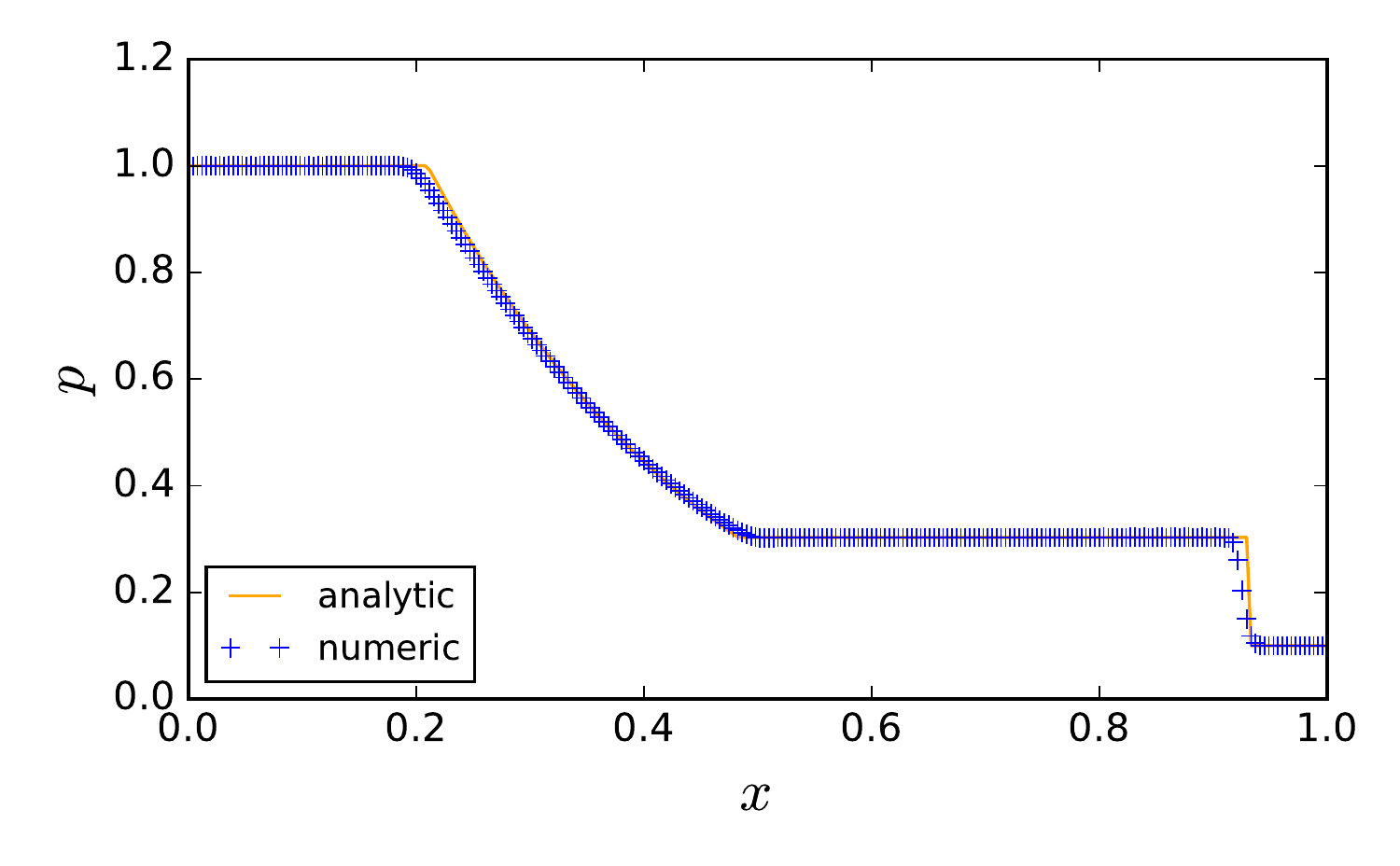}}}\\%
{\resizebox*{7.25cm}{!}{\includegraphics[trim=0.10cm 0.5cm 0.5cm 0.4cm,clip=true]{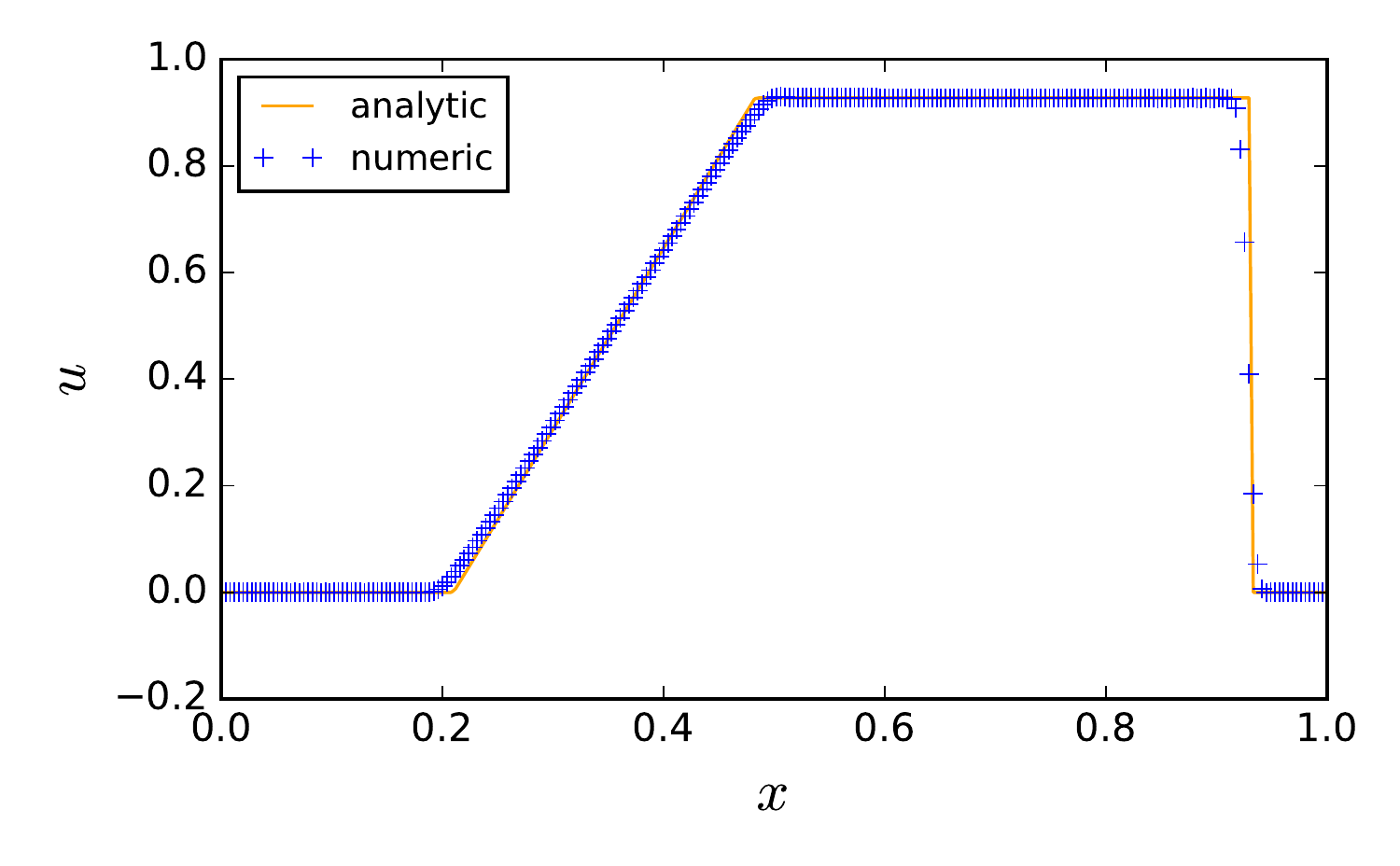}}}%
{\resizebox*{7.25cm}{!}{\includegraphics[trim=0.10cm 0.5cm 0.5cm 0.4cm,clip=true]{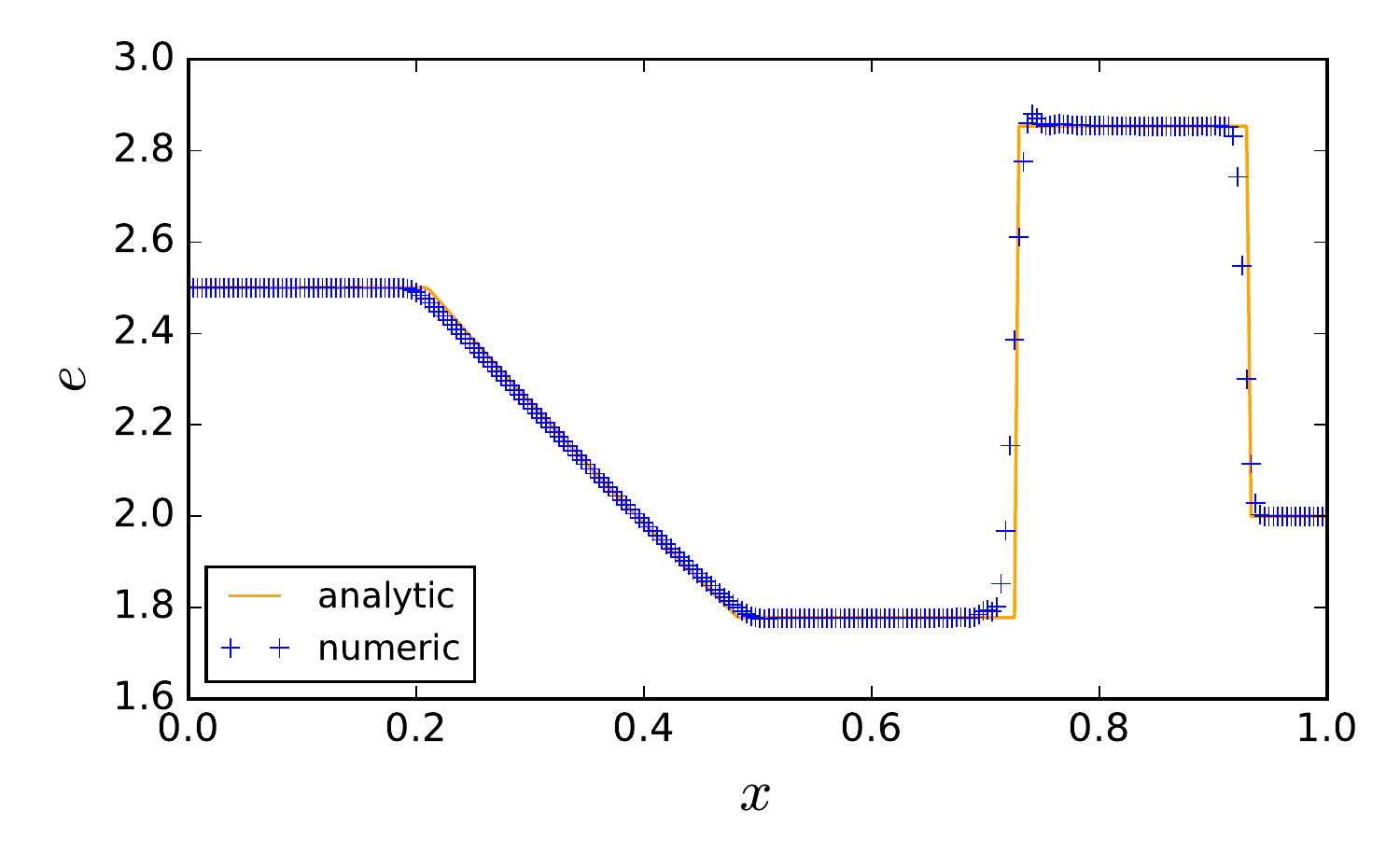}}}%
\caption{
Weak Riemann shock tube test \citep{Sod78} with 256 grid points at $t=0.245$, where the density and pressure on the left is initially set to $\rho,\,p=1.0,\,1.0$, and on the right $\rho,\,p=0.125,\,0.1$. Figures show density, pressure, velocity $u$, internal energy $e$, and the analytic solution (orange line). The diffusivity coefficients used are ${c_{\rm shock}  }=[0,{4},0.5]$. (Colour online)
}%
\label{fig:weak-sod}
\end{minipage}
\end{center}
\end{figure}

\subsection{Weak and moderate shocks}\label{sect:weak}

\begin{figure}
\begin{center}
\begin{minipage}{150mm}
{\resizebox*{7.25cm}{!}{\includegraphics[trim=-0.15cm 1.7cm 0.5cm 0.4cm,clip=true]{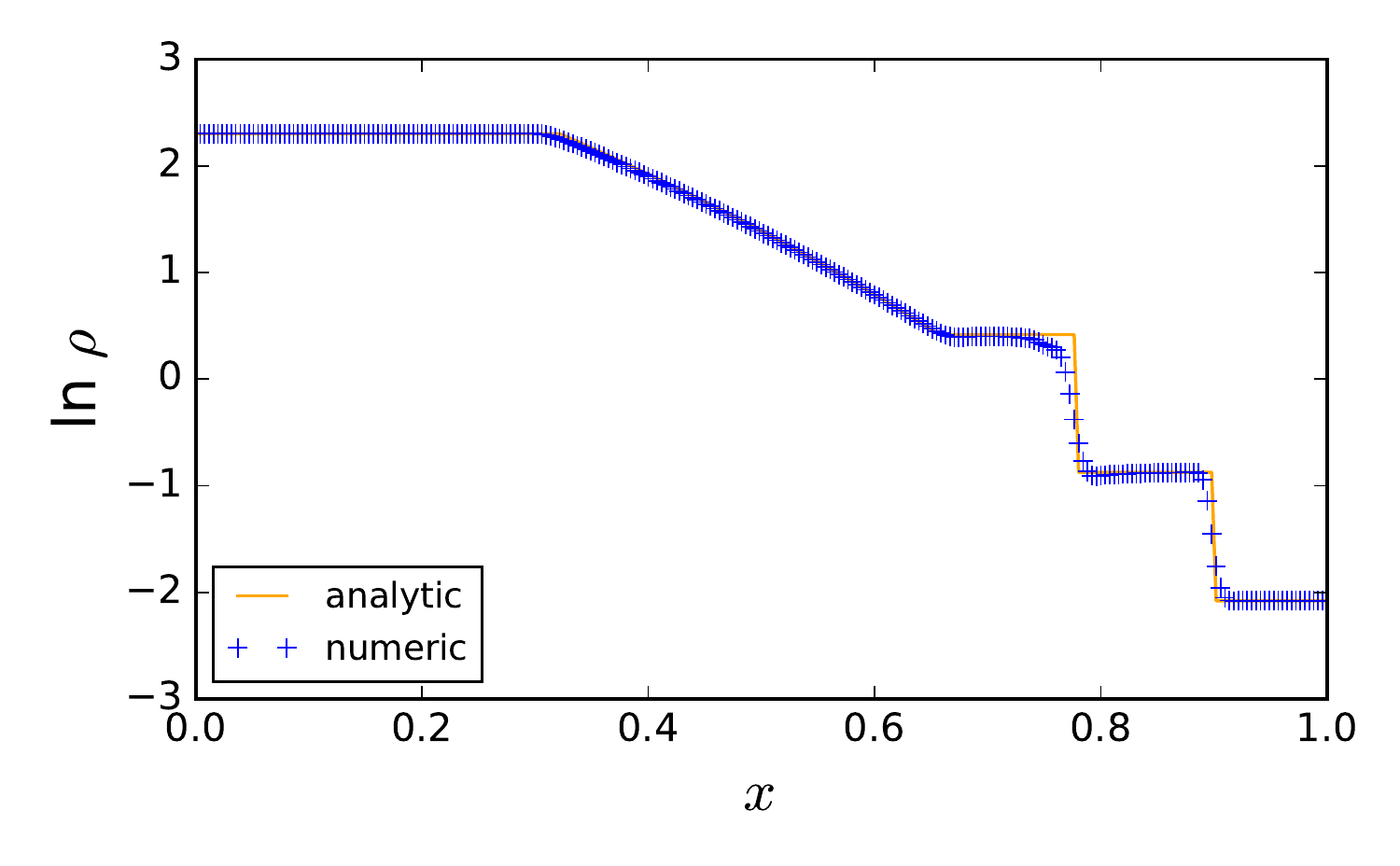}}}%
{\resizebox*{7.25cm}{!}{\includegraphics[trim=-0.02cm 1.7cm 0.5cm 0.4cm,clip=true]{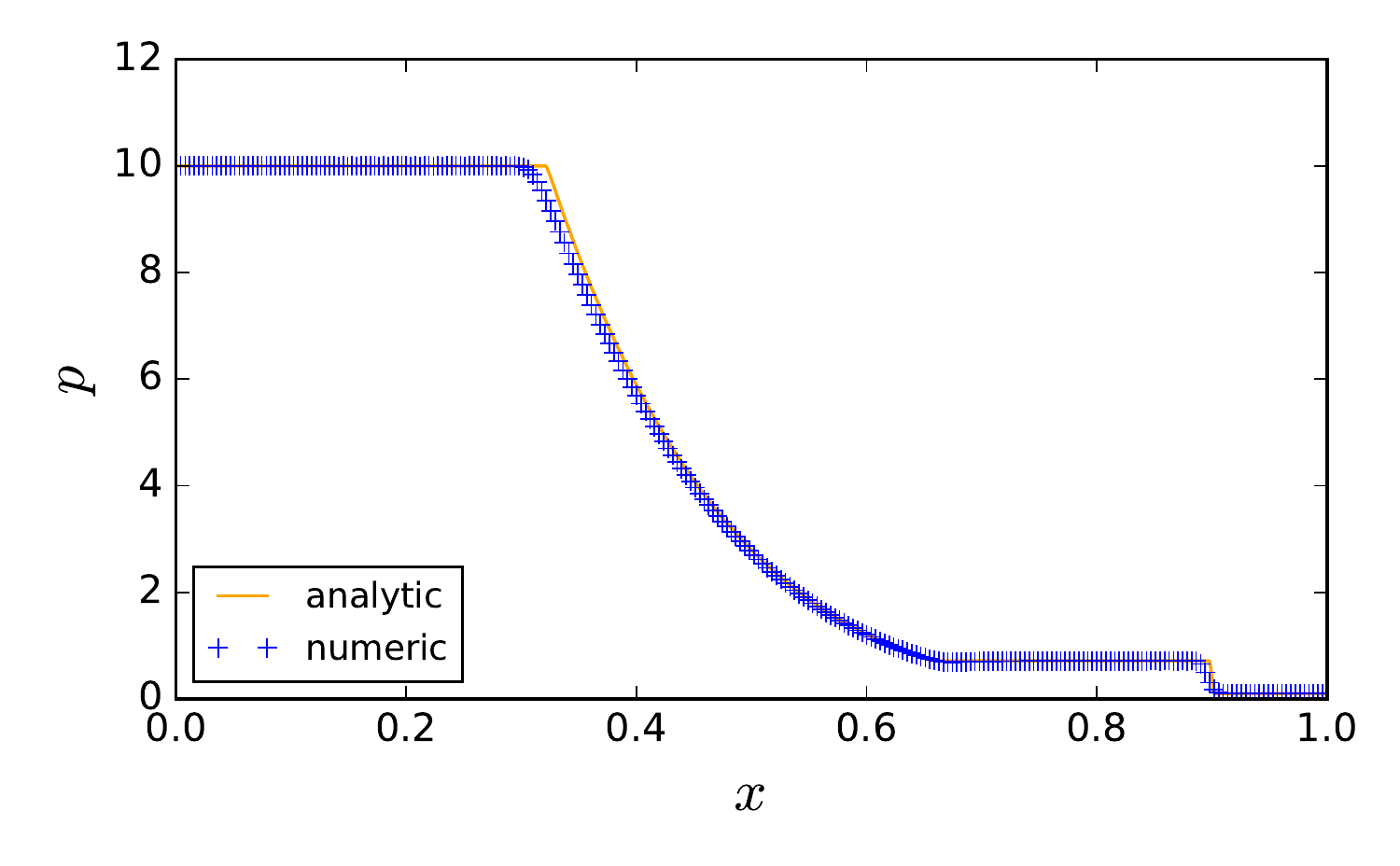}}}\\%
{\resizebox*{7.25cm}{!}{\includegraphics[trim= 0.20cm 0.5cm 0.5cm 0.4cm,clip=true]{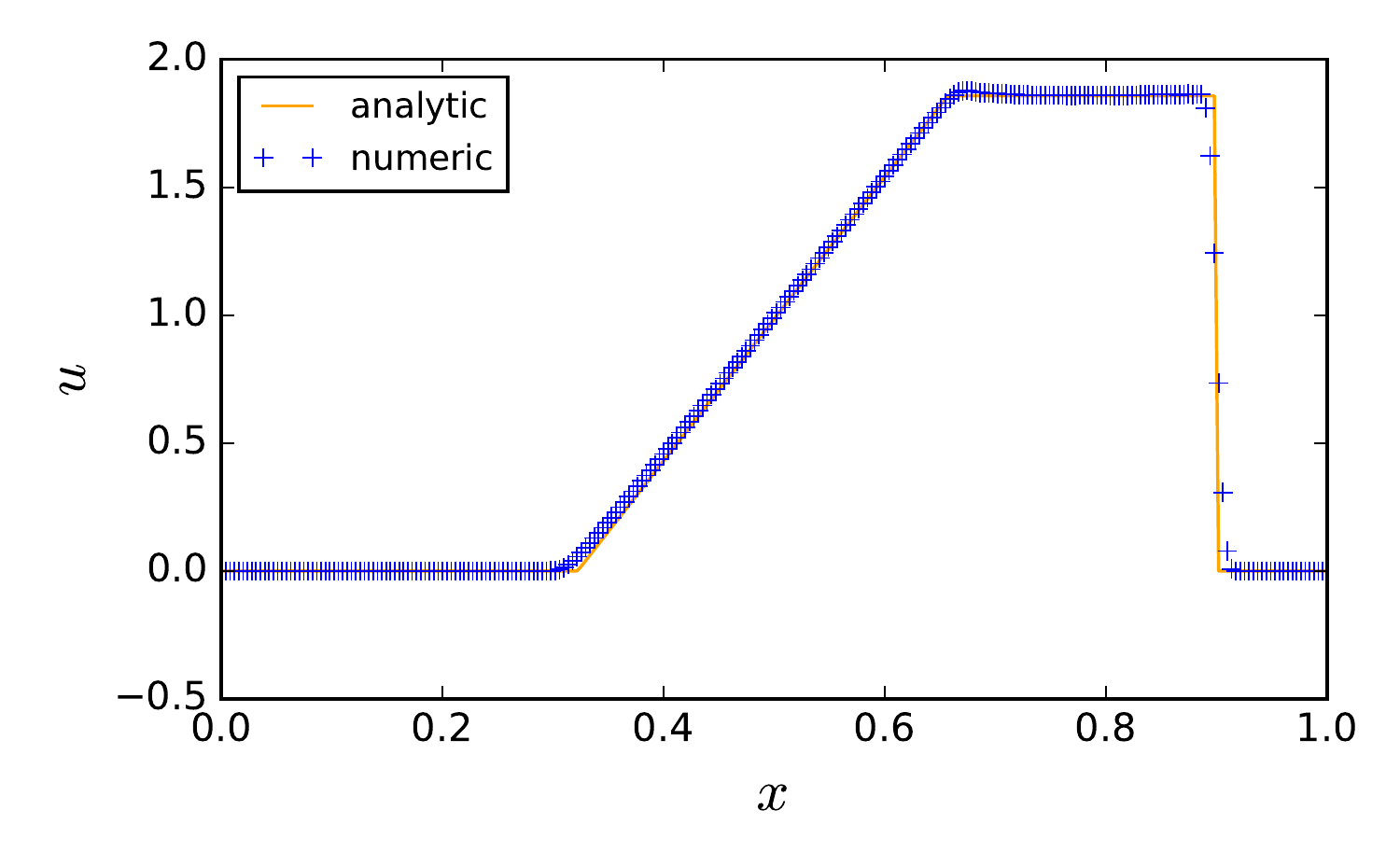}}}%
{\resizebox*{7.25cm}{!}{\includegraphics[trim= 0.20cm 0.5cm 0.5cm 0.4cm,clip=true]{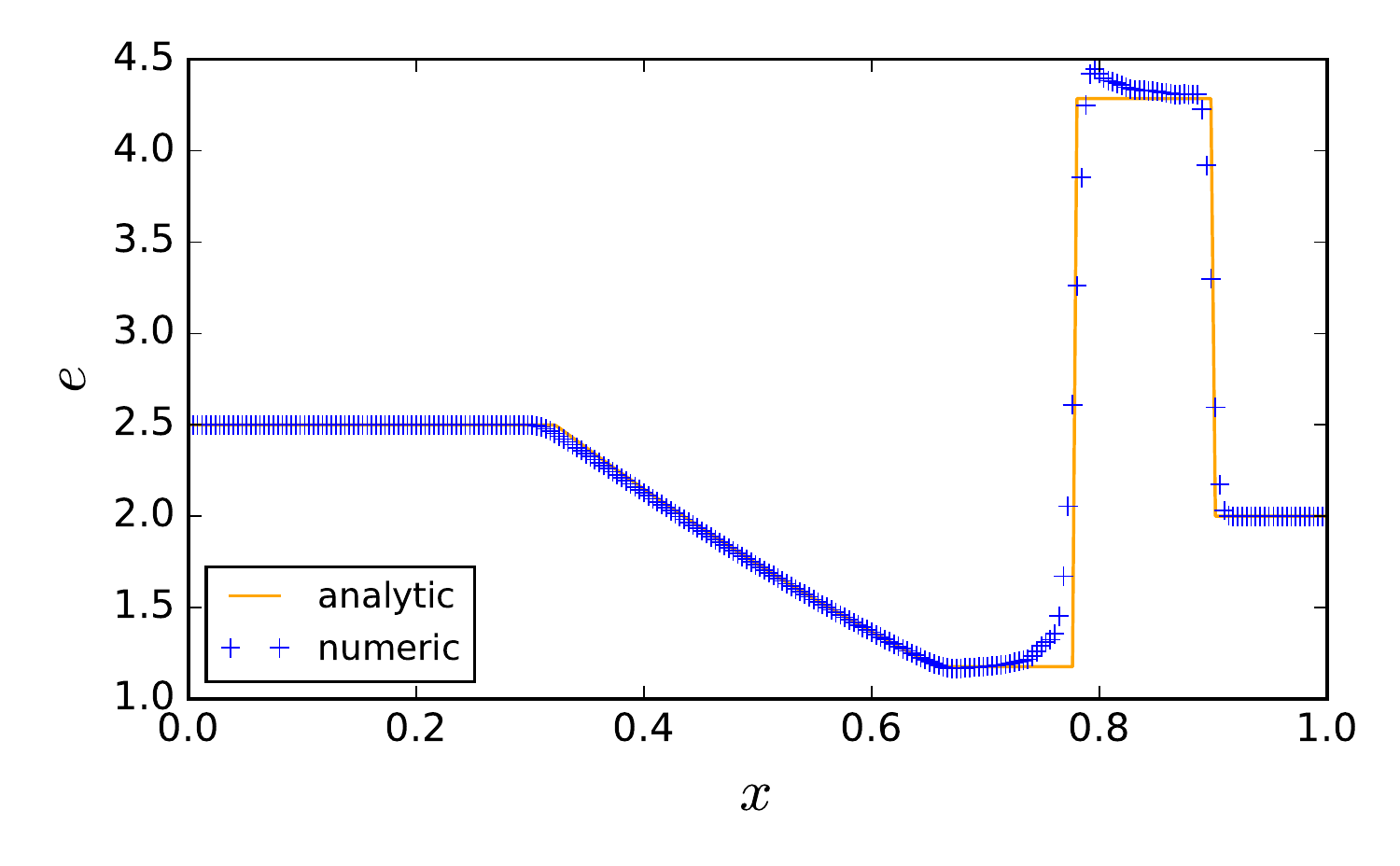}}}%
\caption{
Moderate shock tube test \citep{Sod78} with 256 grid points at $t=0.150$, where the density, pressure on the left is initially set to $\rho,\,p=10.0,\,10.0$, and on the right $\rho,\,p=0.125,\,0.1$. Figures show density, pressure, velocity, internal energy, and the analytic solution (orange line). The diffusivity coefficients used are ${c_{\rm shock}}=[0,{4},0.5]$. (Colour online)
}%
\label{fig:mod-sod}
\end{minipage}
\end{center}
\end{figure}
\begin{figure}
\begin{center}
\begin{minipage}{150mm}
{\resizebox*{7.25cm}{!}{\includegraphics[trim= 0.25cm 1.665cm 0.5cm .2cm,clip=true]{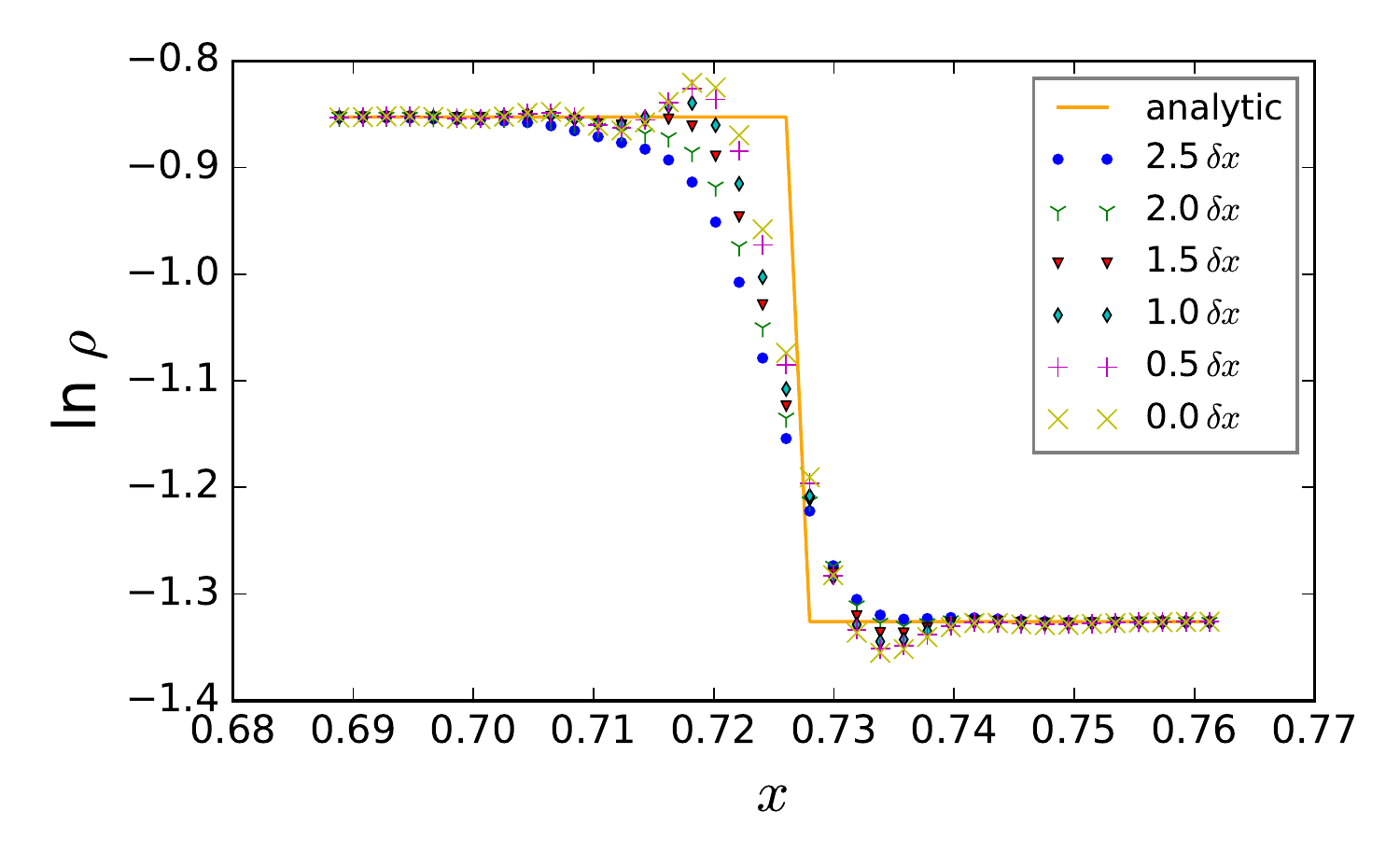}}}%
{\resizebox*{7.25cm}{!}{\includegraphics[trim= 0.25cm 1.665cm 0.5cm .2cm,clip=true]{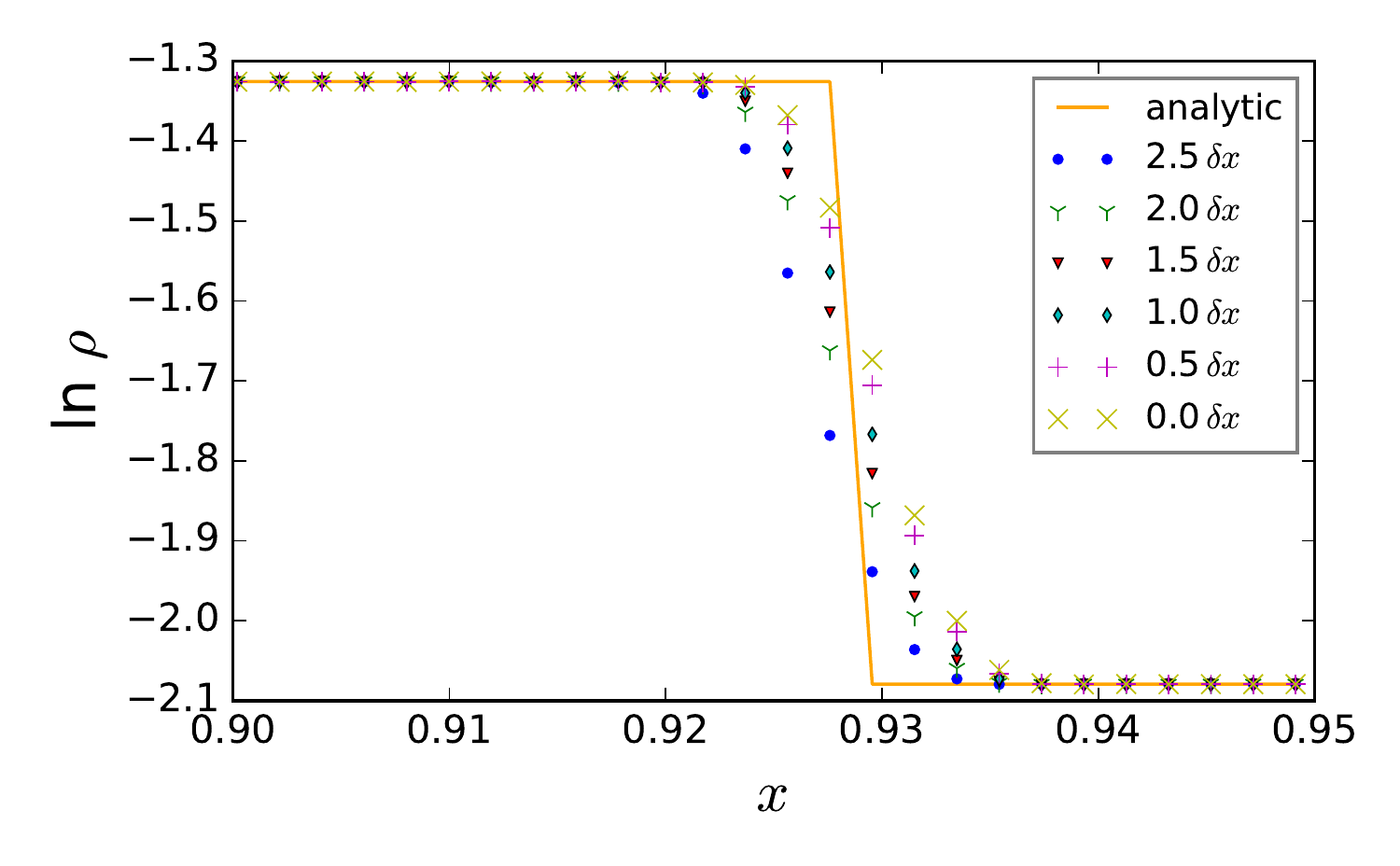}}}\\%
{\resizebox*{7.25cm}{!}{\includegraphics[trim=-0.30cm 0.550cm 0.5cm .2cm,clip=true]{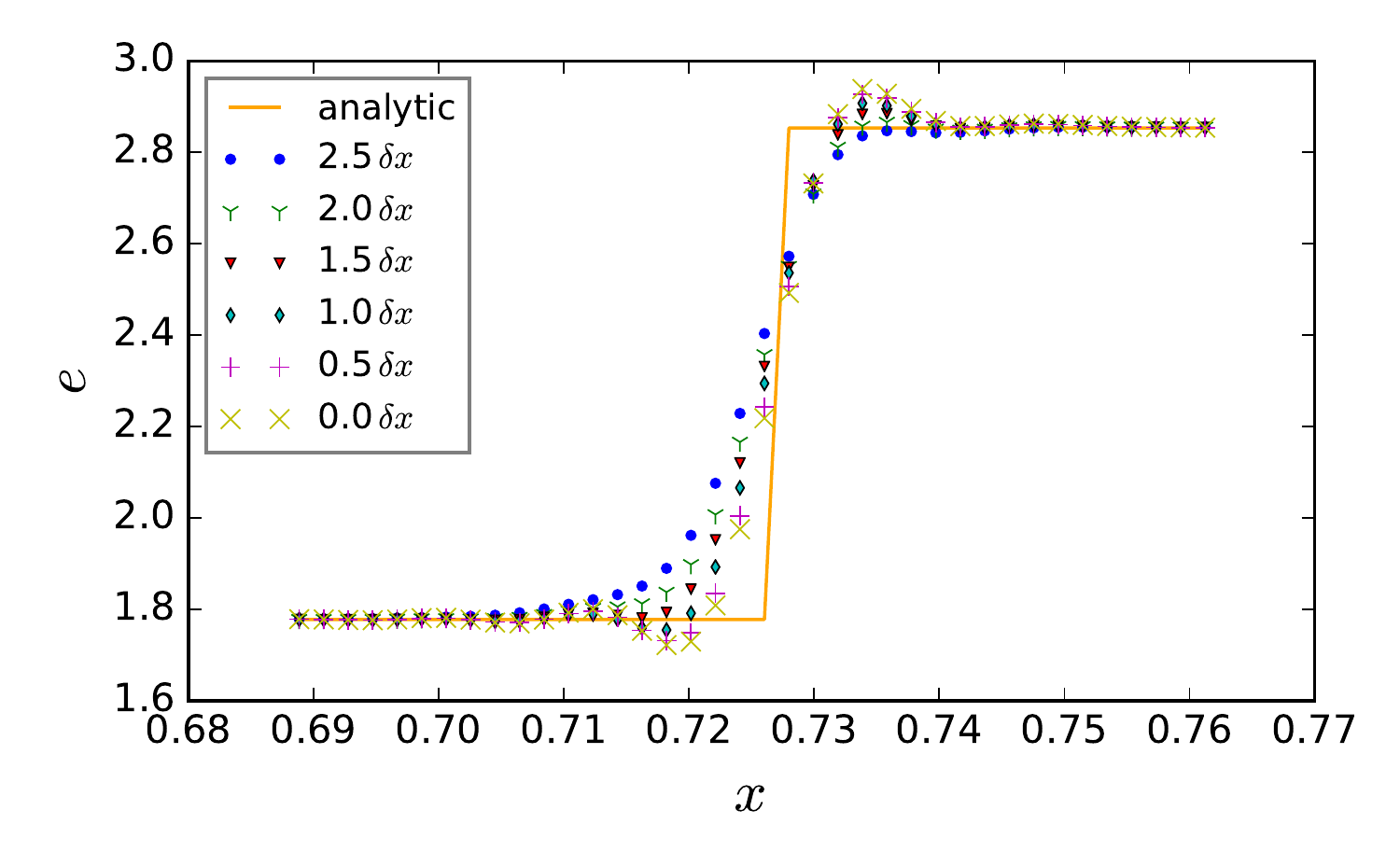}}}%
{\resizebox*{7.25cm}{!}{\includegraphics[trim=-0.30cm 0.550cm 0.5cm 0.2cm,clip=true]{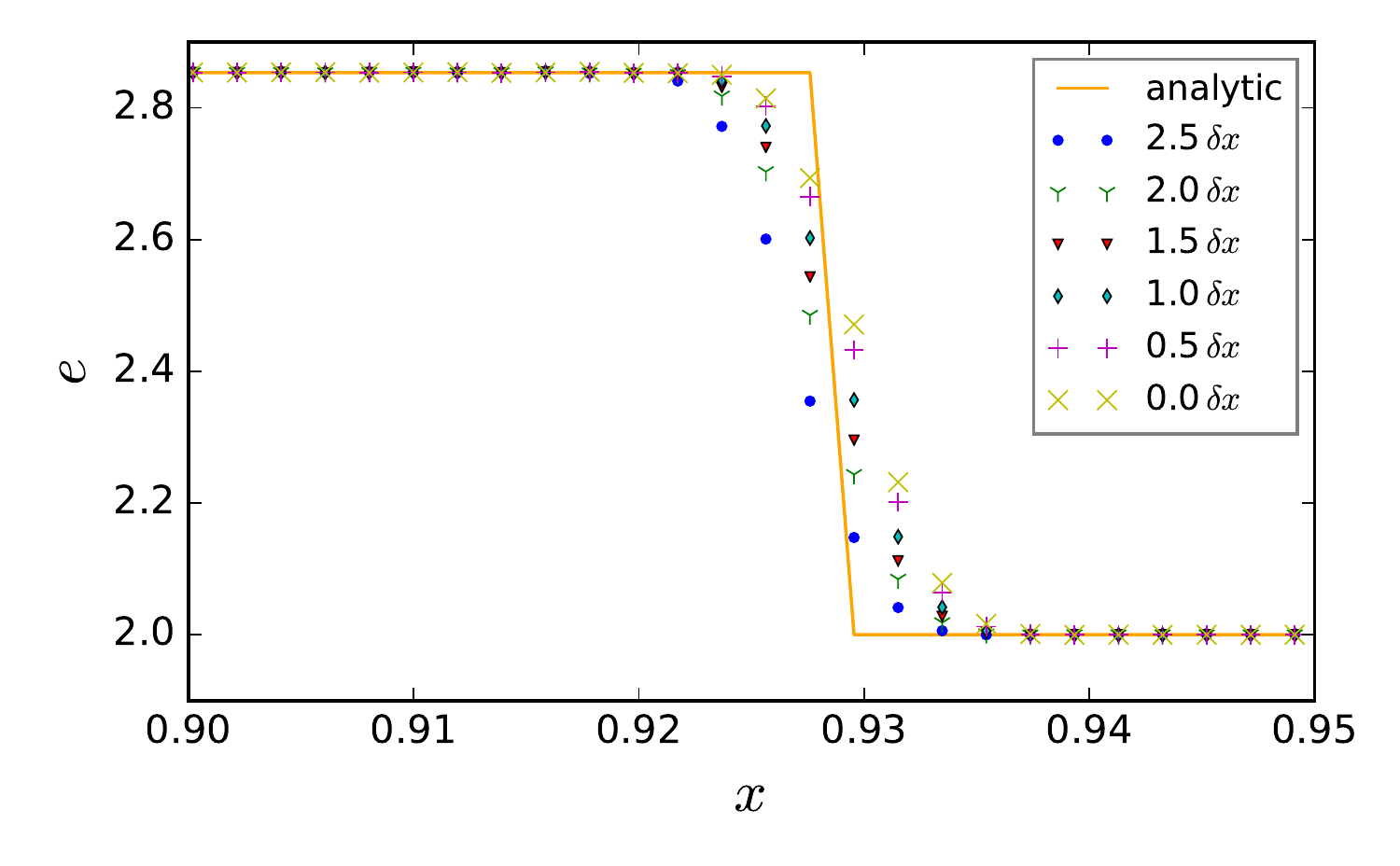}}}%
\caption{
The weak shock tube test \citep{Sod78} detailed in figure\,\ref{fig:weak-sod} with ${c_{\rm shock}}=[0,{4},0.5]$ at $t=0.245$ resolved with 512 grid points.Zoomed-in profiles show contact discontinuity (left) and shock front (right) for initial discontinuity smoothing scale $\ell\in[0,2.5]\delta x$, for gas density (upper) and internal energy (lower). (Colour online)
}%
\label{fig:init-comp}
\end{minipage}
\end{center}
\end{figure}

To assess the quality of the shock handling scheme we consider the Riemann shock tube test using the standard setup described by \citet{Sod78}, based on the exact analytical solution obtained by, e.g., \citet{HSW84}. Results from a weak and a moderate shock test in a one dimensional grid over 256 points with closed boundaries are depicted in figures\,\ref{fig:weak-sod} and \ref{fig:mod-sod}, respectively. For direct comparison with \citet[][their figures 11 and 12]{CK01} we use the adiabatic index $\gamma=1.4$. In each case an initial discontinuity in density and energy is located at $x=0.5$ with zero velocity, and on the right the dimensionless density $\rho = 0.125$ and pressure $p = 0.1$. In the weak shock in figure\,\ref{fig:weak-sod} the density and pressure on the left are both 1.0, and for the moderate shock in figure\,\ref{fig:mod-sod} they are 10. The analytic solutions are {included for comparison}. For these parameters, a reasonable solution can be obtained with the artificial diffusivities $\nu_{\rm shock}={4.0}$ and $\chi_{\rm shock}=0.5$, and with no mass diffusion ($D_{\rm shock}=0$).

However, if the initial discontinuity in mass and energy is not smoothed, then significant oscillations occur in the wake of the shock, which in stronger shocks can lead to artificial hot zones forming due to wall heating, which crash the code. The initial discontinuity profile has the form, here for density,
\begin{equation}
\rho(x)\,=\,\rho_{\rm left}+
          \left(\frac{\rho_{\rm left}-\rho_{\rm right}}{2}\right)
          \left[1 + \tanh\left(\frac{x}{n\delta x}\right)\right],
\end{equation}
where the smoothing length $\ell = n \delta x$ In figure\,\ref{fig:init-comp} we show the results at the shock front and contact discontinuity of applying $\ell\in[0,2.5]\delta x$. The shock front profiles in the right panels retain a similar shape, but delayed with increased initial smoothing. This is where the artificial viscosity is present throughout the evolution. In contrast, the left panels show stronger high frequency wiggles forming at the contact discontinuity for the unsmoothed initial profiles. An optimal smoothing scale is $\ell=1.5\delta x$, and this is used for the weak and moderate shock-tube tests in this paper. In simulations of SN driven turbulence with SN shocks introduced to a highly nonuniform ambient ISM, we have no such fine tuning over the level of discontinuity smoothing, but we apply a 3D-gaussian profile for the initial injection of energy in the SN experiments, rather than steeper or discontinuous profiles, to miminise such numerical instabilities forming in the contact discontinuities behind the shock fronts. 

\begin{figure}
\begin{center}
\begin{minipage}{150mm}
{\resizebox*{7.25cm}{!}{\includegraphics[trim=0.15cm 1.665cm 0.5cm 0.4cm,clip=true]{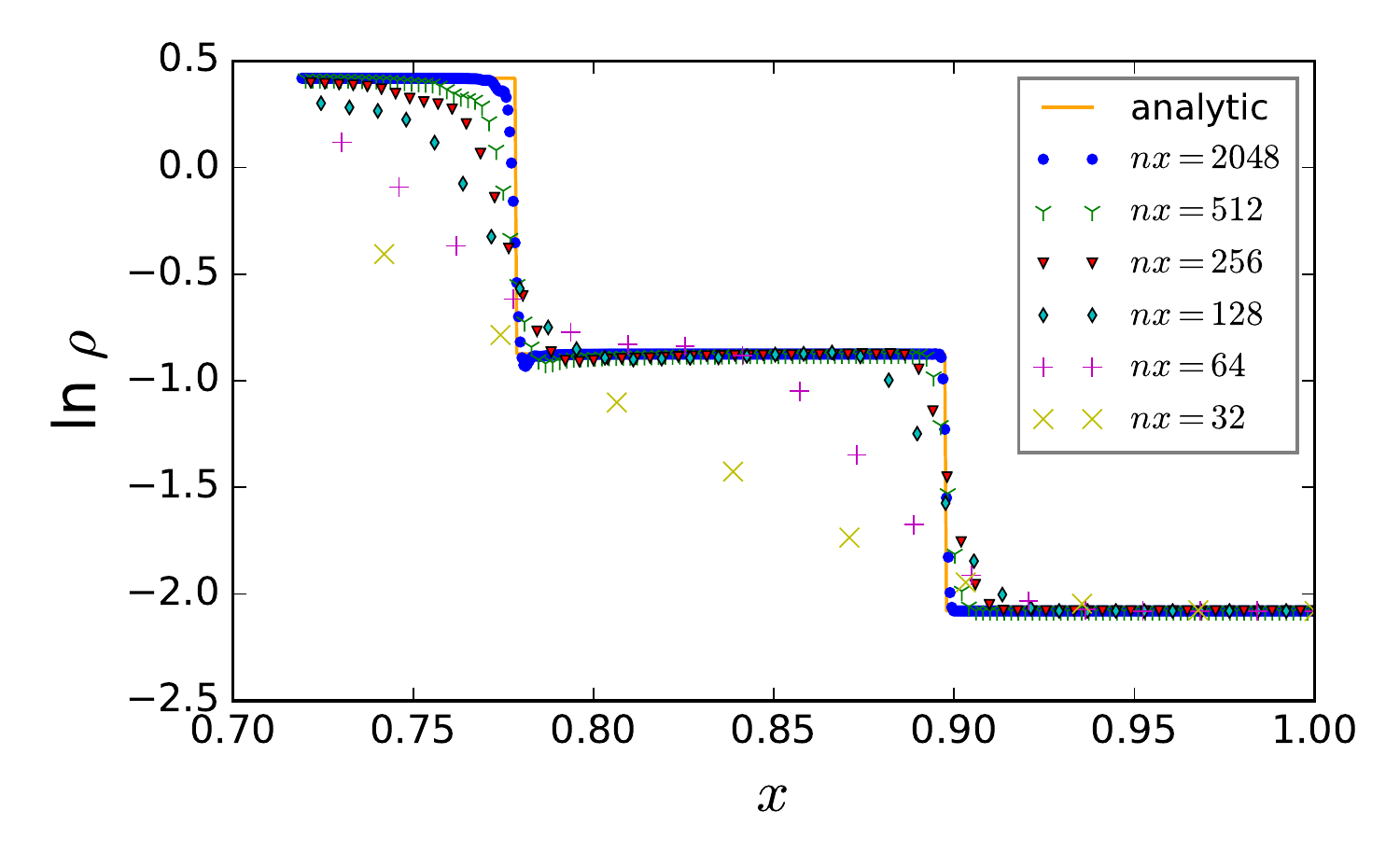}}}%
{\resizebox*{7.25cm}{!}{\includegraphics[trim=0.05cm 1.665cm 0.5cm 0.4cm,clip=true]{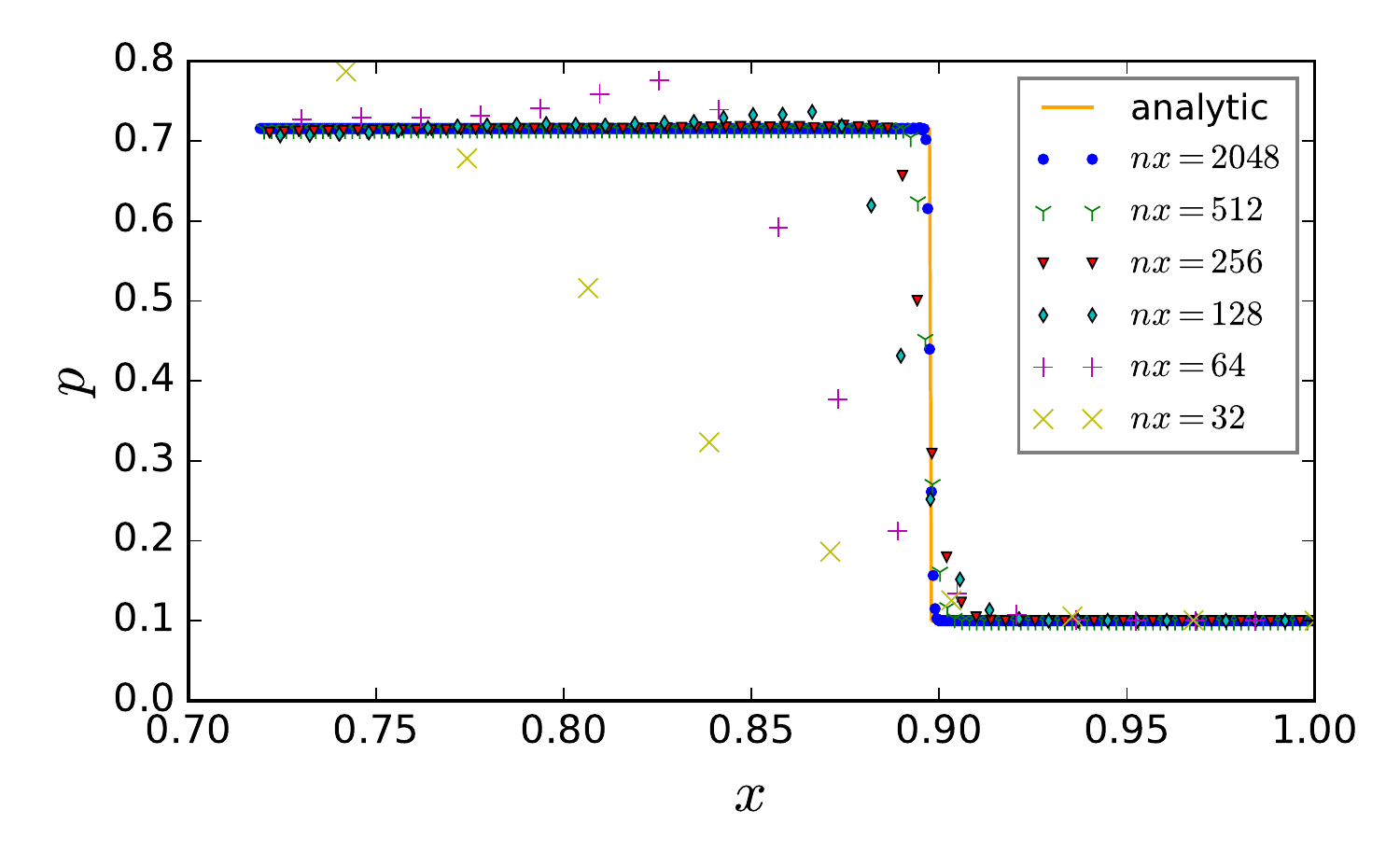}}}\\%
{\resizebox*{7.25cm}{!}{\includegraphics[trim=0.00cm 0.500cm 0.5cm 0.4cm,clip=true]{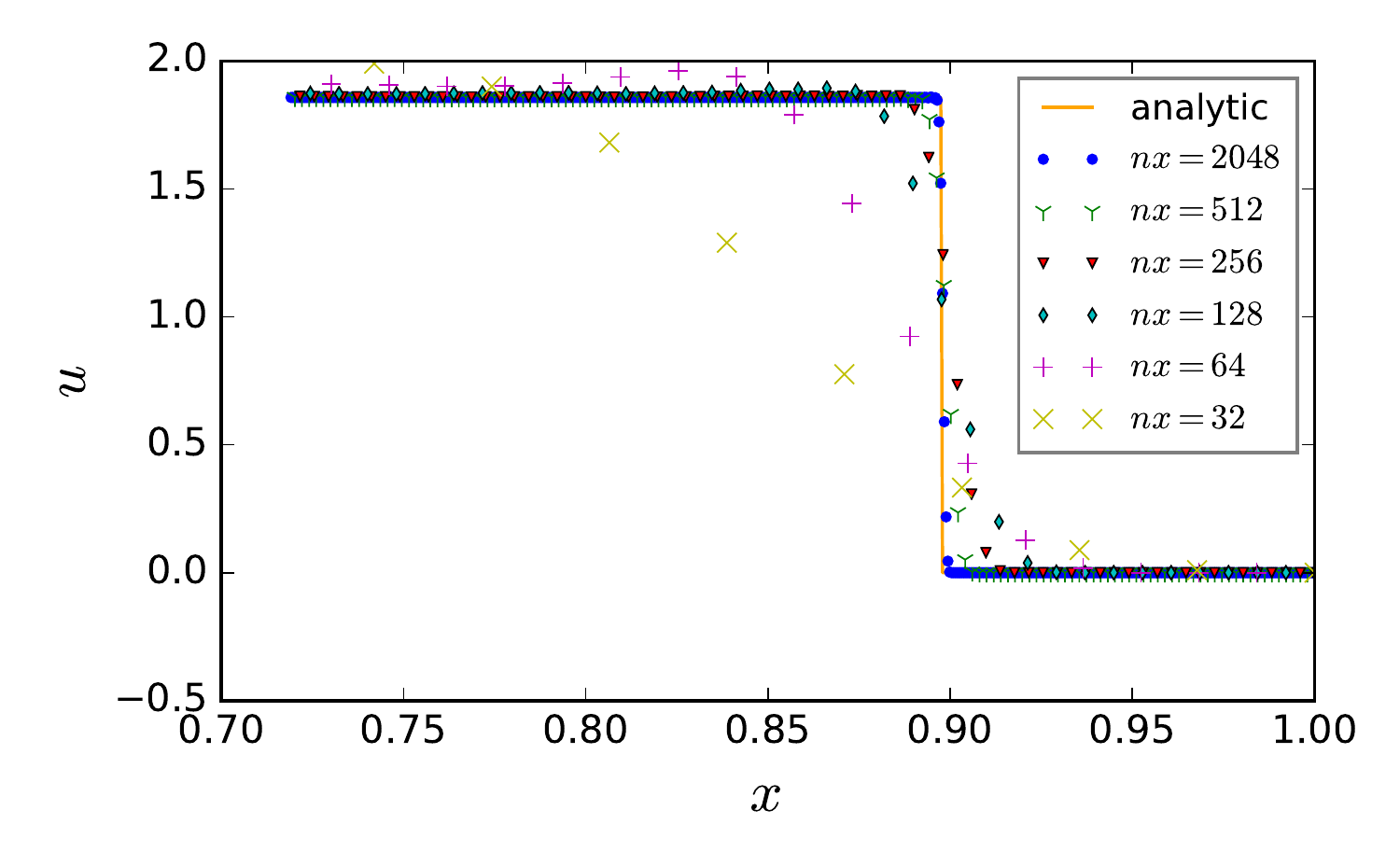}}}%
{\resizebox*{7.25cm}{!}{\includegraphics[trim=0.00cm 0.500cm 0.5cm 0.4cm,clip=true]{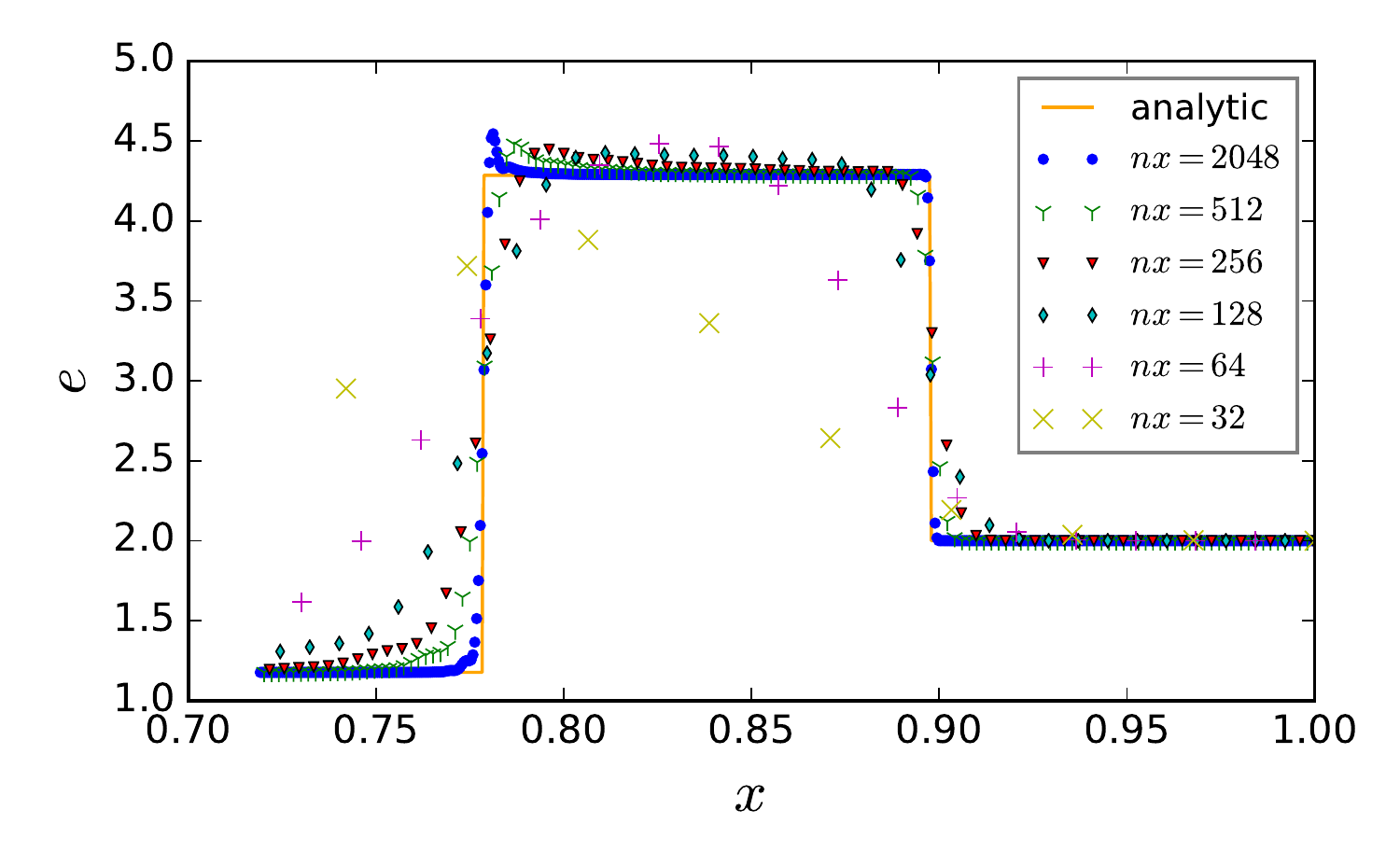}}}%
\caption{
The moderate shock tube test \citep{Sod78} detailed in figure\,\ref{fig:mod-sod} at $t=0.150$ with grid resolution between 32 and 2048. Zoomed-in shock front near $x=0.9$ applying shock-dependent artificial diffusion with diffusivity coefficients ${c_{\rm shock}}=[0,4,0.5]$, initial discontinuity profile smoothing scale $\ell=1.5\delta x$ and hyperdiffusion in the form of upwind differencing. Figures show gas density, pressure, velocity and internal energy. (Colour online)
}%
\label{fig:mod-res-comp}
\end{minipage}
\end{center}
\end{figure}
\begin{figure}
\begin{center}
\begin{minipage}{150mm}
{\resizebox*{7.25cm}{!}{\includegraphics[trim=0.40cm 1.7cm 0.5cm 0.4cm,clip=true]{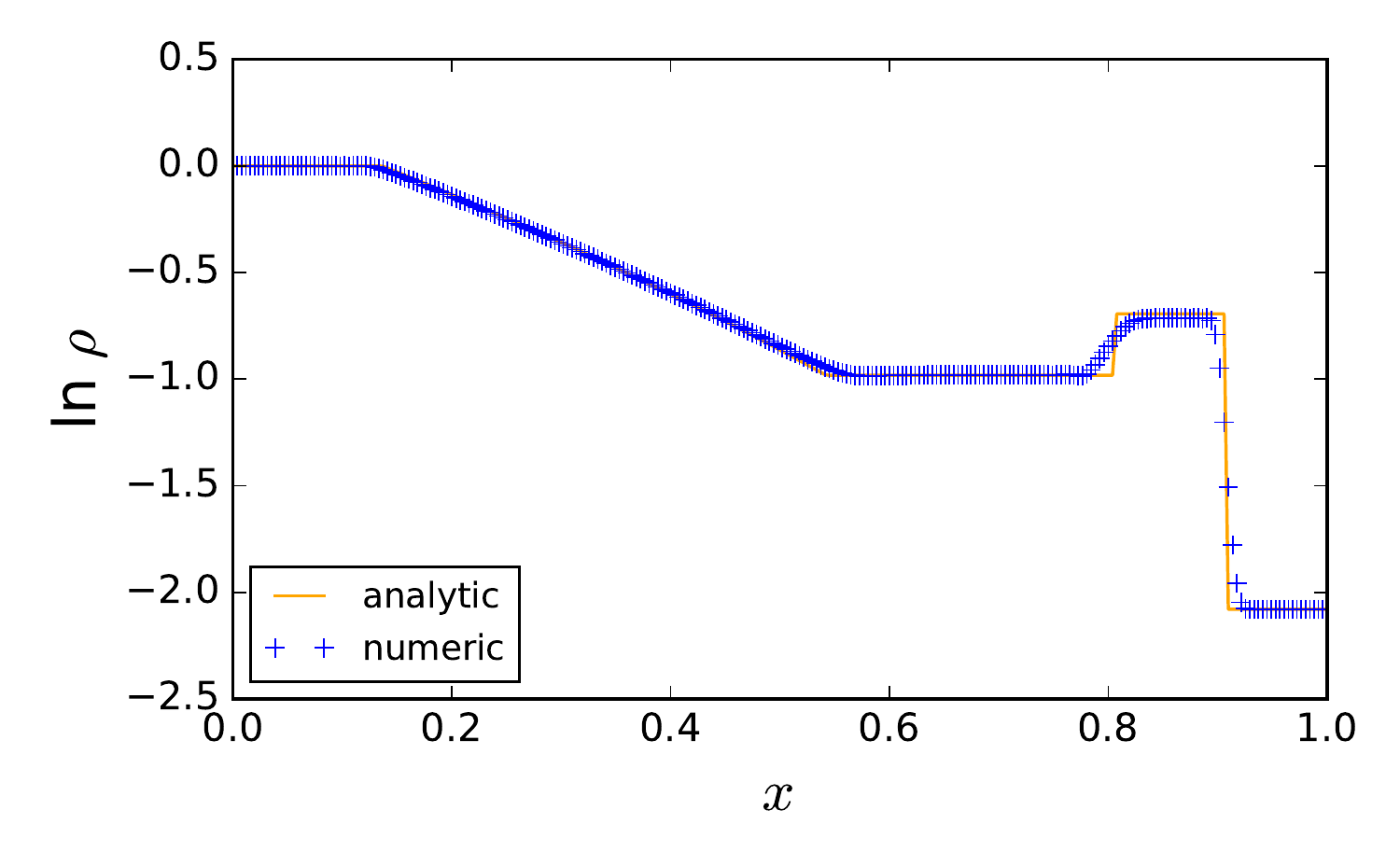}}}%
{\resizebox*{7.25cm}{!}{\includegraphics[trim=0.20cm 1.7cm 0.5cm 0.4cm,clip=true]{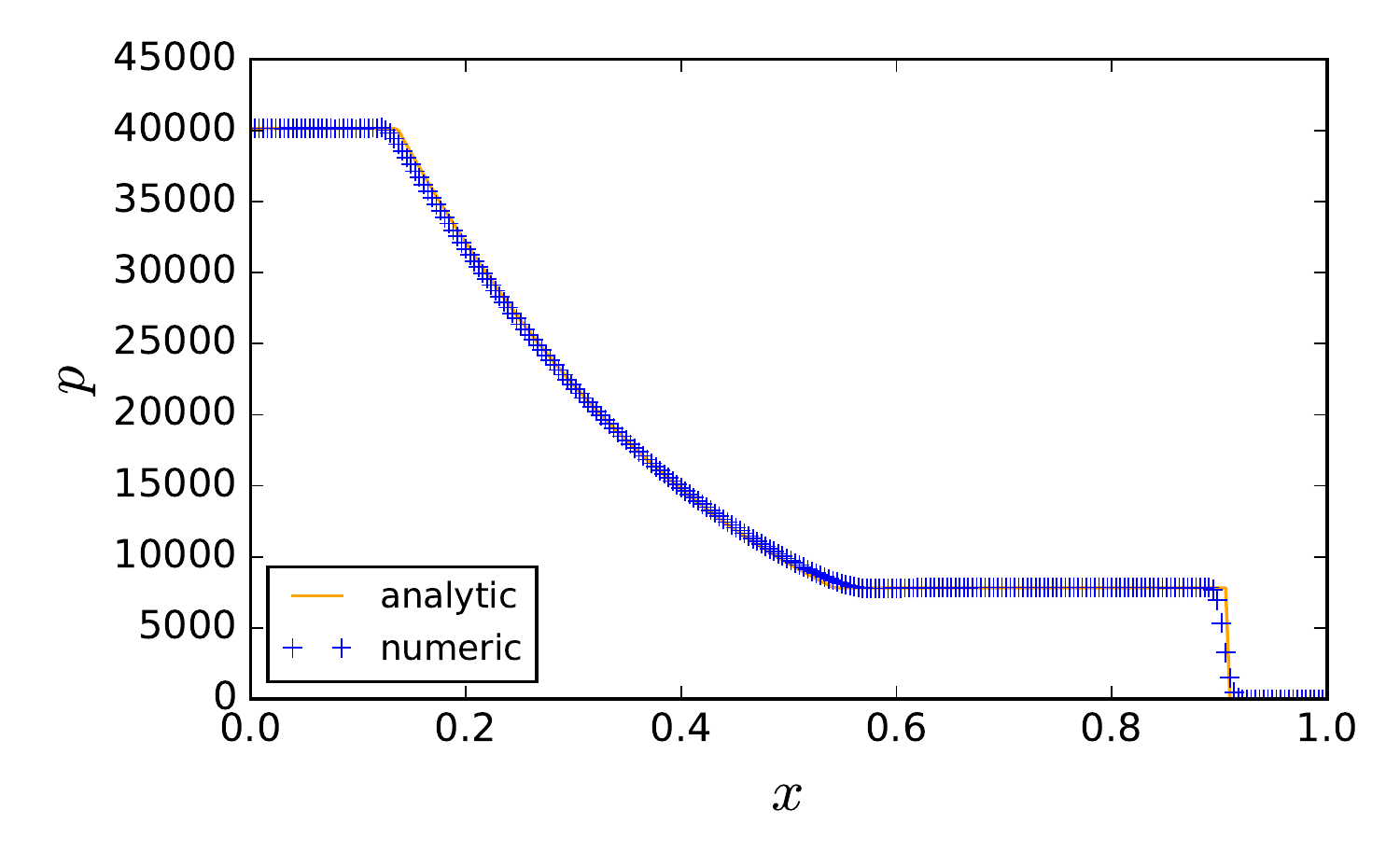}}}\\%
{\resizebox*{7.25cm}{!}{\includegraphics[trim=0.10cm 0.5cm 0.5cm 0.4cm,clip=true]{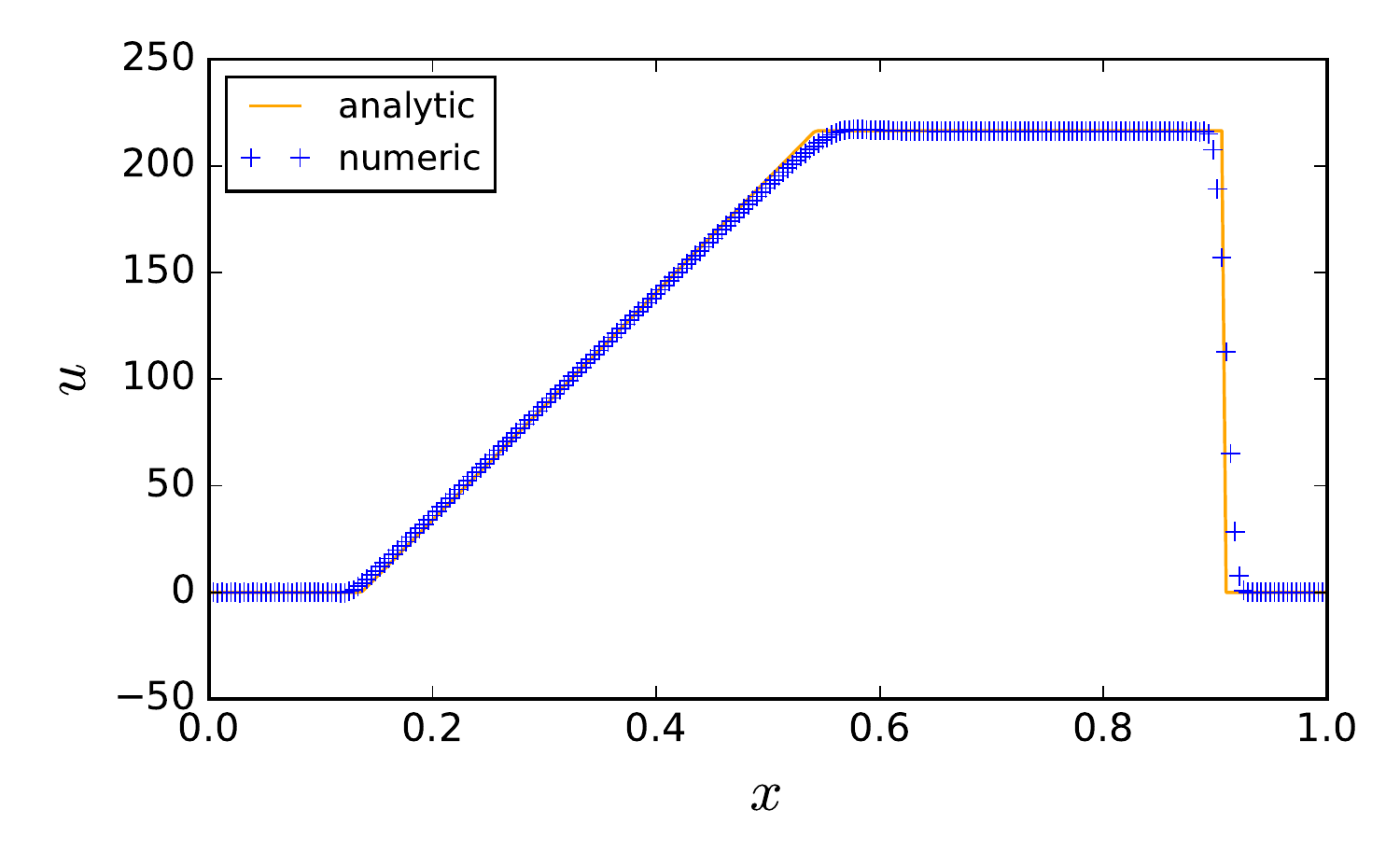}}}%
{\resizebox*{7.25cm}{!}{\includegraphics[trim=0.20cm 0.5cm 0.5cm 0.4cm,clip=true]{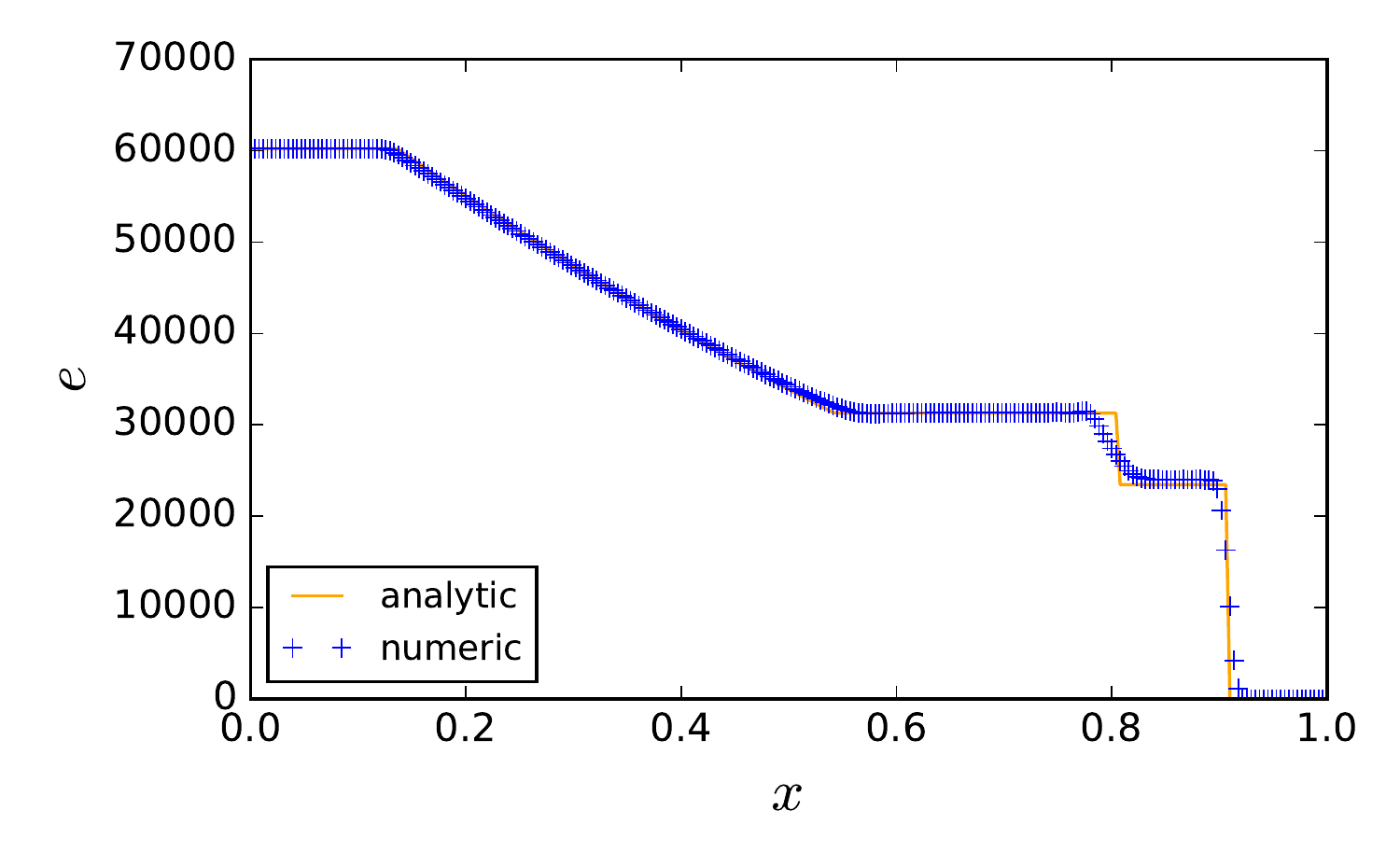}}}%
\caption{High Mach number shock tube test \citep{Sod78} with 256 grid points at {$t=0.0014$}, where the density, pressure on the left is initially set to $\rho,\,p=1.0,\,6.4\cdot10^{-10}$, and on the right $\rho,\,p=0.125,\,4.8\cdot10^{-15}$. Figures show gas density, pressure, velocity, internal energy, and the analytic solution (orange line). The diffusivity coefficients used are ${c_{\rm shock}}=[1,{6,2}]$. (Colour online)
}%
\label{fig:strong-sod}
\end{minipage}
\end{center}
\end{figure}
\begin{figure}
\begin{center}
\begin{minipage}{150mm}
{\resizebox*{7.25cm}{!}{\includegraphics[trim=0.20cm 1.665cm 0.5cm 0.4cm,clip=true]{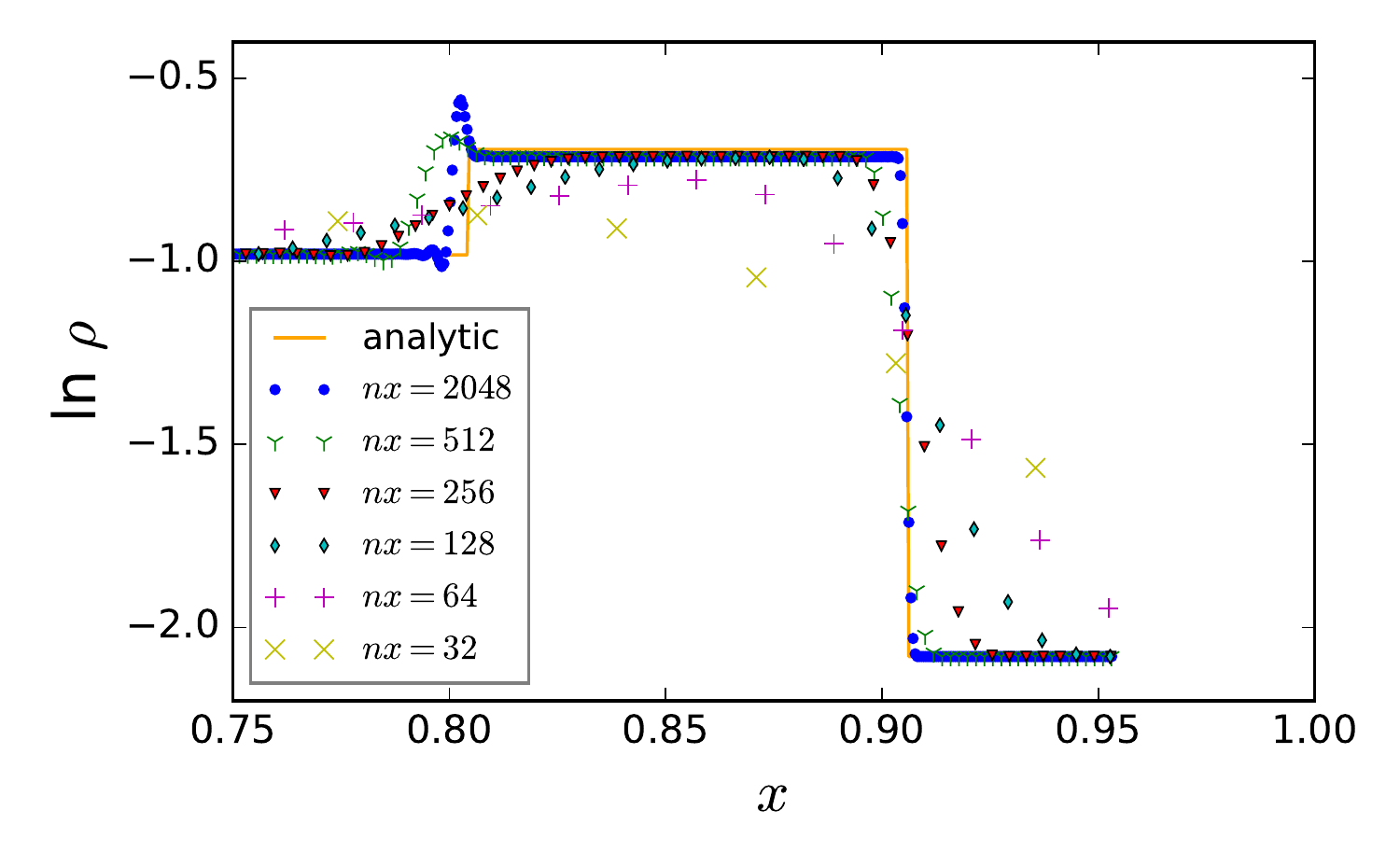}}}%
{\resizebox*{7.25cm}{!}{\includegraphics[trim=0.00cm 1.665cm 0.5cm 0.4cm,clip=true]{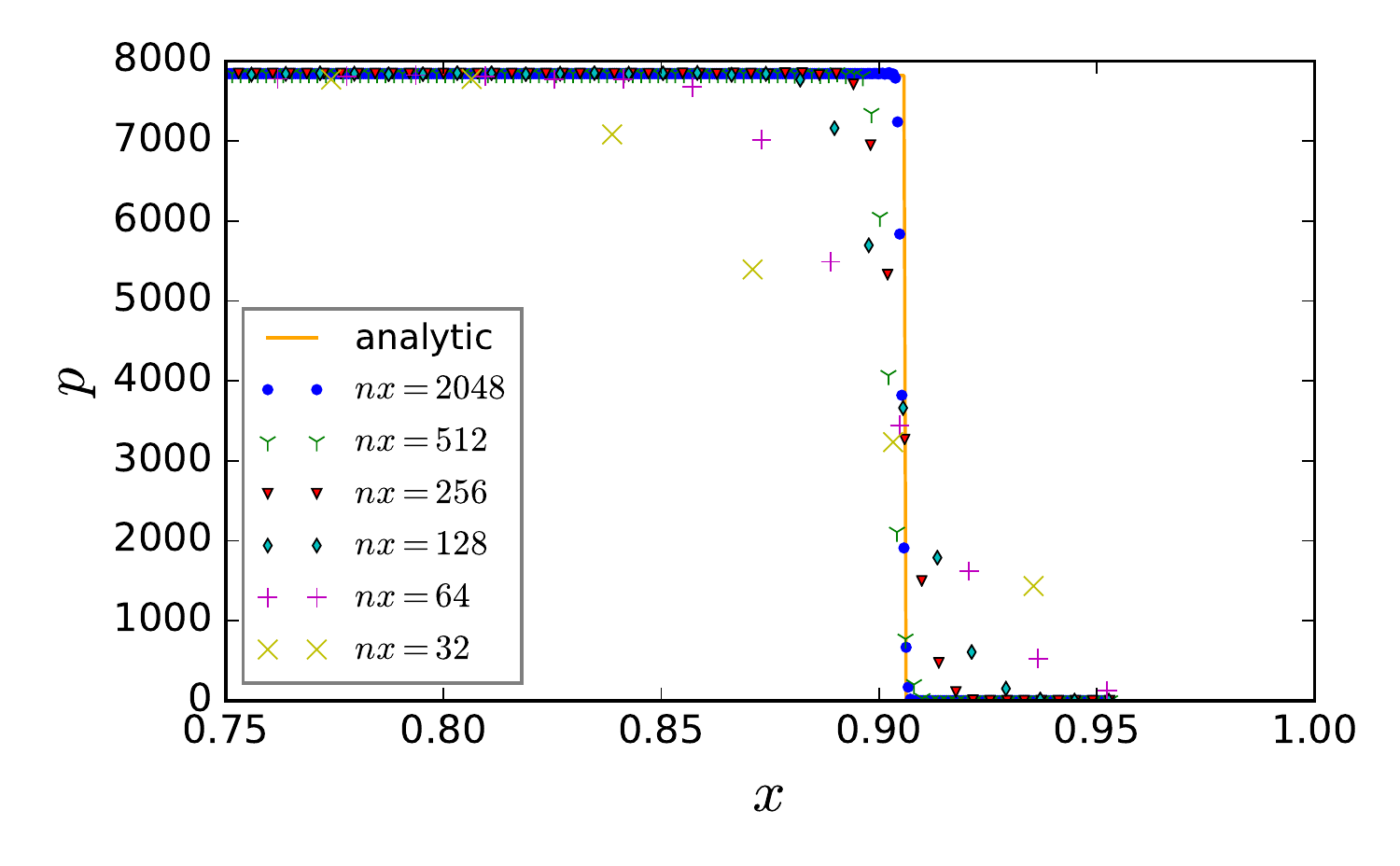}}}\\%
{\resizebox*{7.25cm}{!}{\includegraphics[trim=0.00cm 0.500cm 0.5cm 0.4cm,clip=true]{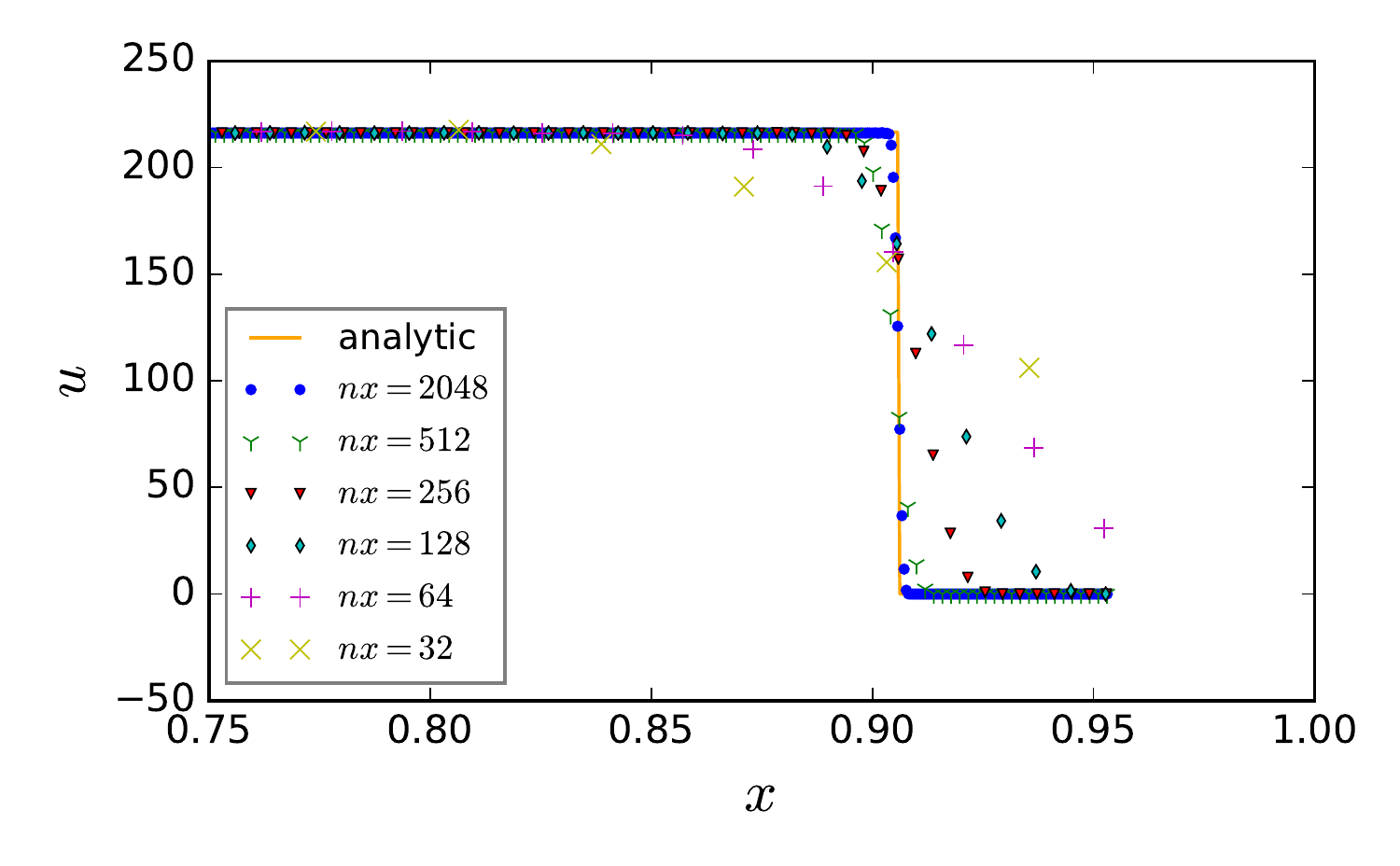}}}%
{\resizebox*{7.25cm}{!}{\includegraphics[trim=0.25cm 0.500cm 0.5cm 0.4cm,clip=true]{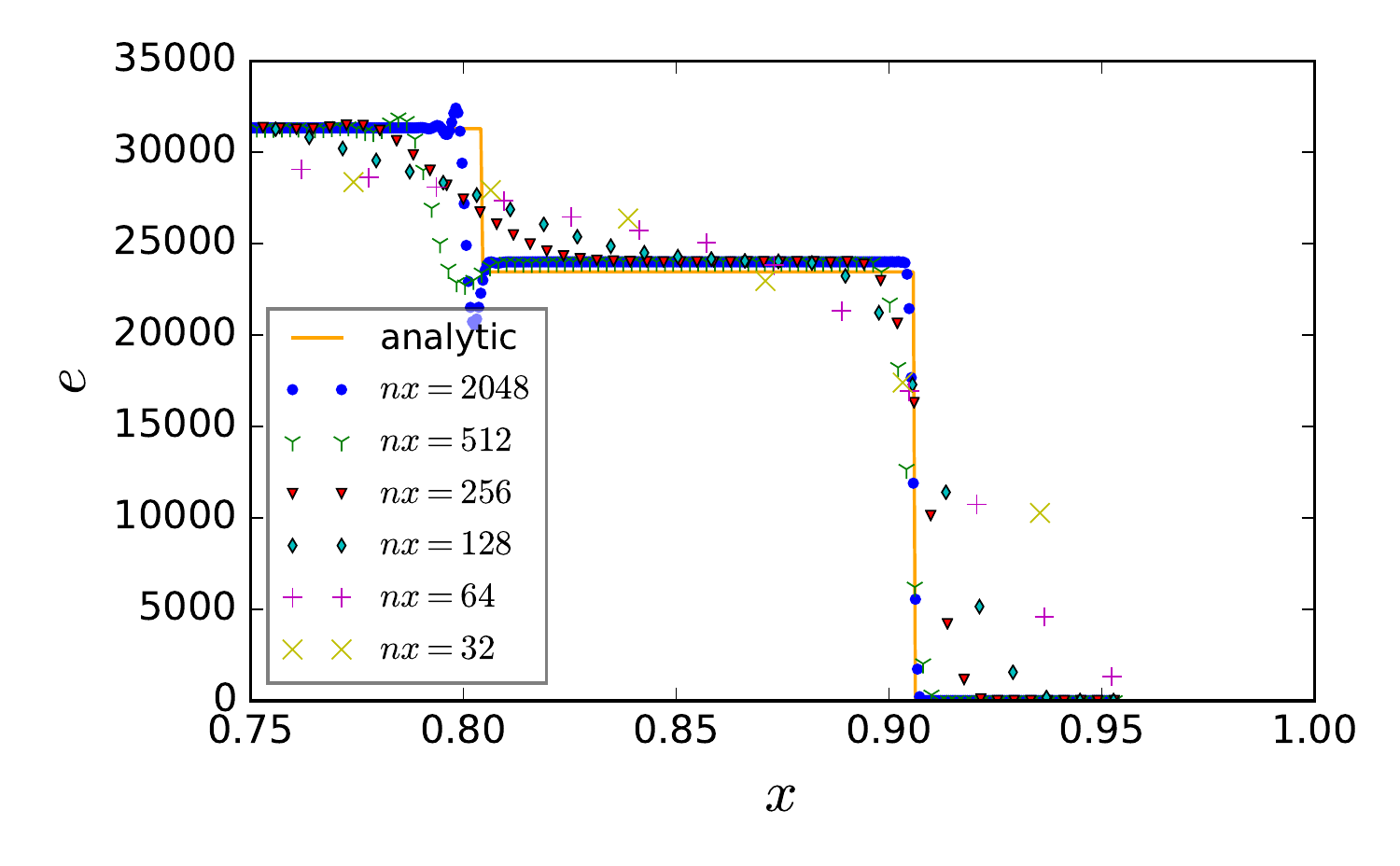}}}%
\caption{
Results for various grid resolutions {(32--2048 grid points)} for the high Mach number {($>100$)} Riemann shock-tube test shown in figure\,\ref{fig:strong-sod}. Figures show gas density, pressure, velocity and internal energy, and the analytic solution. The diffusivity coefficients used are ${c_{\rm shock}}=[1,{6,2}]$.  (Colour online)
}%
\label{fig:res-comp}
\end{minipage}
\end{center}
\end{figure}

In Figures\,\ref{fig:weak-sod} and \ref{fig:mod-sod}, with Mach numbers about 1 and 2.5, respectively, we see the main deviation from the analytic solution is smoothing at the transitions, particularly at the contact discontinuities between $x=0.6$ and 0.8, most evident in the plots for gas density and internal energy. With increasing resolution, which we show in figure\,\ref{fig:mod-res-comp}, the numerical solution asymptotically approaches the analytic solution. We consider grid sizes between 32 and 2048 for the moderate shock, and zoom in on the shock front and the contact discontinuity.

\subsection{High mach number shock}\label{sect:highMach}

The standard shock tube tests indicate that the code can adequately cope with weakly compressible flows. For shocks associated with SN driven turbulence, however, this is not sufficient. Simulations of the ISM commonly include minimal temperatures near 100\,K (sound speed 0.5\,km\,s$^{-1}$) and maximal velocities above 1000\,km\,s$^{-1}$. Even if we exclude the improbable extrema of Mach~2000 associated with SN explosions deep within molecular clouds, we regularly encounter Mach~100 shocks in these simulations, depending on the ambient temperature around each SN location.

In figure\,\ref{fig:strong-sod} we show the results of a Riemann shock test exceeding Mach\,100. The adiabatic index is 5/3. The pressure discontinuity is about five orders of magnitude. As well as increasing the artificial viscosity and thermal diffusivity coefficients to $\nu_{\rm shock}={6.0}$, $\chi_{\rm shock}={2.0}$, we also include an artificial diffusion to the continuity equation, as described in Section~\ref{sec:mass-diff}. In this example we set $D_{\rm shock}=1.0$.

In the density profile of figure\,\ref{fig:strong-sod} there is a small overshoot in energy behind the shock, accompanied by some numerical oscillation at the contact discontinuity. Whether the spikes or dips are the larger depends on the level of smoothing relative to the strength of the discontinuity profile. As mentioned earlier, we have little control of the structure of the shock injections in the turbulent ISM environment, so we expect some such artifacts to be present, but choose coefficients to optimally dampen such oscillations. The post-shock density does not quite reach the analytic value.

We consider the effect of resolution in figure\,\ref{fig:res-comp}, by zooming in on the shock front at this same moment in its evolution. For resolution below 128 grid points the numerical solution is a poor approximation of the analytic solution. We see that the post-shock density is slightly lower than the analytic solution, with correspondingly higher energy. We shall show in section\,\ref{sect:Dcorr} that this asymptotic disparity with the analytic result is due to the mass diffusion term. Apart from the smoothed profile at the shock front ($x\simeq0.91$) the greatest numerical error arises for density and energy at the contact discontinuity ($x\simeq0.81$), where a further step function evolves. As the resolution increases, the distribution converges to a profile more closely aligned with the analytical solution. Even at the contact discontinuity the errors mostly reduce, except for enhanced extrema nearest the discontinuity. Further investigation is required to address this latter detail.

\begin{figure}
\begin{center}
\begin{minipage}{150mm}
\begin{center}
{\resizebox*{9.25cm}{!}{\includegraphics[trim=0.35cm 0.45cm 0.2cm 0.3cm,clip=true]{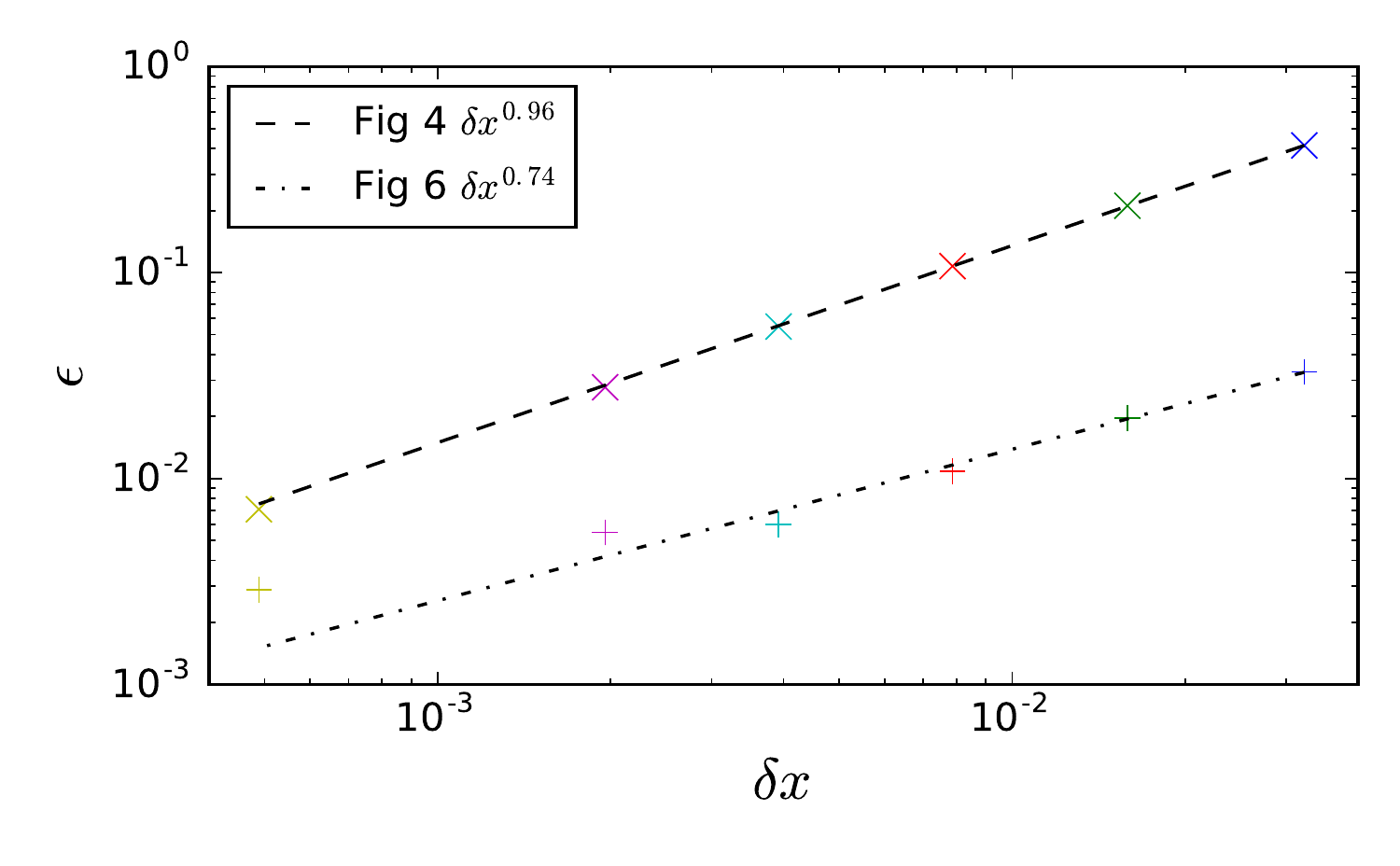}}}%
\caption{
Convergence study for the moderate shock tube test shown in figure\,\ref{fig:mod-res-comp} and the strong shock tube test shown in figure\,\ref{fig:res-comp}. Fit values are shown in legend.  Points with same color were run at same resolution in the two different models. (Colour online)
}%
\label{fig:errors}
\end{center}
\end{minipage}
\end{center}
\end{figure}

To test the convergence of the scheme, we compare for each resolution the $L_1$ error norm, given by \citep[see, e.g.,][]{SN92}
\begin{equation}\label{eq:norm}
\epsilon = \left( 
\sum_{i=1}^N \bigl|q_i-\tilde{q}_i\bigr|
\right)\bigg/N,
\end{equation} 
with $N,q_i$ and $\tilde{q}_i$ denoting the number of gridpoints, the numerical solution and analytical solution, respectively. The errors are shown in figure~\ref{fig:errors} for the moderate shock-tube test, figure\,\ref{fig:mod-res-comp}, and the high Mach number shock-tube resolution comparisons displayed in figure\,\ref{fig:res-comp}. For the moderate solution we find the convergence rate $\epsilon \propto \delta x^{0.96}$, while for the high Mach number solution this reduces to $\delta x^{0.74}$. So the higher order accuracy of the Pencil Code is restricted to first order accuracy for the shock handling. However, this is very localised and the modelling of the turbulence, for which these methods are intended, still mainly benefits from the higher order capabilities of the code.

\subsection{Dependence on diffusivity and mass diffusion 
correction\label{sect:Dcorr}}

The analytic solutions to the weak and moderate shock tube tests are reasonably satisfied with modest artificial viscosity and artificial thermal diffusivity, without any requirement to introduce artificial mass diffusion to the continuity equation. First we shall consider the effects of the two former applications of diffusivity on the numerical solution and then we shall discuss the motivation and consequences of adopting mass diffusion.

\begin{figure}
\begin{center}
\begin{minipage}{150mm}
{\resizebox*{7.25cm}{!}{\includegraphics[trim=0.20cm 1.665cm 0.5cm 0.4cm,clip=true]{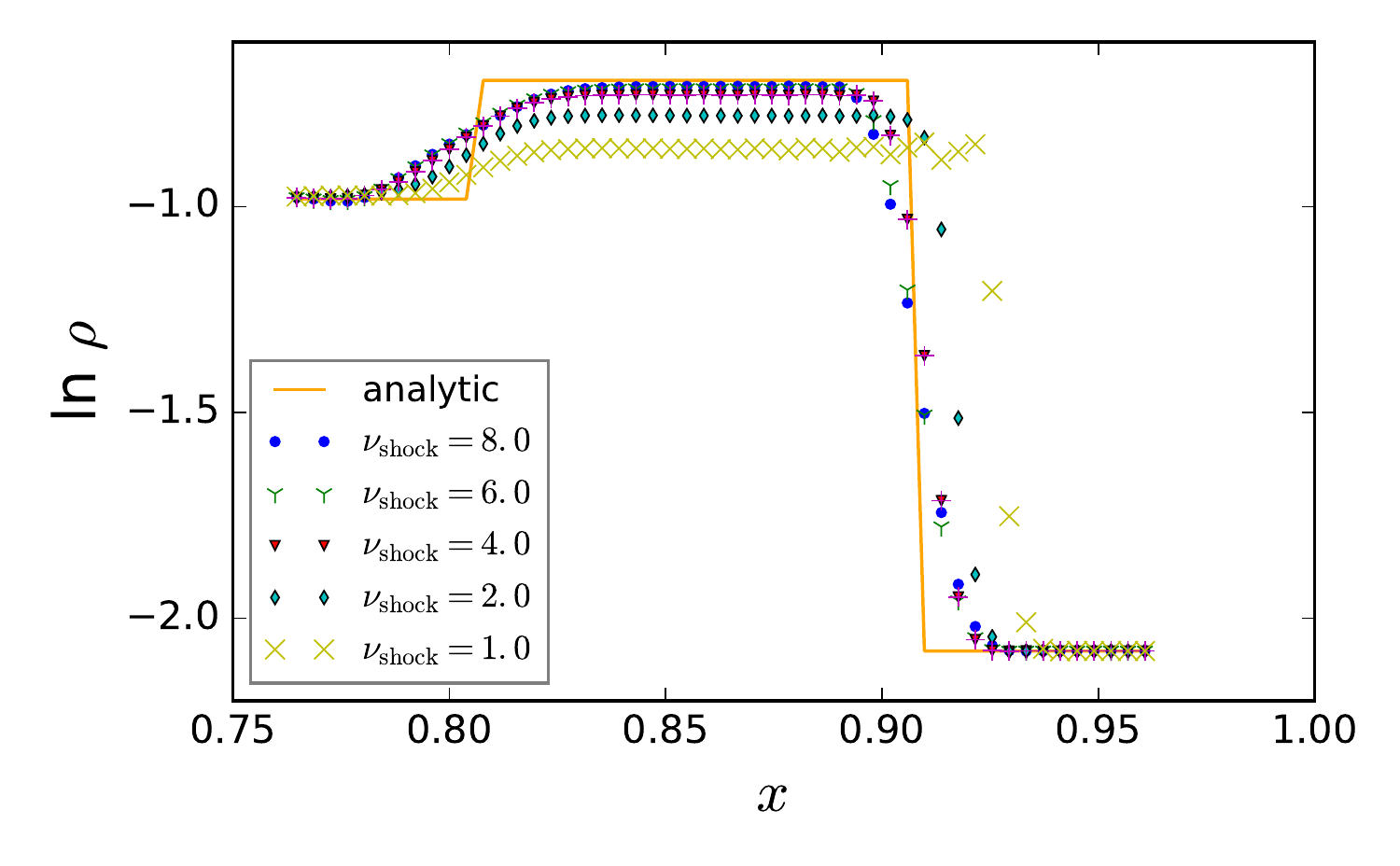}}}%
{\resizebox*{7.25cm}{!}{\includegraphics[trim=0.00cm 1.665cm 0.5cm 0.4cm,clip=true]{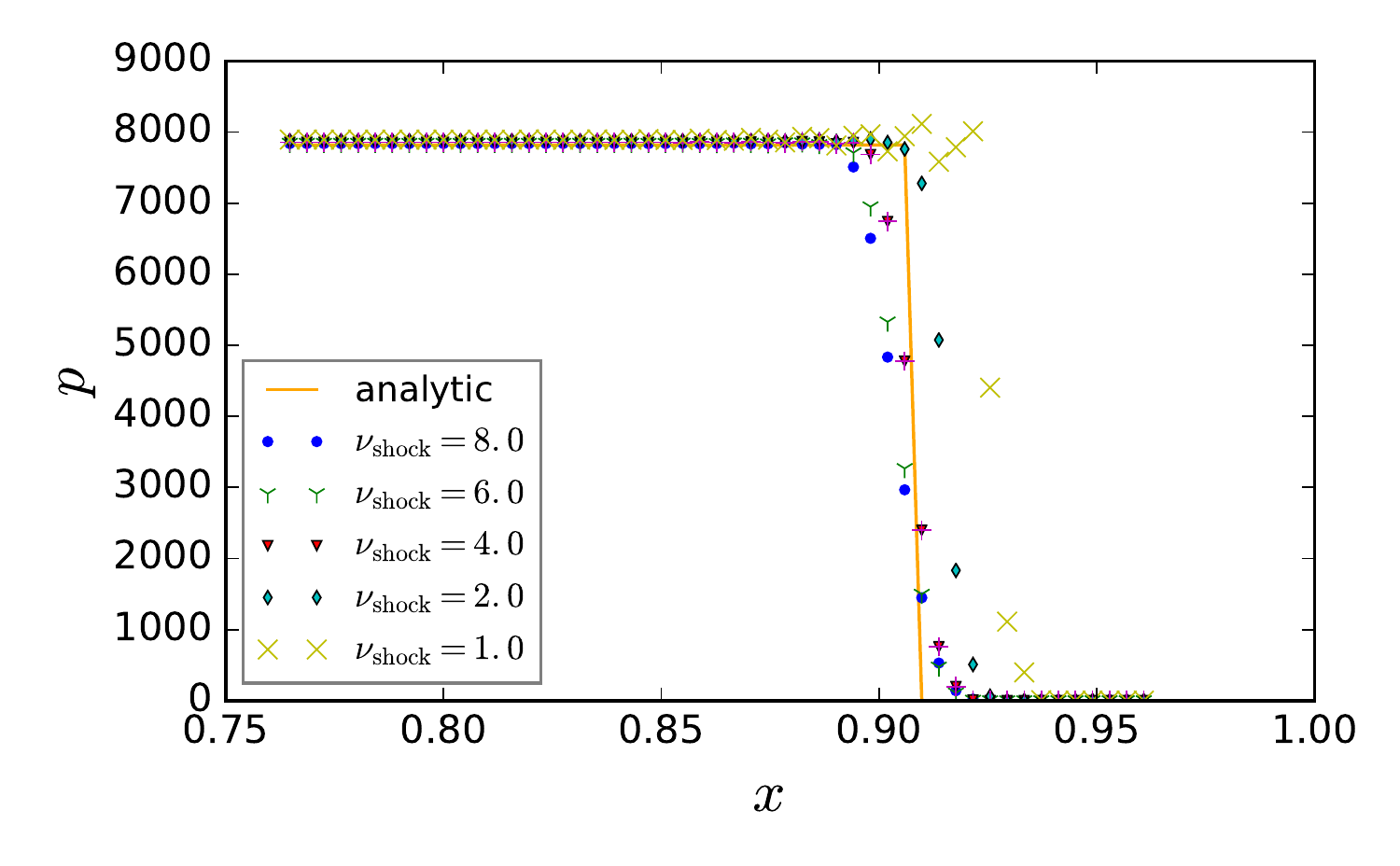}}}\\%
{\resizebox*{7.25cm}{!}{\includegraphics[trim=0.00cm 0.500cm 0.5cm 0.4cm,clip=true]{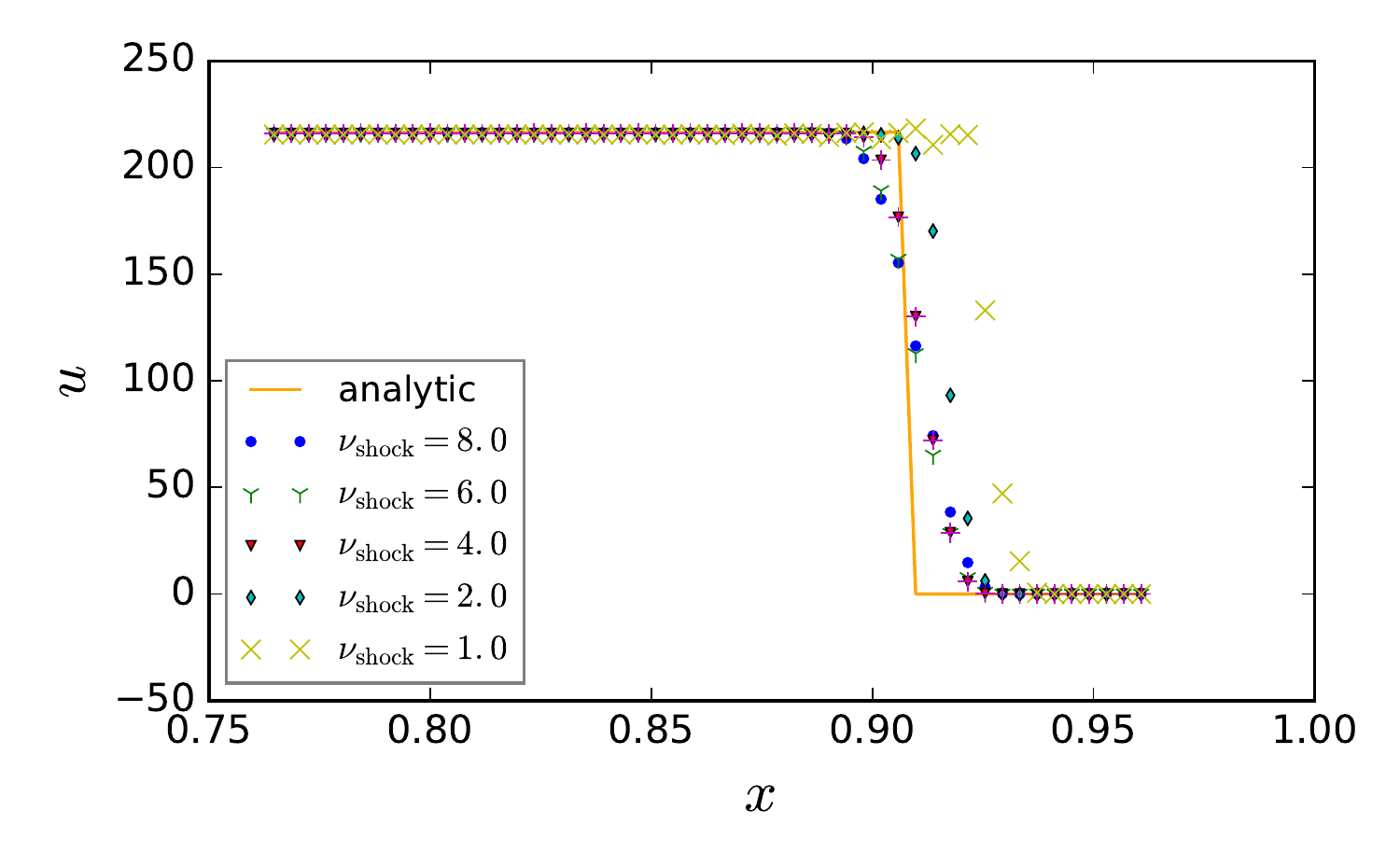}}}%
{\resizebox*{7.25cm}{!}{\includegraphics[trim=0.25cm 0.500cm 0.5cm 0.4cm,clip=true]{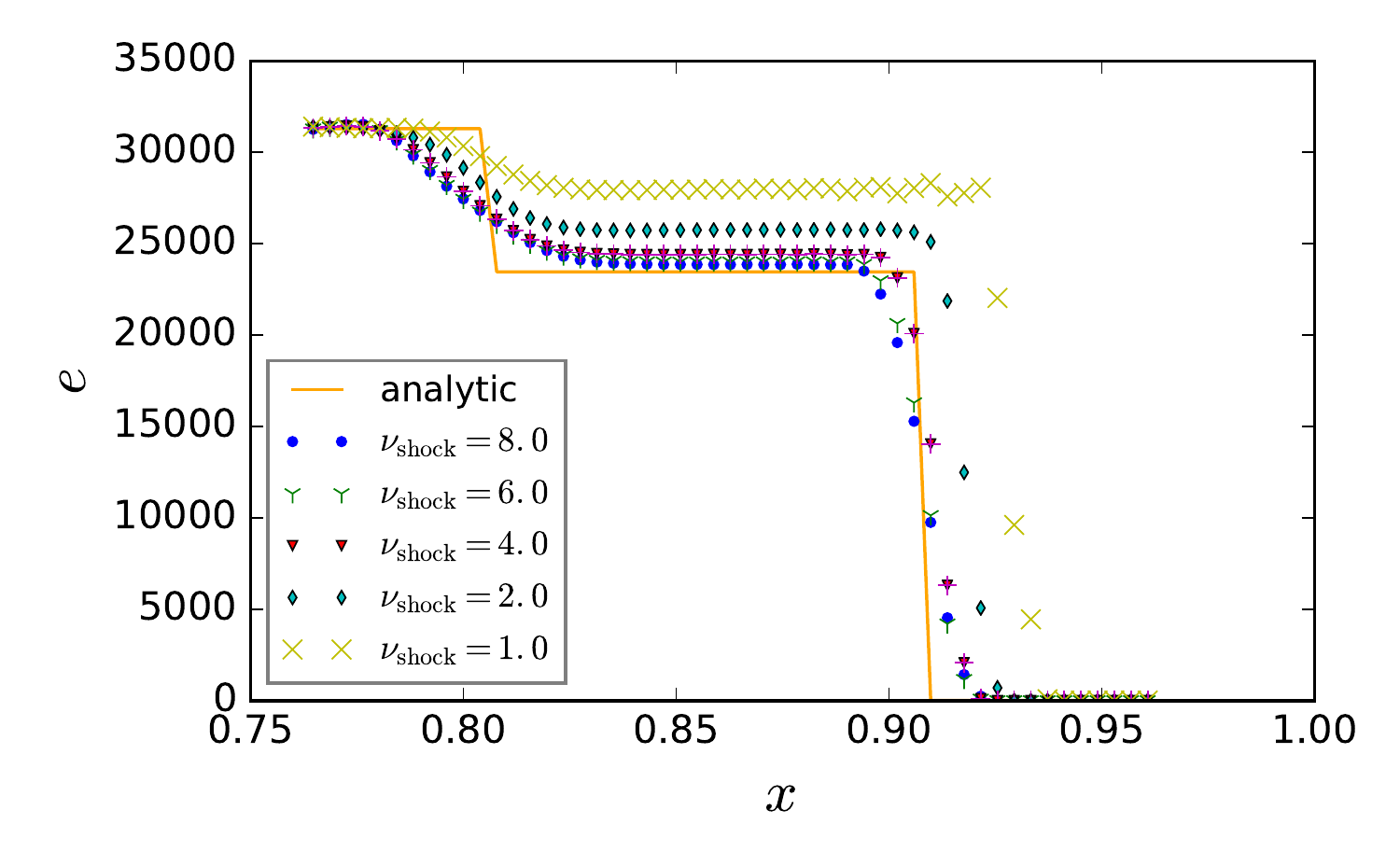}}}%
\caption{
Variation of solution for high Mach number shock tube (figure\,\ref{fig:strong-sod}) with varying shock viscosity coefficient $\nu_{\rm shock}\in[1,2,4,{6},8]$, compared to the analytic solution zoomed in near the shock front, at t = 0.0014 and 256 zone resolution. Figures show gas density, pressure, velocity, internal energy, and the analytic solution (orange line). The diffusivity coefficients are  ${c_{\rm shock}}=[1,{\nu_{\rm shock},2}]$. (Colour online)
}%
\label{fig:nu-comp}
\end{minipage}
\end{center}
\end{figure}

\begin{figure}
\begin{center}
\begin{minipage}{150mm}
{\resizebox*{7.25cm}{!}{\includegraphics[trim=0.20cm 1.665cm 0.5cm 0.4cm,clip=true]{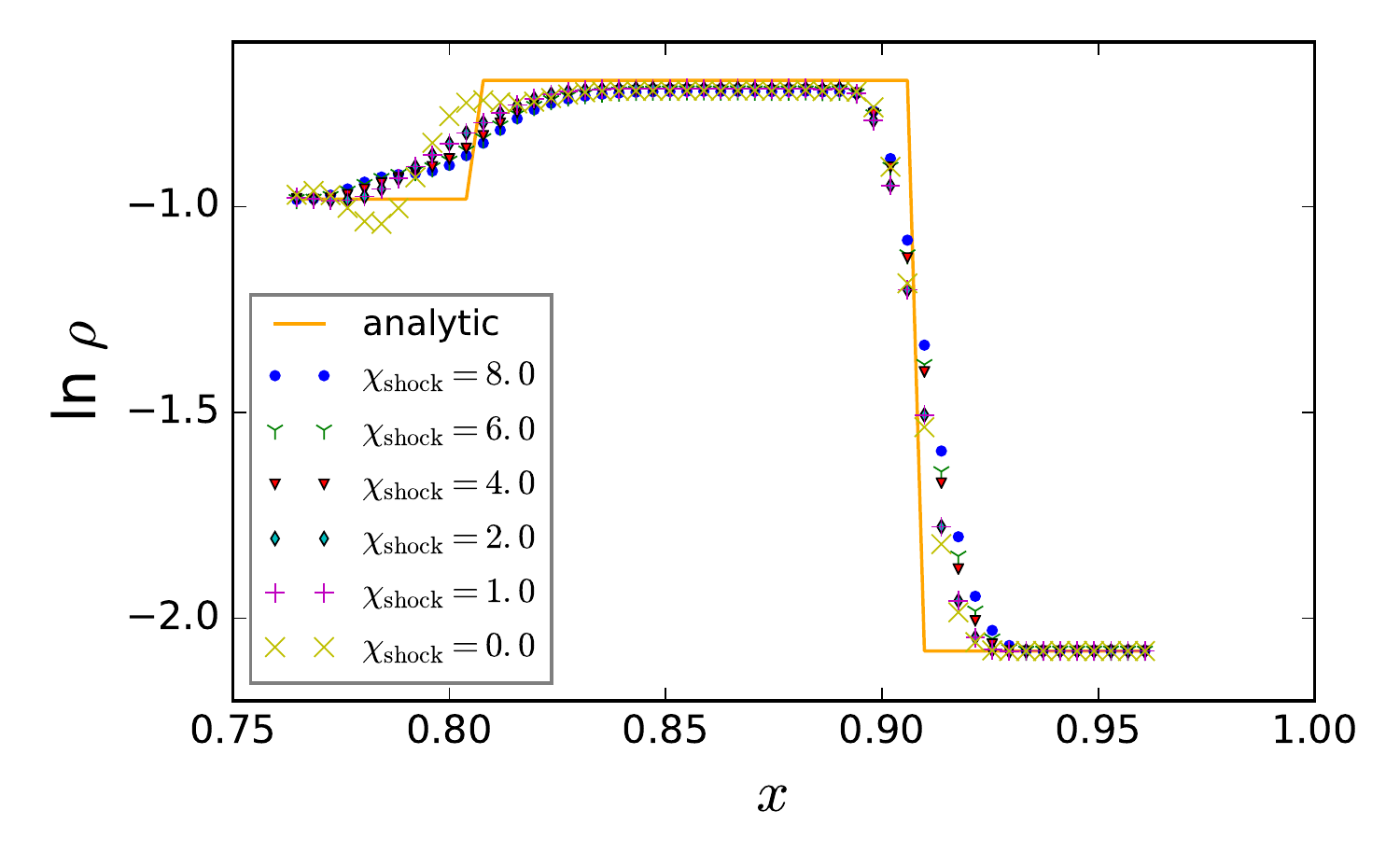}}}%
{\resizebox*{7.25cm}{!}{\includegraphics[trim=0.00cm 1.665cm 0.5cm 0.4cm,clip=true]{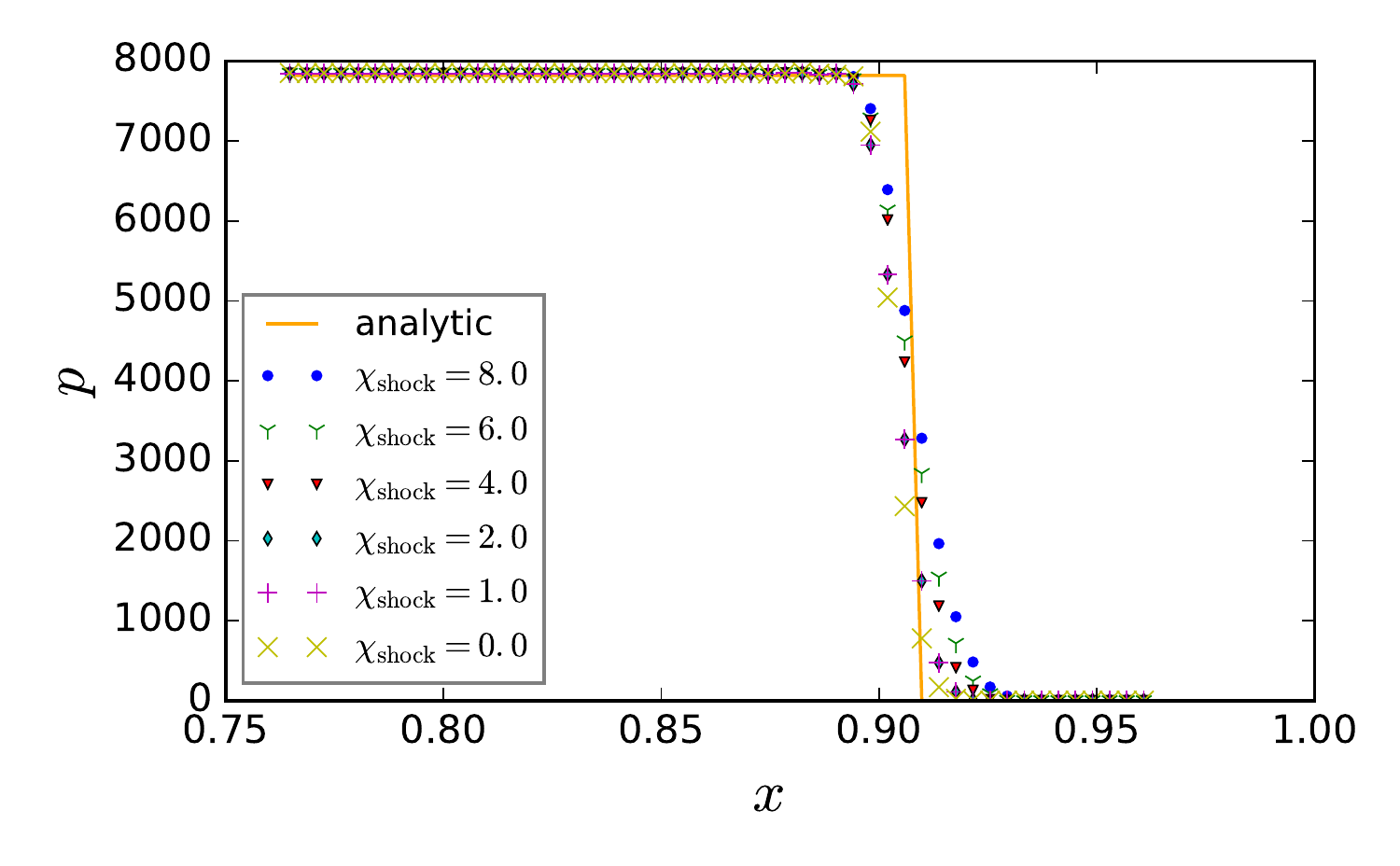}}}\\%
{\resizebox*{7.25cm}{!}{\includegraphics[trim=0.00cm 0.500cm 0.5cm 0.4cm,clip=true]{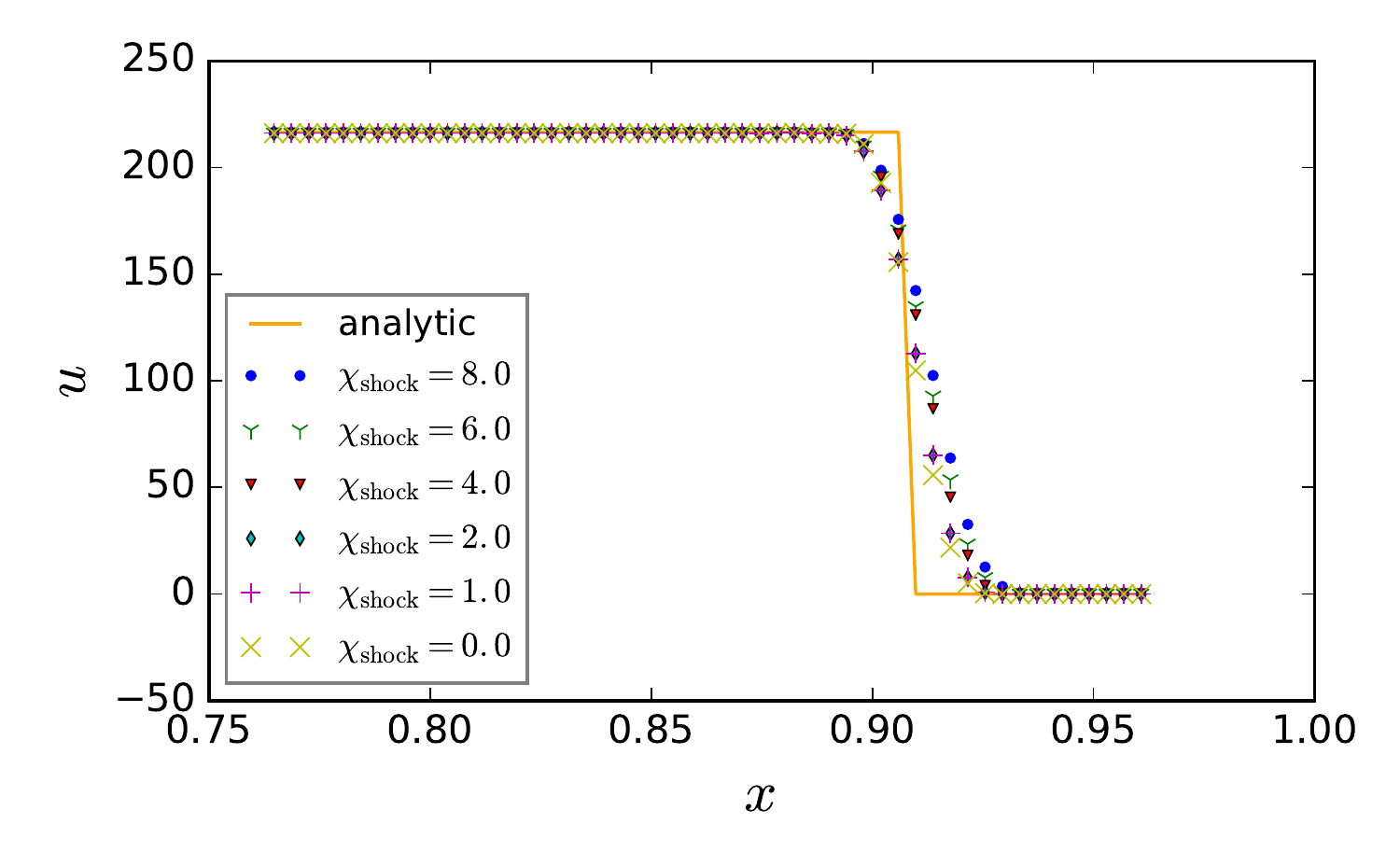}}}%
{\resizebox*{7.25cm}{!}{\includegraphics[trim=0.25cm 0.500cm 0.5cm 0.4cm,clip=true]{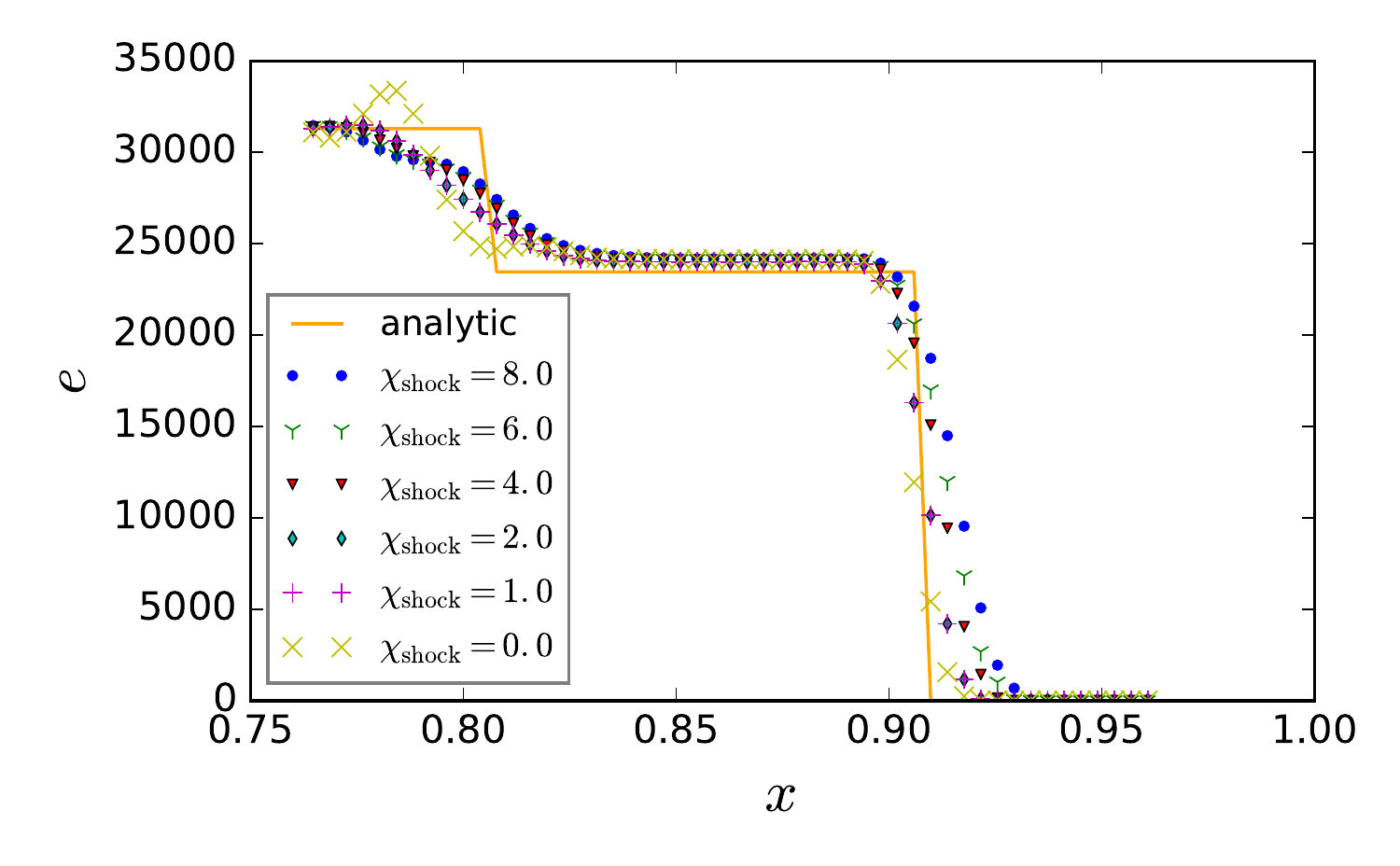}}}%
\caption{
Variation of solution for high Mach number shock tube (figure\,\ref{fig:strong-sod}) with varying shock thermal diffusivity coefficient $\chi_{\rm shock} \in [0,1,2,4,{6},8]$ at t = 0.0014 and 256 zone resolution. Figures show gas density, pressure, velocity, internal energy, and the analytic solution (orange line), zoomed in near the shock front. The diffusivity coefficients are ${c_{\rm shock}}=[1,{6,\chi_{\rm shock}}]$. (Colour online)
}%
\label{fig:chi-comp}
\end{minipage}
\end{center}
\end{figure}
 
When we consider the effect of the artificial viscosity for the strong shock tube test (Mach\,100), illustrated in figure\,\ref{fig:nu-comp}, we see that for $\nu_{\rm shock}=1.0$ an instability appears to overpressure the shock, spreading the faster shock front and reducing its density. As the viscosity coefficient increases the solution approaches the analytic value for the post-shock density and energy, although more smoothed at the corners.  The pressure and velocity shock fronts converge to the analytic position, although also slightly broadened, 
but as seen in the resolution comparisons of figure\,\ref{fig:res-comp}, the solution improves as resolution increases. As can be seen from $\nu_{\rm shock}=6$ and 8, the numerical solution does not continue to broaden, but approaches an asymptotic profile. Further increases do not significantly smooth the shock profiles, but can reduce the timestep and induce nonlinear instabilities due to increased viscous forces or viscous heating, as discussed in section\,\ref{sect:timestep}. These solutions adequately suppress the wiggles in the wake of the shock.

Now consider the dependence of the solution on the shock thermal conductivity $\chi_{\rm shock}$ shown in figure\,\ref{fig:chi-comp}. In all profiles we find weak dependence on the strength of $\chi_{\rm shock}$. The artificial thermal diffusion could reasonably be neglected, except for the density and energy extrema in the wake of the contact discontinuity.

Compared to the dependence on artificial viscosity, increases in the shock thermal diffusion beyond  $\chi_{\rm shock}=4$ or 8 appear to cause less additional diffusion. These values do not appear strongly advantageous over values of 1 and 2, yet we know from experiments with SN turbulence and the higher resolution snowplough tests reported in section\,\ref{sect:snowplough} that the slightly larger oscillations illustrated in the density profile for $\chi_{\rm shock}=0$, are sufficient to lead to numerical instability. We recommend a nonzero value of $\chi_{\rm shock}$, for moderately compressible turbulence $\chi_{\rm shock}\gtrsim 1.0$ and for highly compressible turbulence  $\chi_{\rm shock}\gtrsim {2}.0$. Higher values should be avoided to limit diffusion, particularly in energy.

\begin{figure}
\begin{center}
\begin{minipage}{150mm}
{\resizebox*{7.25cm}{!}{\includegraphics[trim=0.20cm 1.665cm 0.5cm 0.4cm,clip=true]{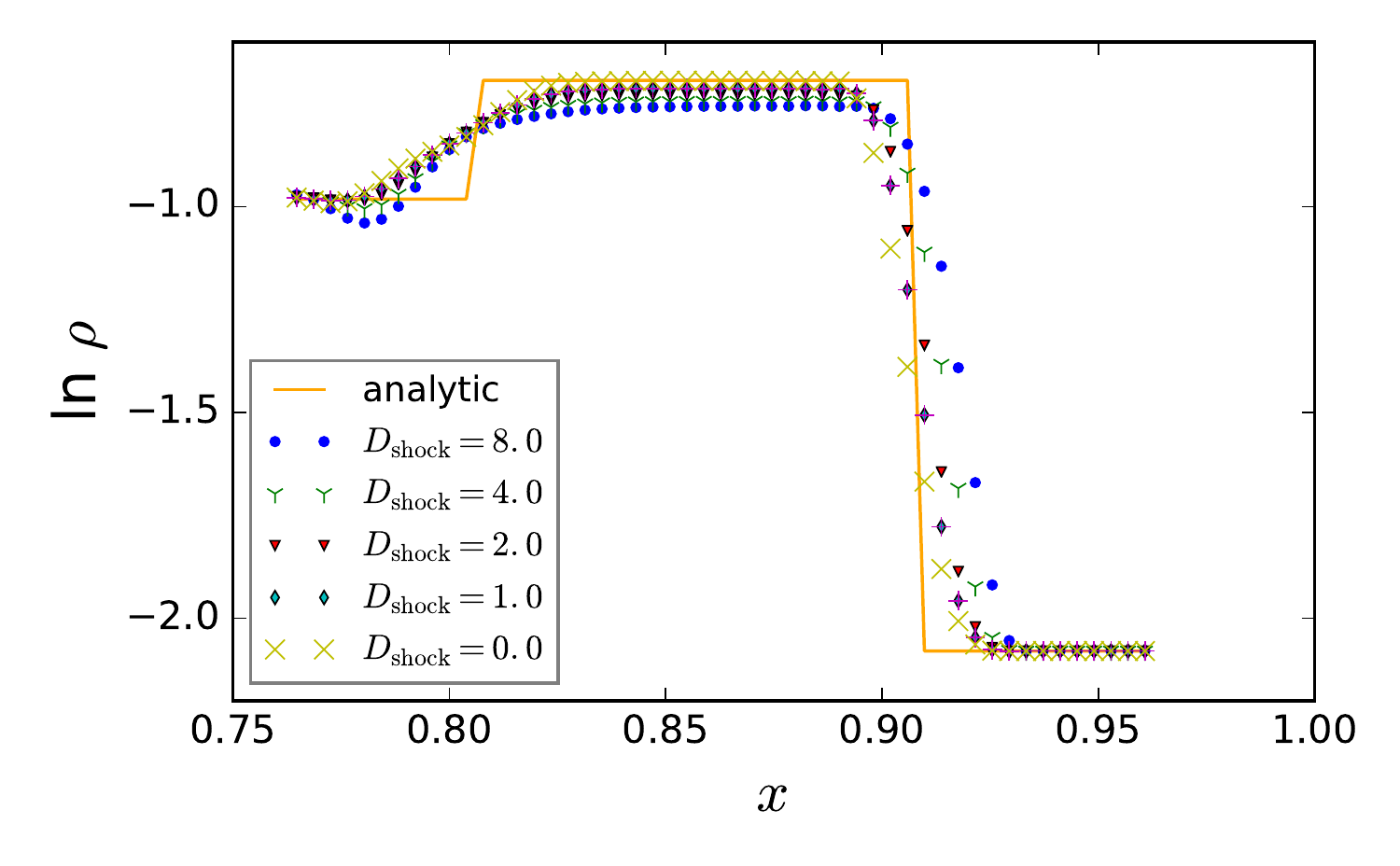}}}%
{\resizebox*{7.25cm}{!}{\includegraphics[trim=0.00cm 1.665cm 0.5cm 0.4cm,clip=true]{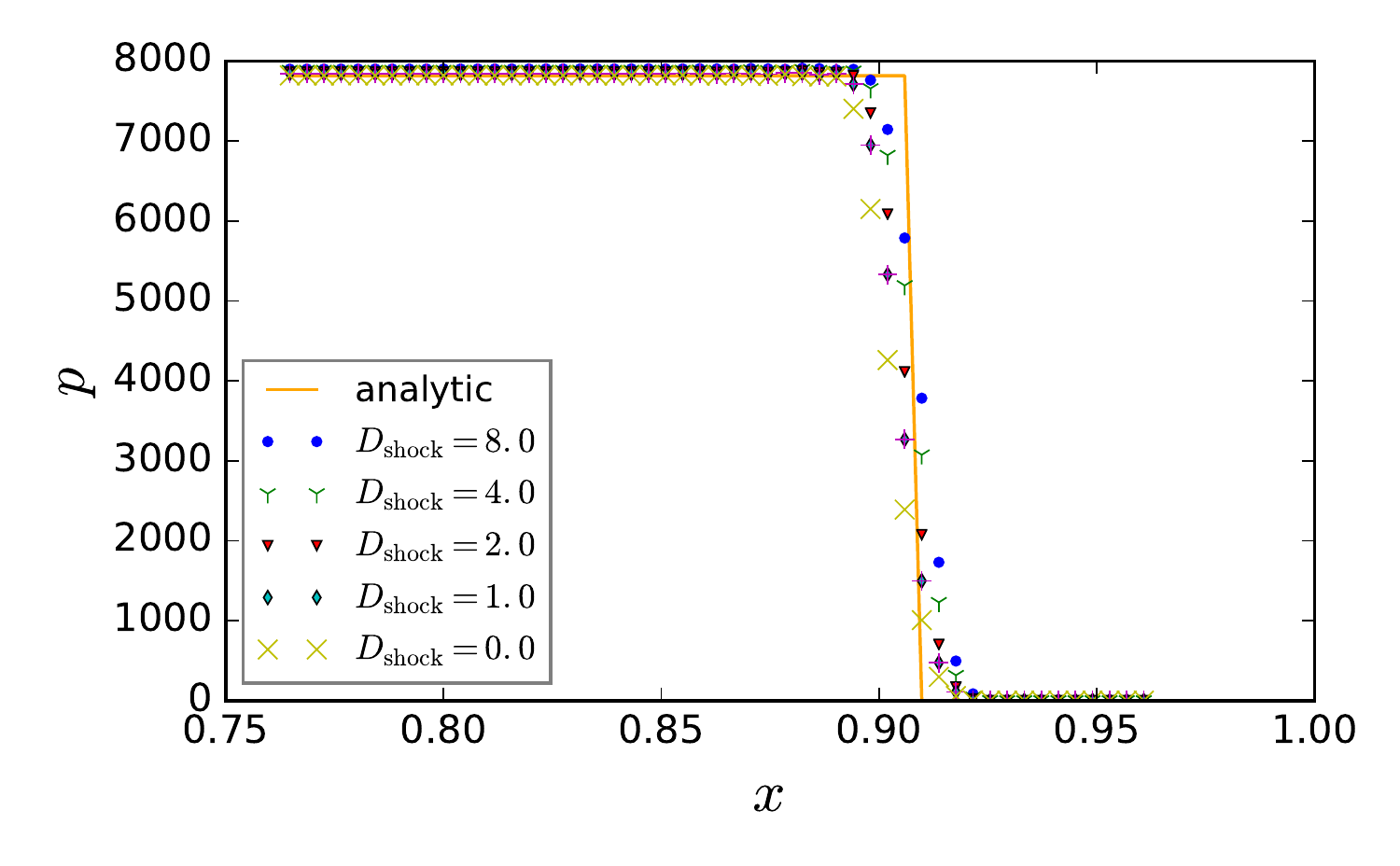}}}\\%
{\resizebox*{7.25cm}{!}{\includegraphics[trim=0.00cm 0.500cm 0.5cm 0.4cm,clip=true]{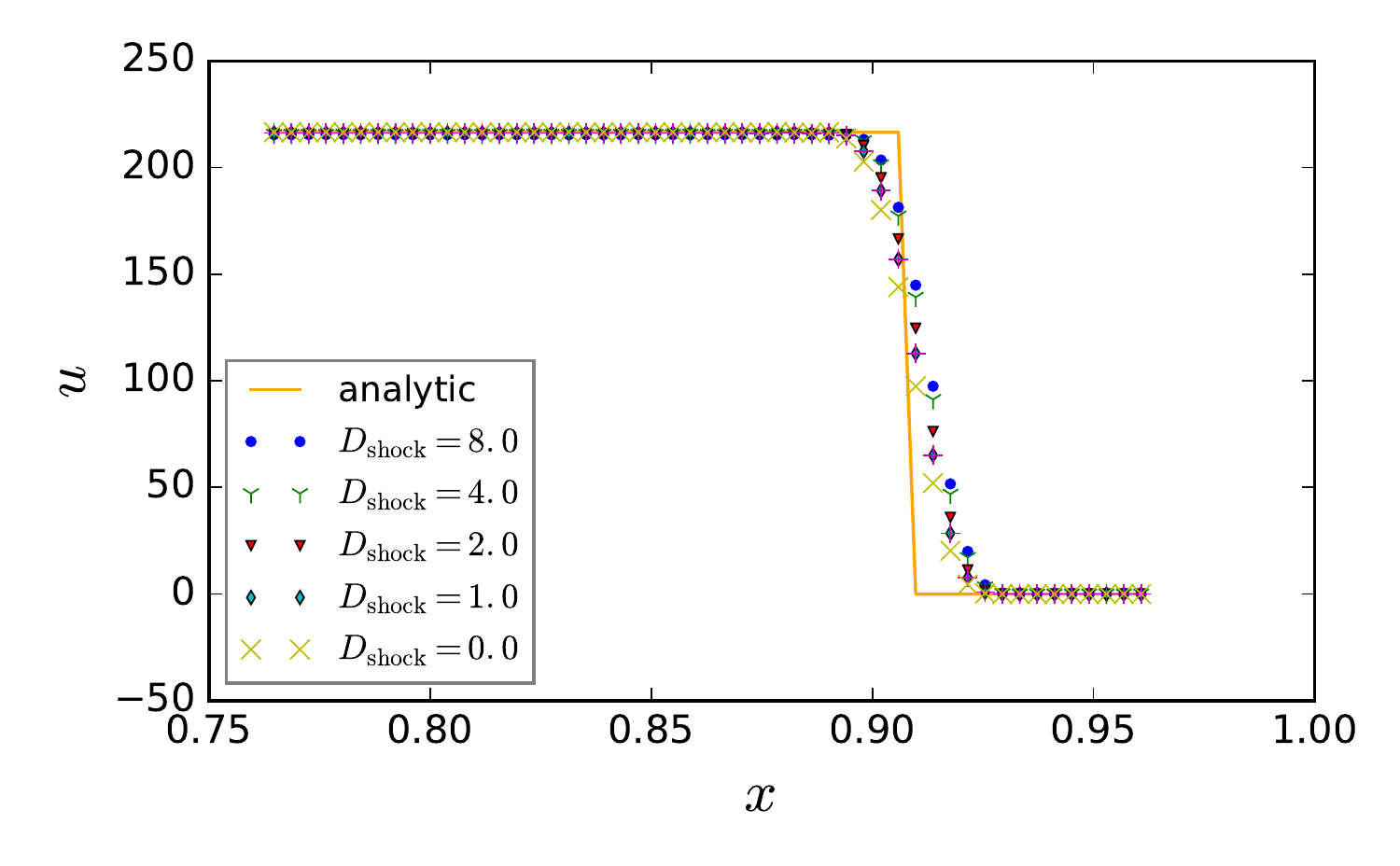}}}%
{\resizebox*{7.25cm}{!}{\includegraphics[trim=0.25cm 0.500cm 0.5cm 0.4cm,clip=true]{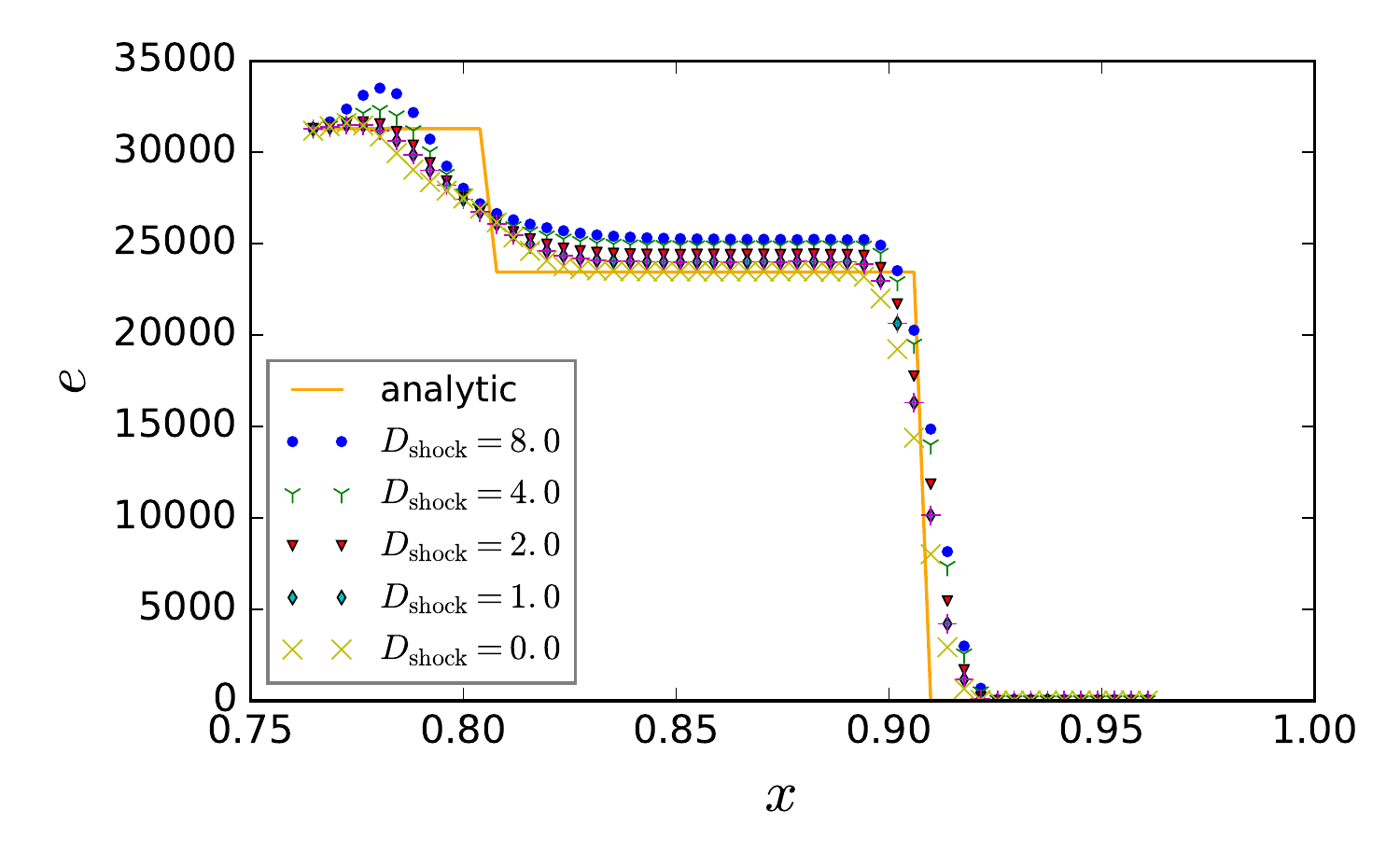}}}%
\caption{
Variation of solution for high Mach number shock tube (figure\,\ref{fig:strong-sod}) with varying shock mass diffusion coefficient $D_{\rm shock} \in [0,1,2,4,8]$ at t = 0.0014, and 256 zone resolution. Figures show gas density, pressure, velocity, internal energy, and the analytic solution zoomed in near the shock front. The diffusivity coefficients are  ${c_{\rm shock}}=[{D_{\rm shock}},{6,2}]$. (Colour online)
}%
\label{fig:D-comp}
\end{minipage}
\end{center}
\end{figure}

\begin{figure}
\begin{center}
\vspace{-0.8cm}
\begin{minipage}{150mm}
{\resizebox*{7.25cm}{!}{\includegraphics[trim= 0.12cm 1.7cm 0.5cm -1.5cm,clip=true]{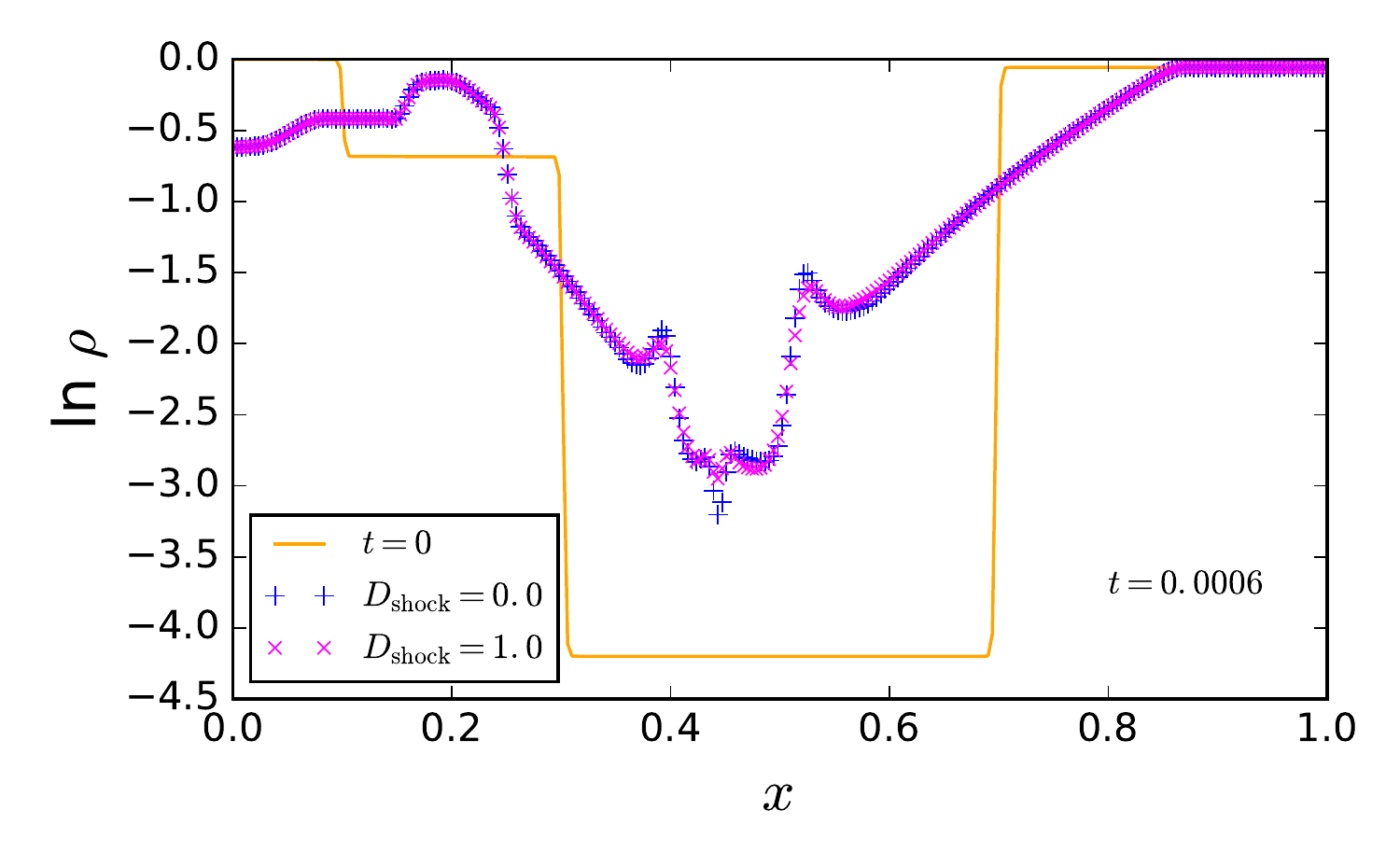}}}%
{\resizebox*{7.25cm}{!}{\includegraphics[trim=-0.25cm 1.7cm 0.5cm  0.4cm,clip=true]{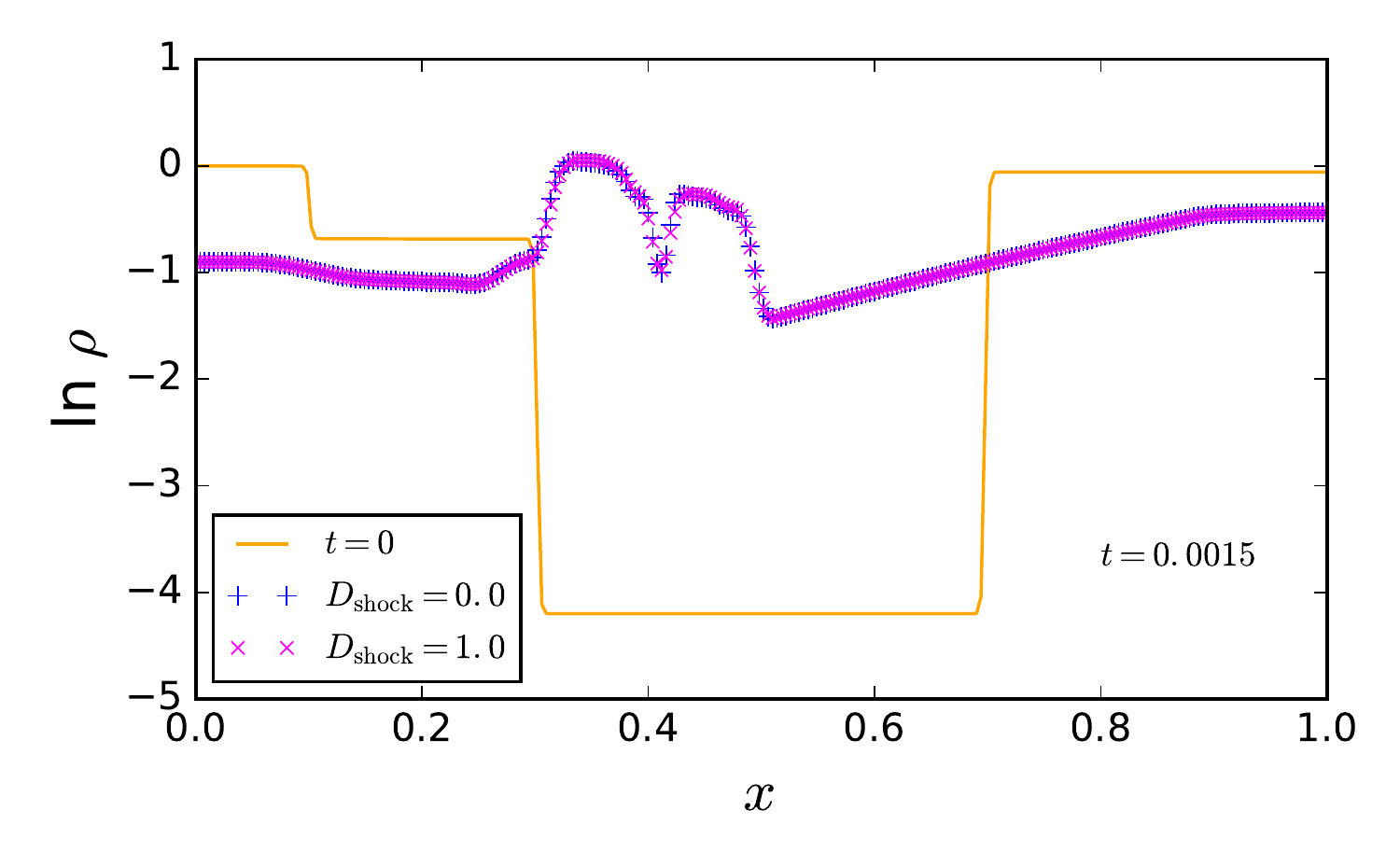}}}\\%
{\resizebox*{7.25cm}{!}{\includegraphics[trim= 0.15cm 1.7cm 0.5cm  0.4cm,clip=true]{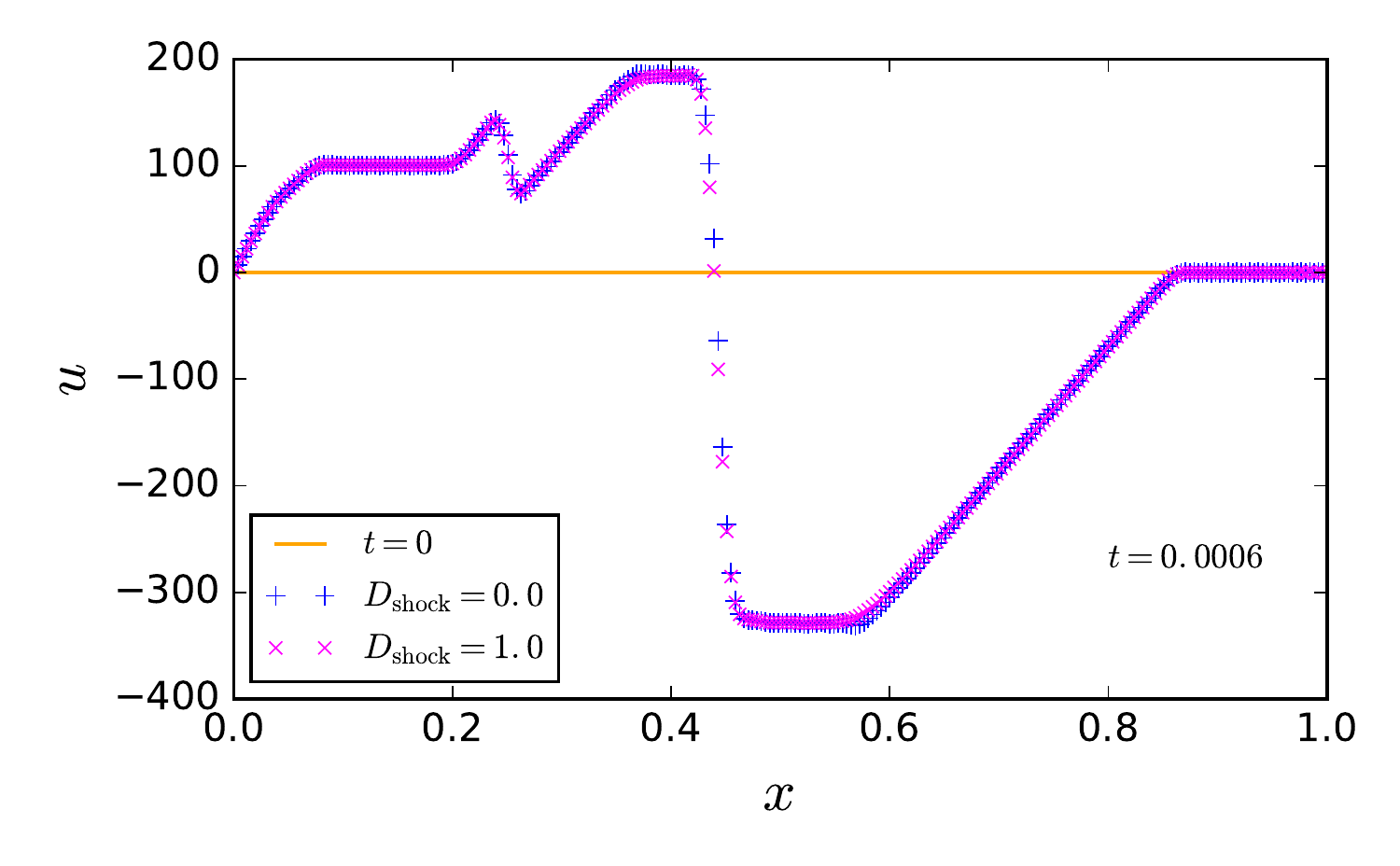}}}%
{\resizebox*{7.25cm}{!}{\includegraphics[trim= 0.15cm 1.7cm 0.5cm  0.4cm,clip=true]{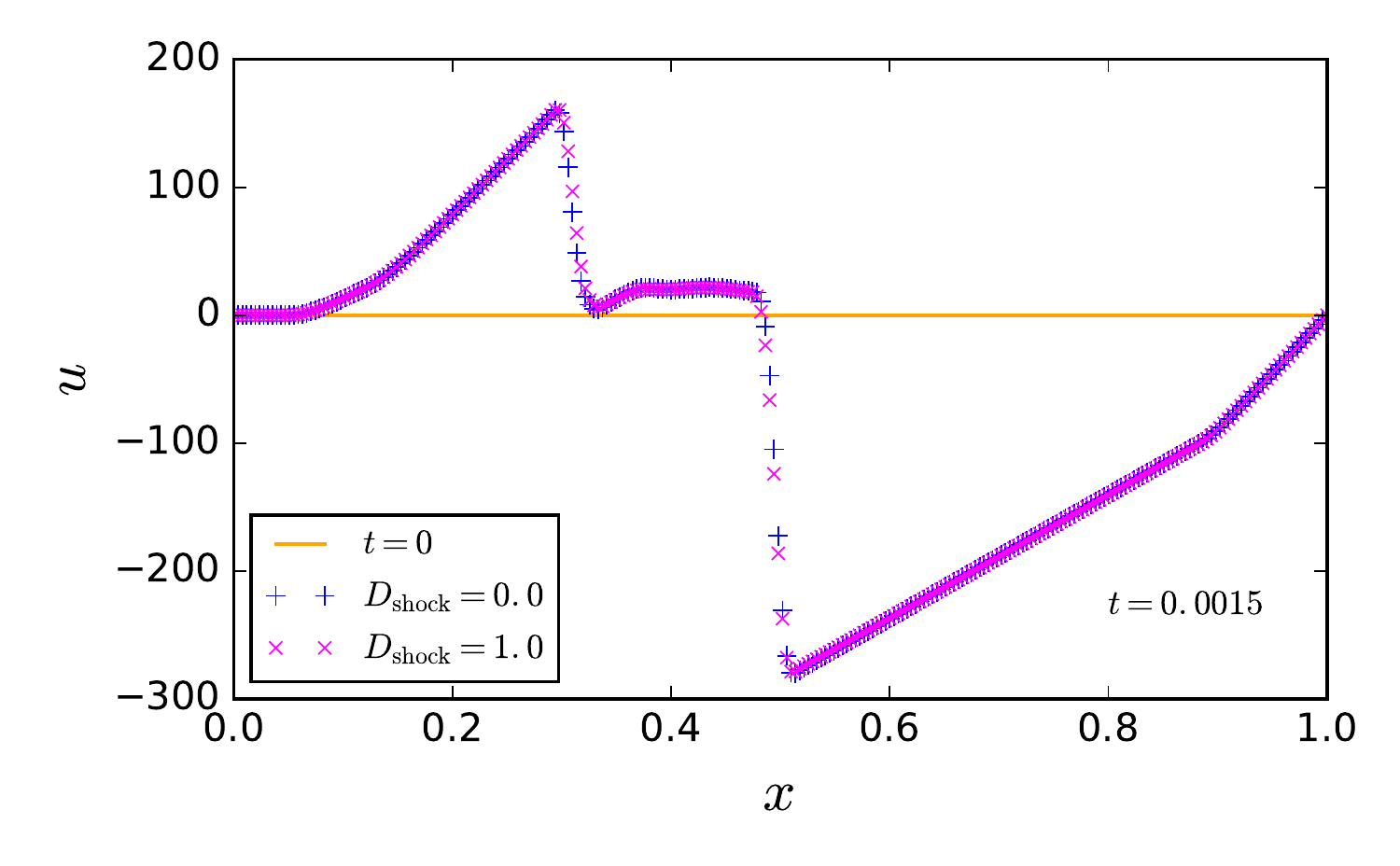}}}\\%
{\resizebox*{7.25cm}{!}{\includegraphics[trim=-0.45cm 0.5cm 0.5cm  0.4cm,clip=true]{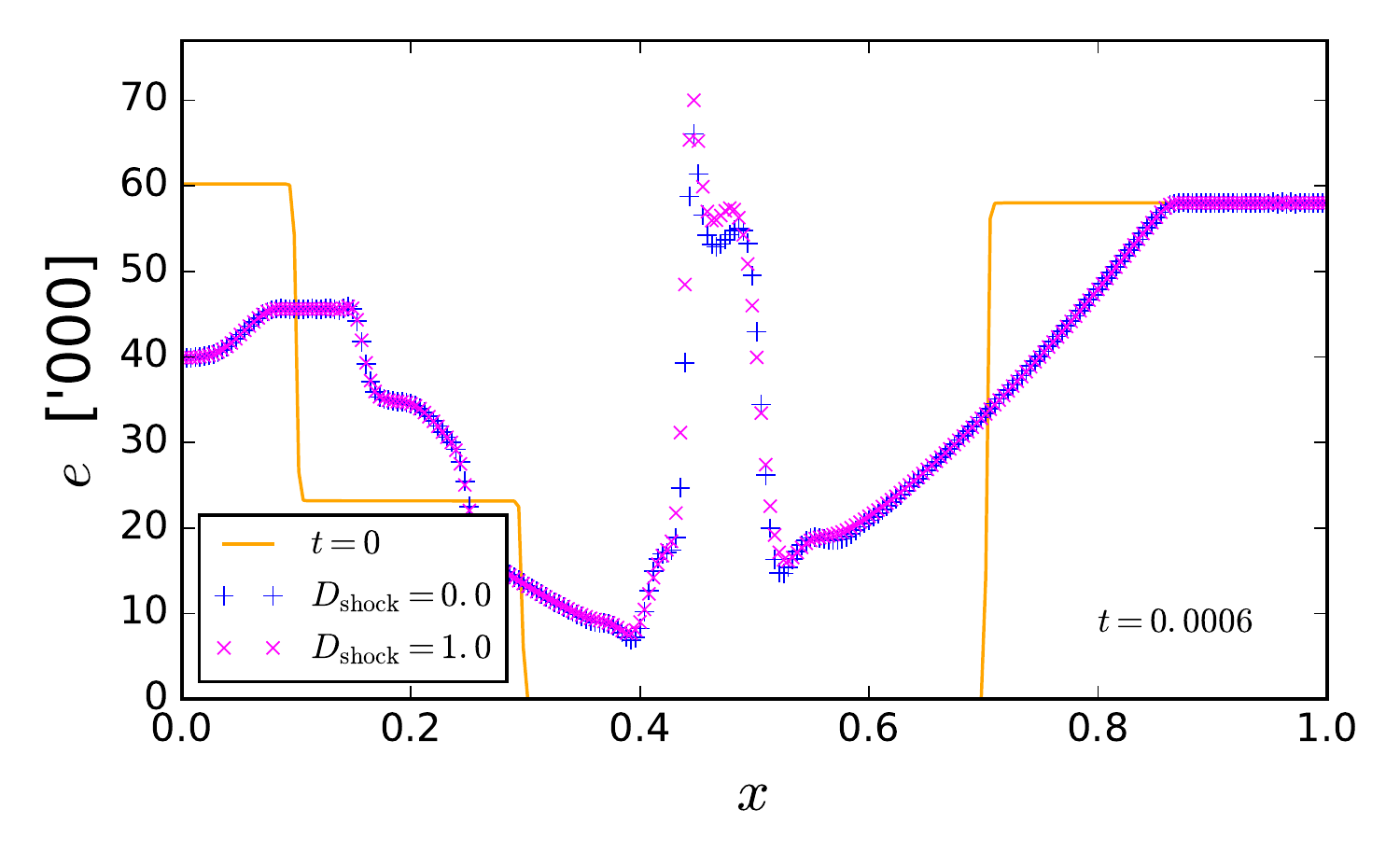}}}%
{\resizebox*{7.25cm}{!}{\includegraphics[trim=-0.25cm 0.5cm 0.5cm  0.4cm,clip=true]{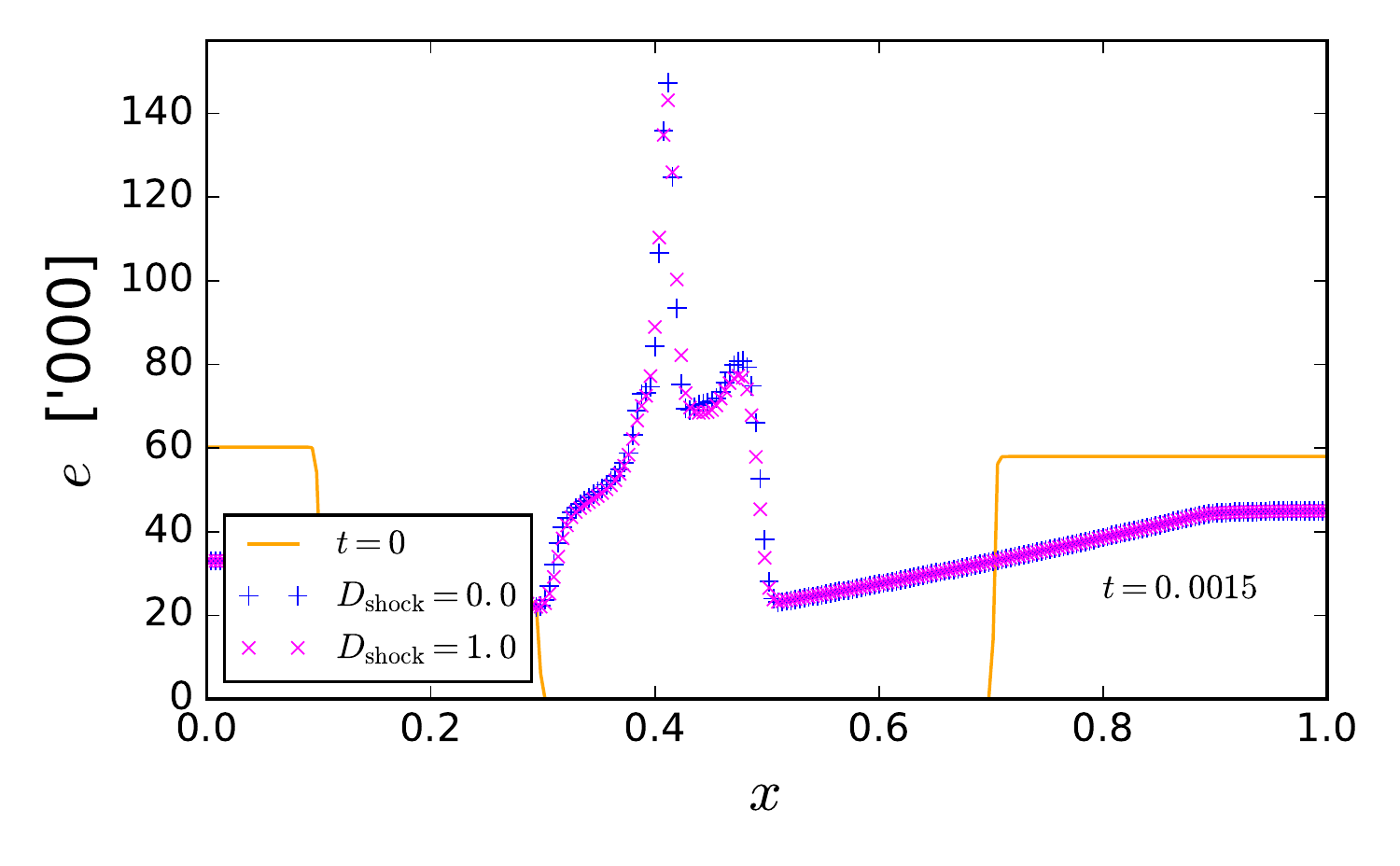}}}%
\caption{
Results with 256 grid points for high Mach number
Riemann shock-tube interacting triplet. Numerical solutions for density, velocity and energy are compared for mass diffusion coefficient $D_{\rm shock}=0$ and 1 at $t=0.0006$ (left panels) and 0.0015 (right panels). ${c_{\rm shock}}=[{D_{\rm shock}},6,2]$. (Colour online)
}%
\label{fig:clash}
\end{minipage}
\end{center}
\end{figure}

In figure\,\ref{fig:D-comp} we display results for varying values of $D_{\rm shock}$. When this is set to zero, the lag in the position of the shock front is largest, but the post-shock density approaches the analytic value most closely. There does not appear to be convergence to an asymptotic profile as $D_{\rm shock}$ increases, with tests using values above 50 showing greater diffusion at both the shock and the contact discontinuity. Larger magnitude extrema at the contact discontinuity may even appear for higher values.

In all the experiments presented in this paper, the mass diffusion does not improve the numerical solutions. However, the SN-driven turbulence experiments are susceptible to numerical instabilities near interacting shocks, and artificial mass diffusion has been found to suppress these effects. To investigate the underlying process, we applied a shock-tube test with three initial discontinuities, designed to induce collisions between shock waves.

The initial density, velocity and energy profiles are displayed in figure\,\ref{fig:clash}, alongside their evolved profiles at $t=0.0006$ and 0.0015 for $D_{\rm shock}=0$ and 1. At 0.0006 the energy peak near $x=0.5$ is near the convergent flow and subject to the viscosity applying near the shock. The energy near the compression is not more enhanced without mass diffusion. At $t=0.0015$ the spike in energy at $x=0.4$ is almost an order of magnitude higher than at $t=0.0006$, and is not near the compressive flows at $x=0.5$ and 0.3. The energy spike is significantly enhanced without mass diffusion and is associated with a deeper local minimum in the density.

We conclude from this analysis that modest artificial viscosity and thermal diffusivity allow a reasonable representation of adiabatic shocks with high Mach number. The lag in shock position and thickness of the shock front in the lowest resolution runs are not significantly dependent on the values of the diffusion coefficients, and are both resolution artifacts. The mass diffusion spreads the mass in the shock front beyond the analytic region. Numerical instability in the full turbulent simulations seems to occur at interacting shocks that produce wall heating and density deficits.  Inclusion of mass diffusion likely regularizes these points, inhibiting numerical instability. $D_{\rm shock}\simeq1$ or 2 is recommended and appears sufficient to avoid numerical instability.

\section{3D supernova remnants
\label{sect:snowplough}}

Previous tests of SN modelling with the Pencil Code are reported in \citet[][Appendix A]{Gent:2012}. These included higher shock diffusivity coefficients, explicit shear viscosity and thermal conductivity, and suppression of the cooling near shock fronts. We now use access to greater computational resources to apply the tests across a wider range of ambient densities and grid resolution, with enhanced timestep control, the improved treatment of artificial viscosity described here, and without unphysical suppression of radiative cooling. Numerical results are compared with analytic solutions for an SN remnant expanding into a perfect, homogeneous, monatomic gas at rest. For these experiments we apply adiabatic index $\gamma=5/3$.

The SN energy is injected into the existing density distribution in a sphere with an initial nominal radius of {$R_0$}. The energy injection radial profile follows 
\begin{equation}  
   E(R) = E_0\exp\Bigl(-\left[
   R/{R_0}
\right]^{{2}}\Bigr),
\end{equation}
with normalising coefficient $E_0$ set such that the volume integral of $E(R)$ is $10^{51}\erg$. The remnant origin is located on a grid point, and $R_0\gtrsim5$ grid zones. This provides a sufficiently smooth initial shock front, which can also be handled in a highly nonuniform turbulent injection site, while the remnant formed has a reasonably uniform internal temperature.

Although the minimum initial radius is {at least 5 grid zones}, a further constraint is to expand the injection radius to ensure at least ${50}\, M_\odot$ is present to limit extreme heating of the gas and corresponding drops in the time step, as well as numerical instability. Consequently the $0.001\cmcube$ model has an initial radius of 78\,pc. For these tests, the low density models can cope with smaller injection radii, but in the turbulent system we need to ensure there is enough total mass to avoid local numerical instability. Some authors avoid the additional complications of turbulent injection sites by smoothing the gas to a uniform density. For example, \citet{Joung:2006} adjust the radius to enclose $60\, M_\odot$, then smooth the volume to a uniform density. To handle explosions in high density regions, {where delayed evacuation of the remnant interior induces excess} cooling {that} can inhibit the power of the SN, one solution is simply to delete enough mass inside the injection site to allow high enough temperature  or to move the mass to the remnant shell at injection. So far, we have been able to avoid such measures and, particularly when evolving the dynamo, would prefer not to unphysically remove the gas from the magnetic field or consider also rearranging its ambient vector potential field.

\subsection{Adiabatic remnant} 

The early stages of SN evolution are approximately adiabatic. For a uniform ambient medium they are well described by the Sedov-Taylor analytic solution \citep{Taylor50,Sedov59},
\begin{equation}
\label{eq:sed}
R\,= \left(\kappa\frac{E\SN}{\rho_{0}}\right)^{\!{1}/{5}}t^{{2}/{5}},
\end{equation}
where $R$ is the remnant radius, $E\SN$ the explosion energy, $\rho_{0}$ the ambient gas density, and the dimensionless parameter $\kappa\approx2.026$ for $\gamma=5/3$ \citep{Ostriker88}.

\begin{figure}
\begin{center}
\begin{minipage}{150mm}
\begin{center}
{\resizebox*{9.25cm}{!}{\includegraphics[trim=0.35cm 0.45cm 0.2cm 0.3cm,clip=true]{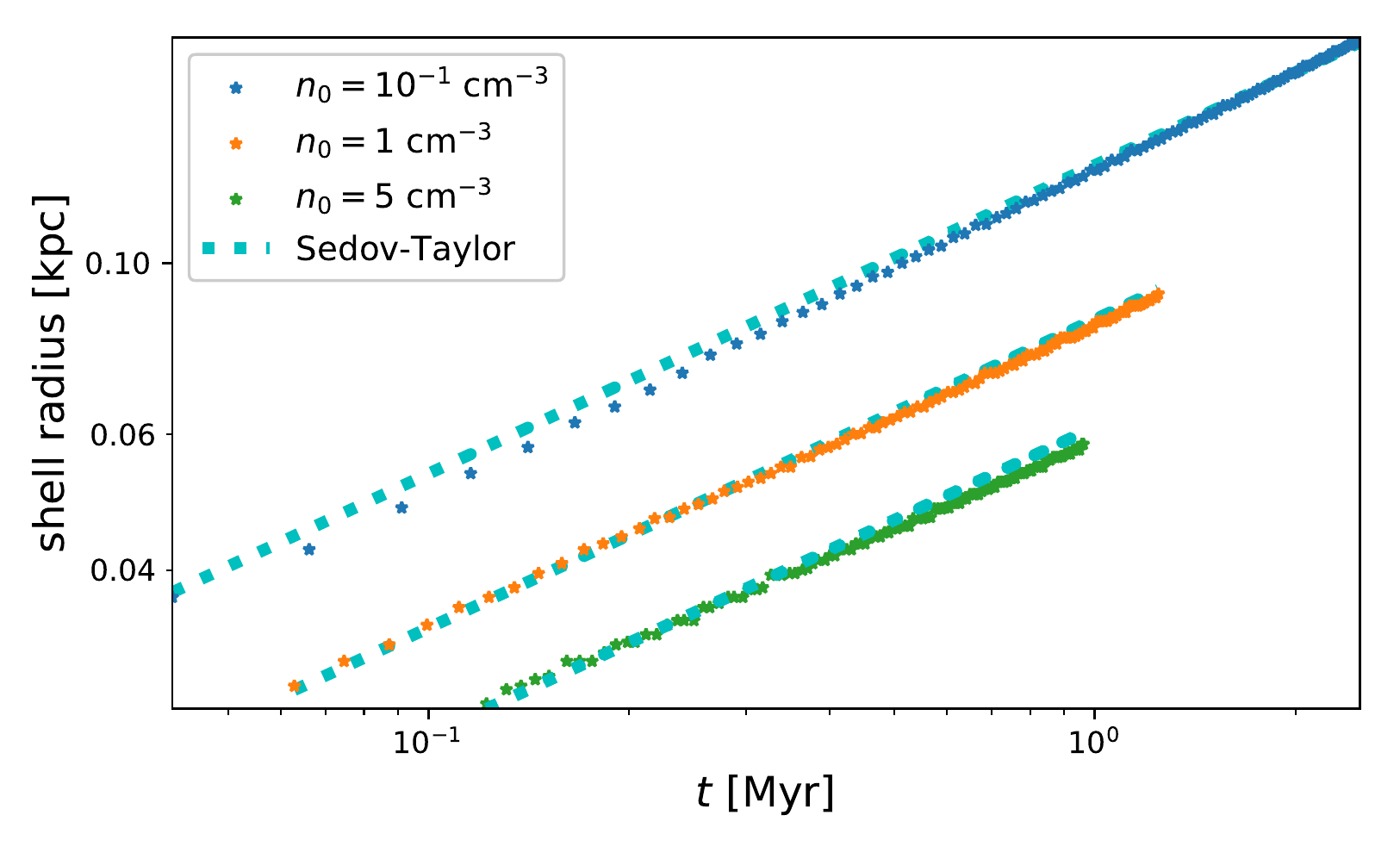}}}%
\caption{Time evolution of remnant shell radius in the adiabatic Sedov-Taylor  regime  for ambient gas number density between 0.1 and 5 cm$^{-3}$ with resolution along each side of 4\,pc. The diffusivity coefficients used are ${c_{\rm shock}}=[1,{6,2}]$ and viscosity $\nu=0.004c_s$ kpc km s$^{-1}$. (Colour online)
}%
\label{fig:ST}
\end{center}
\end{minipage}
\end{center}
\end{figure}

In figure\,\ref{fig:ST} for a range of ambient ISM densities we compare the radial evolution of the remnant shell between our numerical models and the analytic solution described by (\ref{eq:sed}). The coefficients used are ${c_{\rm shock}}={[1,6,2]}$.
As explained in section\,\ref{sect:cs_visc}, we also include for all SN remnant experiments viscosity $\nu\simeq c_s\,\delta x$ to verify that the numerical solution remains valid for the parameters relevant to modelling SN-driven turbulence. The power law growth of the radius is in reasonable agreement with the analytic prediction, while only slightly more retarded as the ambient density increases. From (\ref{eq:fvisc}) we see a contribution to the viscous forces from the gradients of $\ln\rho$, $\nu$ and $\vect\nabla{\bm \cdot}\vect u$. The net effect is negative, becoming more relevant as density increases.

  \subsection{Momentum conserving, pressure driven and momentum driven snowplough} 
When radiative cooling processes are included the SN evolution changes. The Pencil Code currently has two implementations of radiative cooling associated with SN turbulence, both based on piecewise power law dependence of the cooling coefficient on temperature. These are described in \citet[][see their figure\,1]{Gent:2013a} and are based on \citet{RBN93} {(RB)} and a combination of \citet{WHMTB95} and \citet{SW87} {(WSW)}. The contribution from FUV heating follows \citet{WHMTB95} \citep[see][]{Gent:2013a}, which is truncated for temperatures above 10$^{4}$\,K. As the remnant expands and the shock front accumulates more gas from the ambient ISM, cooling becomes more efficient in the increasingly dense shell. With the loss of energy the shell speed falls. The standard momentum-conserving snowplough solution for a radiative SN remnant has the form
\begin{equation}
\label{eq:snpl}
R\,=\,R_{0}\biggl[1+4\frac{\dot{R_{0}}}{R_{0}}(t-t_{0})\biggr]^{{1}/{4}} ,
\end{equation}
where $R_{0}$ is the radius of the SN remnant at the time $t_{0}$ of the transition from the adiabatic stage, and $\dot{R_{0}}$ is the shell expansion speed at $t_{0}$. The transition time is determined by \citet{Woltjer72} to align with half of the SN energy being lost to radiation; this happens when
\begin{equation}
\dot{R_{0}}\,=\,230\kms \left(\frac{n_{0}}{1\cmcube}\right)^{\!{2}/{17}}\left(\frac{E\SN}{10^{51}\erg}\right)^{\!{1}/{17}},
\end{equation}
with $n_0$ the gas number density of the ambient ISM. The transitional expansion speed thus depends very weakly on parameters.
  
\citet{Cioffi88} obtained numerical and analytical solutions for an expanding SN remnant with special attention to the transition from the Sedov--Taylor stage to the radiative stage. These authors adjusted an analytical solution for the pressure-driven snowplough stage to fit their numerical results to an accuracy of within 2\%
and 5\% in terms of $R$ and $\dot R$, respectively. (Their numerical resolution was $0.1\,{\rm pc}$ in the interstellar gas and $0.01\,{\rm pc}$ within ejecta.)  They thus obtained
\begin{equation}
\label{eq:pds}
R\,=\,R_{\rm{p}} \left(\frac{4}{3}\frac{t}{t_{\rm{p}}}-\frac{1}{3}\right)^{\!3/10},
\end{equation}
where the subscript ${\rm{p}}$ denotes the radius and time for the transition to the pressure driven stage. The estimated time of this transition is 
\begin{equation}
t_{\rm{p}}\simeq13\Myr\left(\frac{E\SN}{10^{51}\erg}\right)^{\!3/14}\left(\frac{n_0}{1 \cmcube}\right)^{-4/7}. 
\end{equation}
This continues into the momentum driven stage with
\begin{equation}
\label{eq:mcs}
\left(\frac{R}{R_{\rm{p}}}\right)^{\!4}
=\,\frac{3.63~\left(t-t_{\rm{m}}\right)}{t_{\rm{p}}}\left[1.29-\left(\frac{t_{\rm{p}}}{t_{\rm{m}}}\right)^{\!0.17}\,\right]
+ \left( \frac{R_{\rm{m}}}{R_{\rm{p}}}\right)^{4},
\end{equation}
where subscript ${\rm{m}}$ denotes the radius and time for this second transition,
\begin{equation}
R_{\rm{m}}^4 \,
=\, 4.66 \,\frac{t}{t_{\rm{p}}}\left[1-0.939\left(\frac{t}{t_{\rm{p}}}\right)^{-0.17}
+0.153\left( \frac{t}{t_{\rm{p}}}\right)^{-1}\right]
\end{equation}
and
\begin{equation}
t_{\rm{m}}\,\simeq\, 61\, t_\mathrm{p} \left(\frac{\dot{R}_{\rm{ej}}}{10^3\kms}\right)^{\!3}
                 \left(\frac{E\SN}{10^{51}\erg}\right)^{-3/14}\left(\frac{n_0}{1\cmcube}\right)^{-3/7},
\end{equation}
where $\dot{R}_{\rm{ej}}\simeq5000\kms$ is the initial velocity of the $4M_\odot$ ejecta. The shell momentum in the latter solution tends to a constant, and the solution thus converges with the momentum-conserving snowplough (see (\ref{eq:snpl})); but, depending on the ambient density, the expansion may become subsonic and the remnant merge with the ISM beforehand. 

To ultimately follow the analytical solution we require the injection radius to be significantly less than $R_{\rm{p}}$, so that the early evolution remains adiabatic. Although, in most cases our remnant injection site is sufficiently compact and diffuse to evolve to follow the analytical solution, for low resolution and high ambient density, $R_0$ may be near or even beyond $R_{\rm{p}}$. In such cases an adjustment needs to be made to compensate for the radiative losses that would have preceeded. We follow the approach of \citet[][their eq.~(16)]{SBHO15} to define a fraction $f_{\rm kin}$ of $E_{\rm SN}$ injected as kinetic energy.
\begin{equation}\label{eq:frackin}
f_{\rm kin}\,=\,\frac{3.97\times10^{-6}\mu}{\delta x^2}\,\frac{n_0}{ E_{\rm SN}}\,\frac{R_{\rm{p}}^7}{t_{\rm{p}}^2 }\,, 
\end{equation}
where $\mu$ is the mean molecular weight, and $R_{\rm{p}}$ and $t_{\rm{p}}$ themselves depend on $n_0$ and $E_{\rm SN}$, and metallicity $Z$, which we assume to be 1. $R_{\rm{p}}$, $t_{\rm{p}}$ and $E_{\rm SN}$ are normalised by pc, $10^3$\,yr and $10^{51}$\,erg, respectively. Following \citet{KO15} we normalise $n_0$ by $\rho_0/(1.4m_{\rm H})$, assuming 10\% helium abundance and $m_{\rm H}$ the hydrogen atomic mass. This fraction can be greater than 1 and a high fraction of kinetic energy makes the model more vulnerable to numerical instability, so we apply a cap of 0.075, and apply only where $R_{\rm{p}}<1.5R_0$ for $\delta x>1$\,pc.

The remnant shell radial profiles are computed for the ambient densities $n_0=(0.001,0.01,0.1,1,5)\cmcube$ used in the tests reported here. Although these semi-analytical models are a useful comparison to examine the accuracy of our numerical models, there are differences to consider. The \citet{Cioffi88} 1D analysis was conducted for ambient ISM with number density $0.1\cmcube$ and ambient temperature 10\,K to ensure the blast wave remained strong; their cooling follows a different piecewise power-law fit \citep{RCS76} than we use, and is truncated below $1.2\cdot10^4$\,K; and they do not include UV heating. They use resolution 0.1\,pc outside the remnant, and 0.01\,pc to resolve the ejecta. Because heating and cooling apply in our models at lower temperatures, for each density and each cooling function the ambient temperature is set at thermal equilibrium, so the external pressure remains constant over time.

\begin{figure}
\begin{center}
\begin{minipage}{150mm}
\begin{center}
{\resizebox*{9.25cm}{!}{\includegraphics[trim=0.35cm 0.45cm 0.2cm 0.3cm,clip=true]{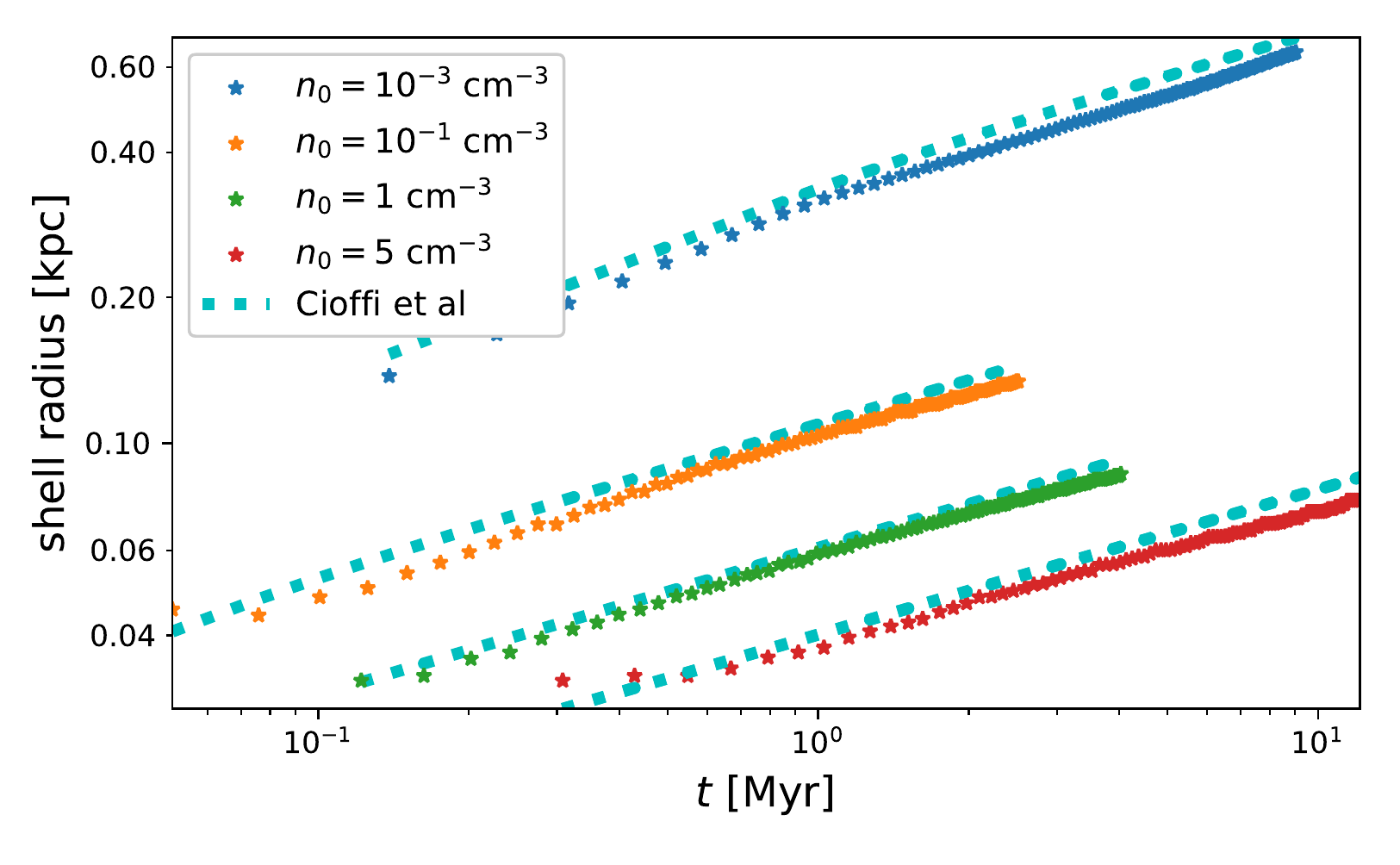}}}%
\caption{Time evolution of remnant shell radius following cooling function of RB for ambient gas number density between 0.001 and 5 cm$^{-3}$ with grid cell resolution of 4\,pc.  The diffusivity coefficients used are ${c_{\rm shock}}=[1,{6,2}]$ and $\nu=0.004c_s$ kpc km s$^{-1}$. (Colour online)
}%
\label{fig:RB}
\end{center}
\end{minipage}
\end{center}
\end{figure}

Results for the  RB cooling curve are illustrated in figure\,\ref{fig:RB}. The power law is a good fit for the ambient density $0.1\cmcube$, most closely matching the \citet{Cioffi88} setup, although the shell radius is somewhat retarded in our model. 

\begin{figure}
  [ht]
\begin{center}
\begin{minipage}{150mm}
{\resizebox*{7.25cm}{!}{\includegraphics[trim=0.35cm 0.4cm 0.0cm 0.4cm,clip=true]{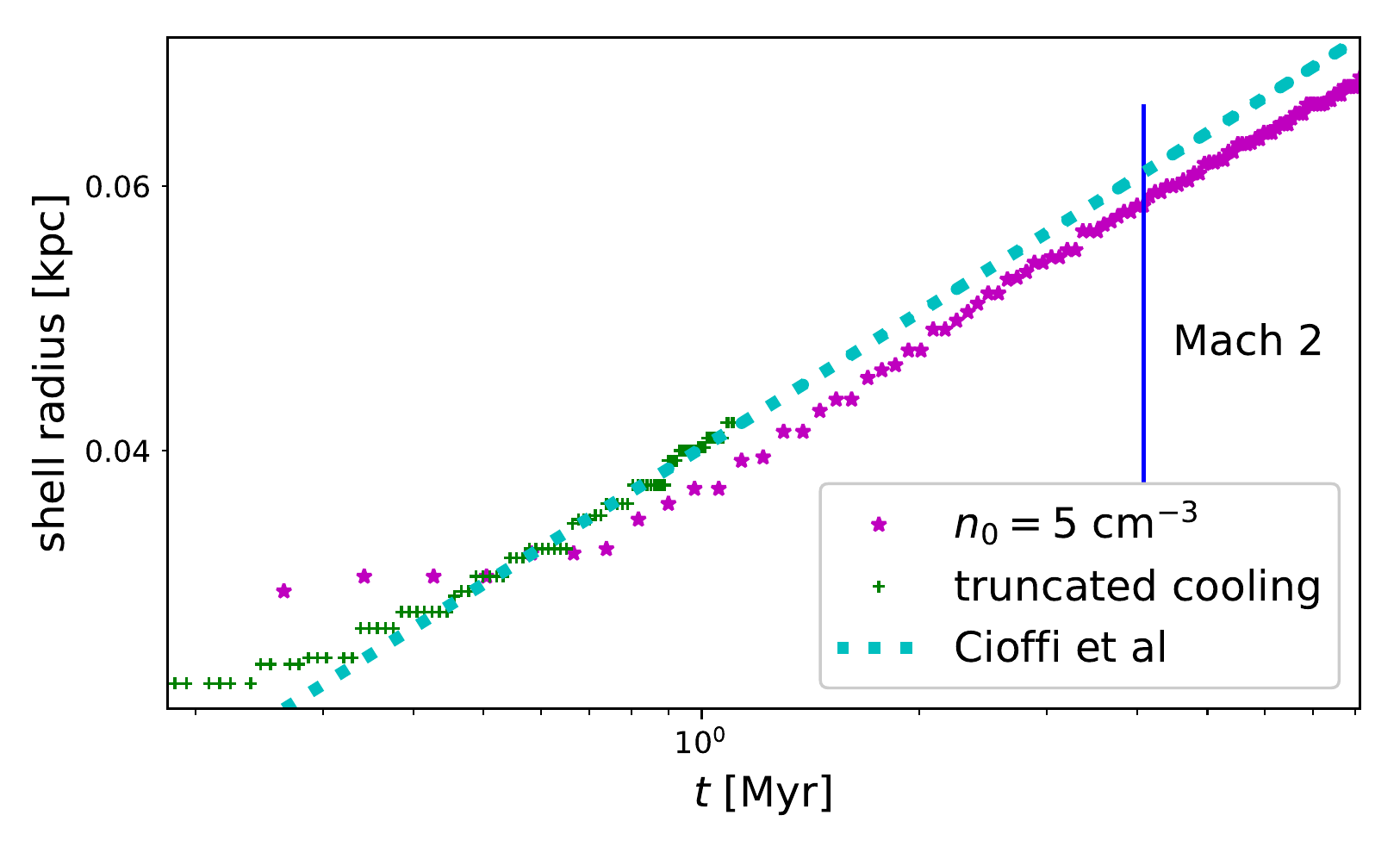}}}%
{\resizebox*{7.25cm}{!}{\includegraphics[trim=0.35cm 0.4cm 0.0cm 0.4cm,clip=true]{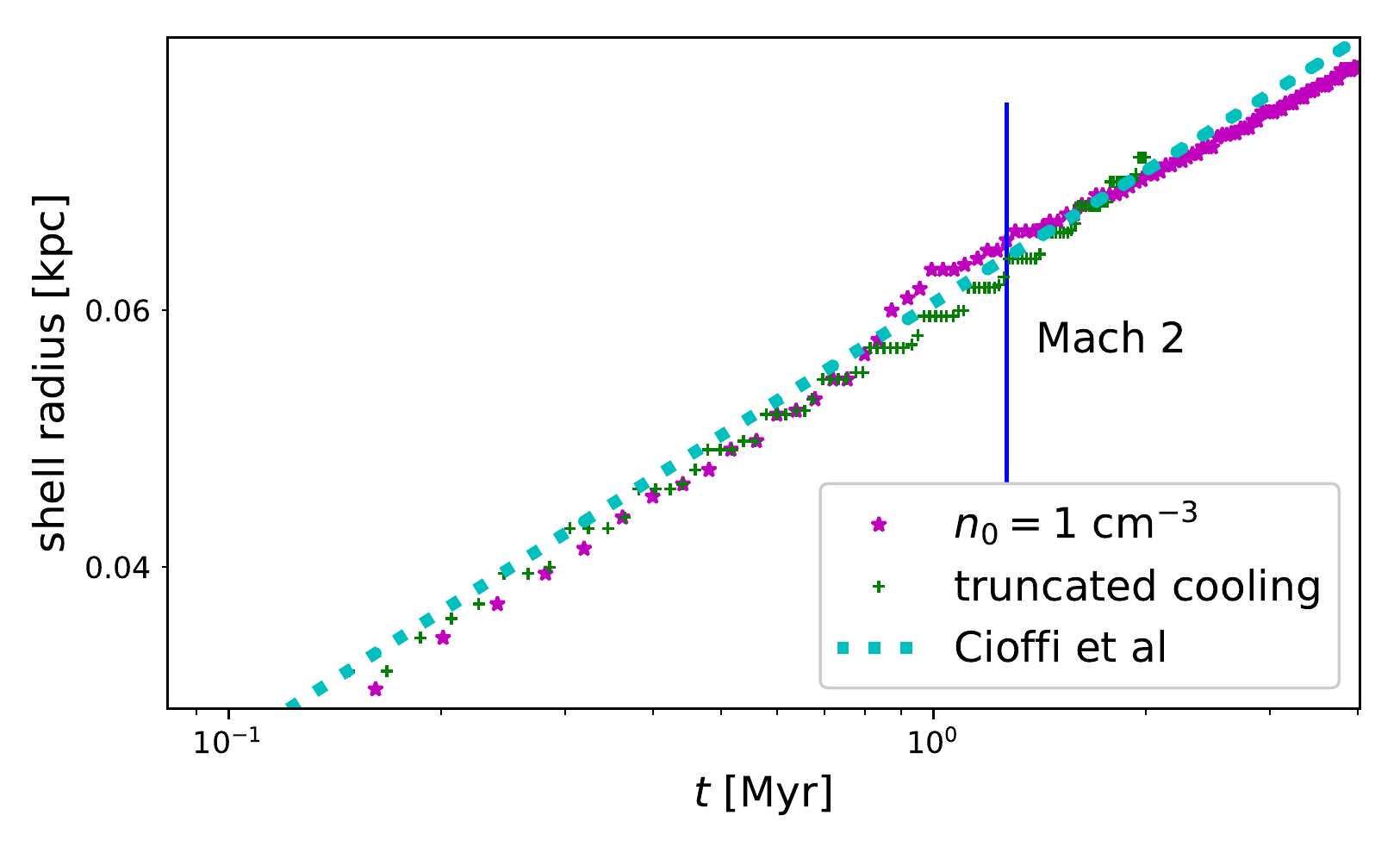}}}\\%
{\resizebox*{7.25cm}{!}{\includegraphics[trim=0.35cm 0.4cm 0.0cm 0.4cm,clip=true]{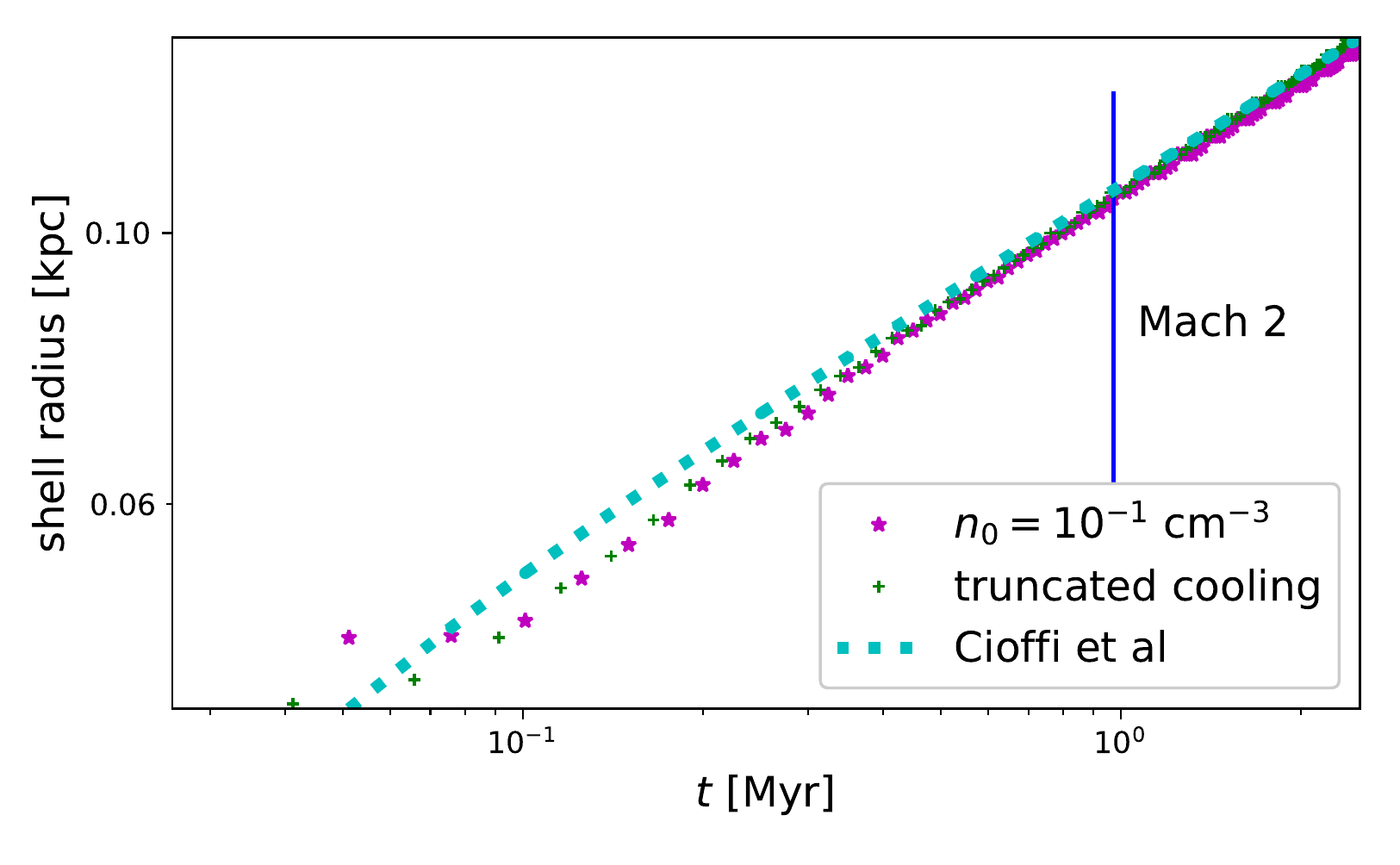}}}%
{\resizebox*{7.25cm}{!}{\includegraphics[trim=0.35cm 0.4cm 0.0cm 0.4cm,clip=true]{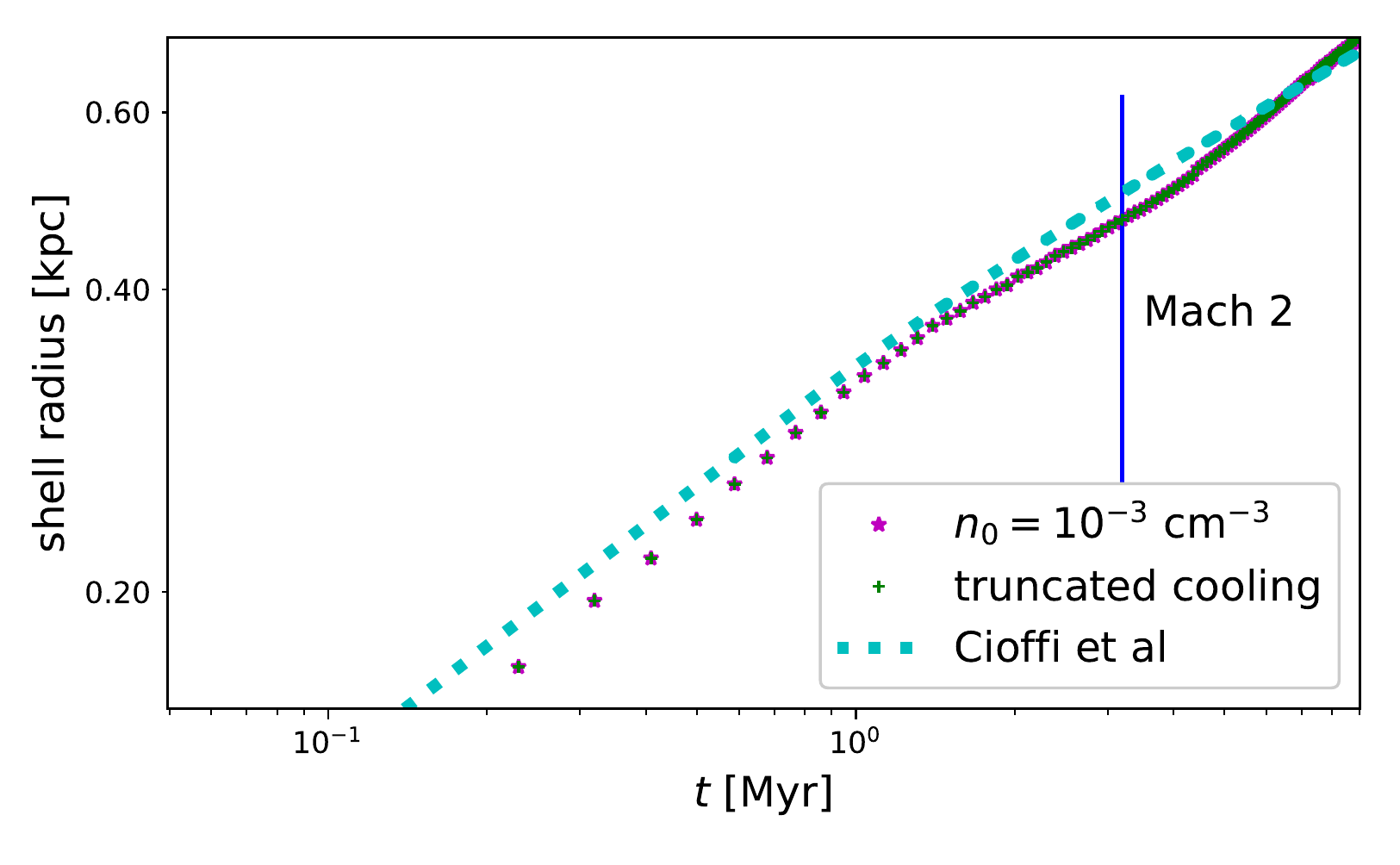}}}%
\caption{Time evolution of remnant shell radius following cooling function of WSW, and for the same cooling function, but truncated below $T=1.2\times10^4$\,K for ambient gas number density between $10^{-3}$ and 5 cm$^{-3}$ at grid resolution of 4\,pc. The \citet{Cioffi88} analytic solution {is} indicated for comparison. The diffusivity coefficients used are ${c_{\rm shock}}=[1,{6,2}]$ and $\nu=0.004c_s$ kpc km s$^{-1}$. The time is marked at which the shell speed has slowed to Mach\,2 for the non-truncated cooling model.
(Colour online)
}%
\label{fig:4pc-radius}
\end{minipage}
\end{center}
\end{figure}

In figure\,\ref{fig:4pc-radius} we see better agreement with \citet{Cioffi88} for the combination cooling model, except for ambient density $n_0=5\cmcube$. This is accounted for by the truncation in the cooling applied by \citet{Cioffi88}. For comparison we tested the combination cooling model with cooling truncated below $1.2\times10^4$\,K, and obtain excellent agreement with the analytic solution at all ambient densities as shown in figure\,\ref{fig:4pc-radius}. The cooling at all temperatures is faster for the RB model than the combination WSW cooling model, so it is understandable that the remnants in the former case would expand slower. We note that the ambient temperatures vary between the model cooling function and its truncated version, since we applied a thermostatic equilibrium to the ambient ISM, and hence the truncated models have merger with the ISM earlier at the higher sound speed. The time at which the remnant reaches Mach\,2 {(non-truncated cooling)} is added to indicate how close the numerical solution is to the analytic solution when the remnant is near to becoming subsonic with respect to the ambient medium.

The relationship between cooling, temperature and density is nonlinear, so it is not clear that the relations derived by \citet{Cioffi88} for $0.1\cmcube$ ambient ISM should  also apply for the other densities. Nonetheless the results at all ambient densities seem reasonable. For higher densities (above $1\cmcube$) at this resolution, the inclusion of $f_{\rm kin}$ is required.

  \subsection{Effects of ambient density, resolution and diffusivity
\label{sect:pars}} 

\begin{figure}
\begin{center}
\begin{minipage}{150mm}
\begin{center}
\subfigure[Ambient density 5         cm$^{-3}$.]{\resizebox*{14.5cm}{!}{\includegraphics[trim=0.75cm 0.45cm 0.5cm 0.3cm,clip=true]{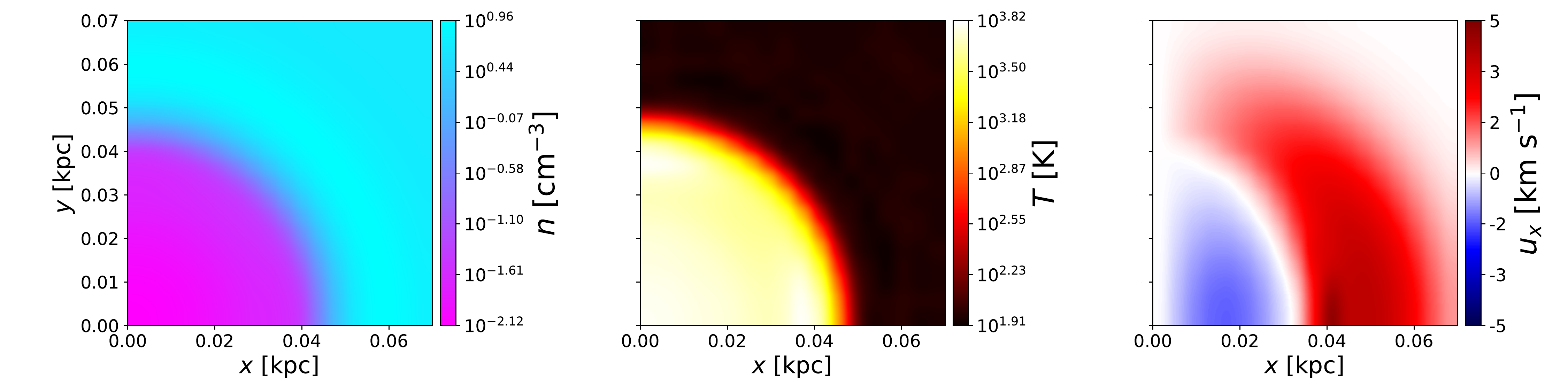}}}\\%
\subfigure[Ambient density 10$^{-1}$ cm$^{-3}$.]{\resizebox*{14.5cm}{!}{\includegraphics[trim=0.75cm 0.45cm 0.5cm 0.3cm,clip=true]{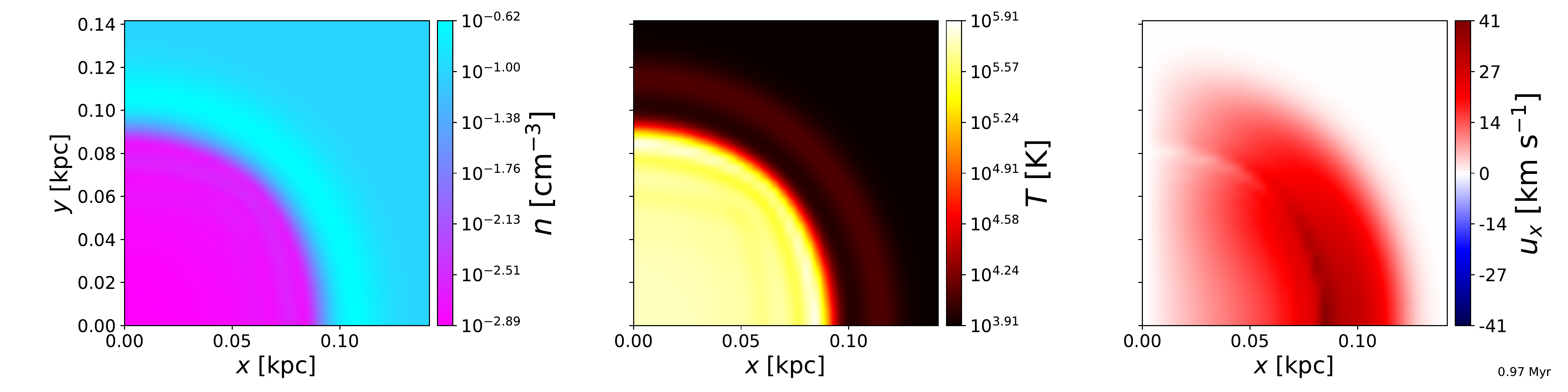}}}\\%
\subfigure[Ambient density 10$^{-3}$ cm$^{-3}$.]{\resizebox*{14.5cm}{!}{\includegraphics[trim=0.75cm 0.45cm 0.5cm 0.3cm,clip=true]{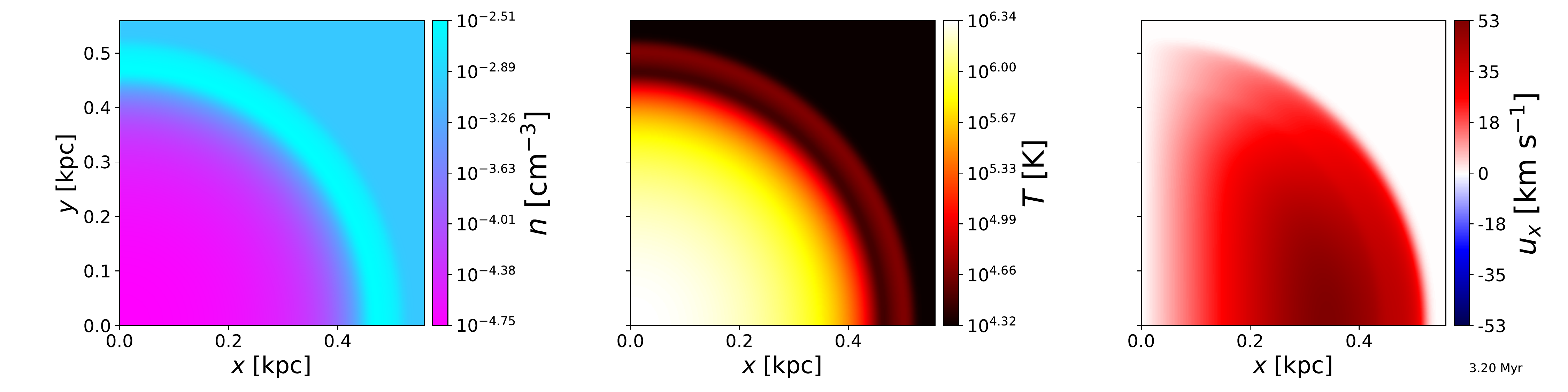}}}%
\caption{
Two-dimensional quarter-slices from the $z=0$ plane depicting gas number density, temperature and $x$-component of velocity. Ambient gas number density is (top) $5 \cmcube$, (middle) $0.1 \cmcube$, and (bottom) $10^{-3}$\,cm$^{-3}$. Each snapshot corresponds to the time at which the remnant shell speed falls to Mach\,{2} for the ambient ISM. Note the resulting order of magnitude change in the size scale (and thus numerical resolution of the shock) from top to bottom. The diffusivity coefficients used are ${c_{\rm shock}}=[1,{6,2}]$ and $\nu=0.004c_s$ kpc km s$^{-1}$. (Colour online)
}%
\label{fig:4pc-shells}
\end{center}
\end{minipage}
\end{center}
\end{figure}

Snapshots of the remnant density, temperature and velocity  distributions at 4\,pc resolution for ambient densities 5, 0.1 and 0.01$\cmcube$ are displayed in figure\,\ref{fig:4pc-shells}. The 3D simulations are on a Cartesian grid, with the SN origin at $\bm0$. As the profile is symmetric, unnecessary duplication is avoided by displaying a quarter plane. The snapshots are at times when the shell expansion is near Mach\,{2}. In the middle column the cooling shell is faintly visible for $5\cmcube$, where its temperature has dropped below the ambient medium, while in the lower density runs the shell has not yet cooled below the ambient ISM although visibly cooler than the external shocked region.

\begin{figure}
\begin{center}
\begin{minipage}{150mm}
\begin{center}
{\resizebox*{9.25cm}{!}{\includegraphics[trim=0.35cm 0.45cm 0.2cm 0.3cm,clip=true]{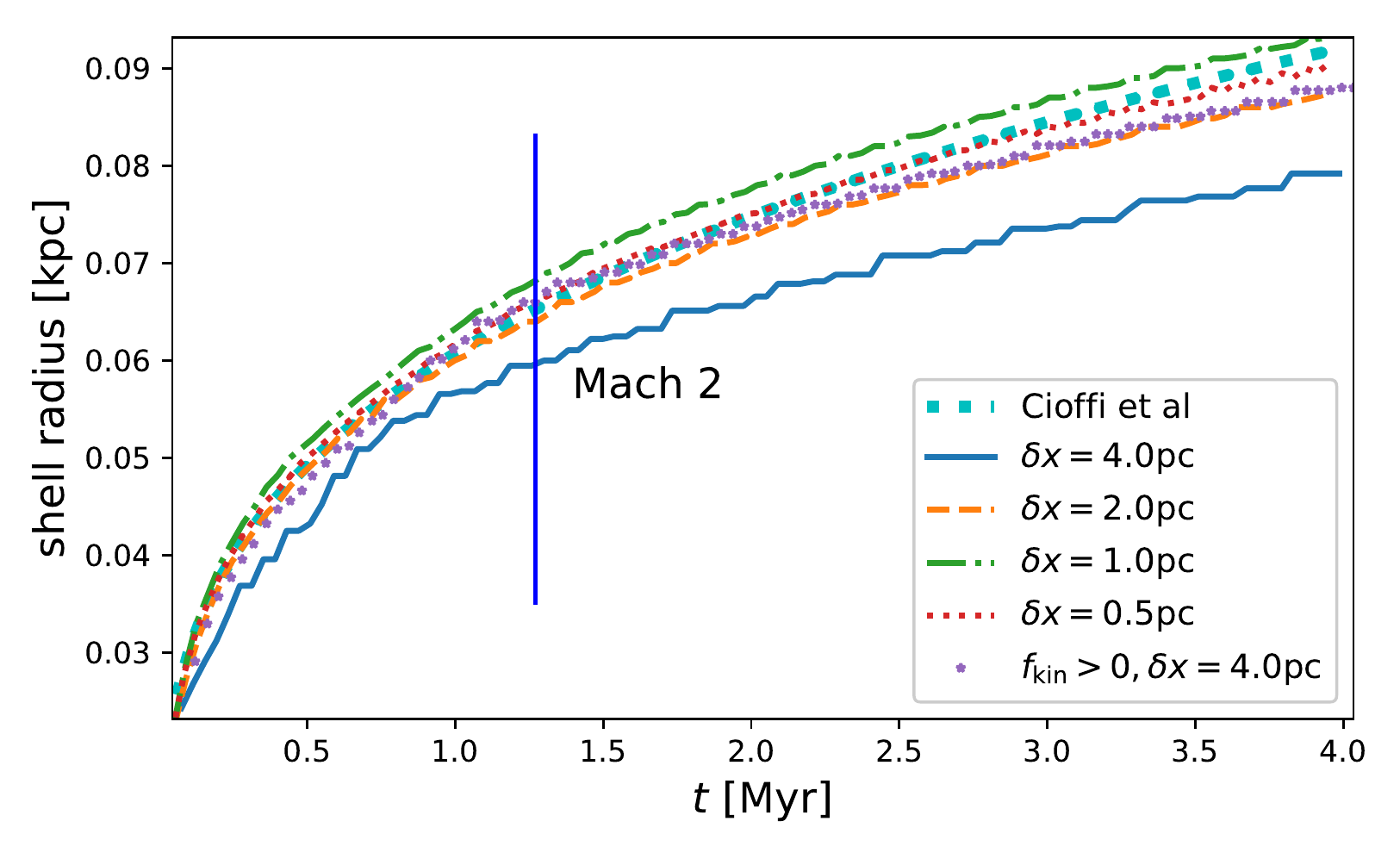}}}%
\caption{
Time evolution of remnant shell radius following the combined cooling function of WSW for unit ambient density at grid resolution of 0.5--4~pc. With increasing resolution the numerical solution converges to the semi-analytic solution of \citet[][]{Cioffi88}. The diffusivity coefficients used are ${c_{\rm shock}}=[1,{6,2}]$ and $\nu=c_s\,\delta x$. The time is marked at which the shell speed has slowed to Mach\,2. For comparison at 4\,pc resolution we show the effect of including a fraction $f_{\rm kin}$ of the SN energy as kinetic. (Colour online)
}%
\label{fig:1D-res}
\end{center}
\end{minipage}
\end{center}
\end{figure}

For ambient density $1\cmcube$ we explore the effects of resolution. At this density the thermal equilibrium with the WSW cooling and heating occurs at about 2185\,K, which is on a thermally unstable branch of the cooling curve. In developing the numerical model this density proved to be more vulnerable to instabilities than either lower or higher densities. It therefore was of most interest for the resolution tests.

In figure\,\ref{fig:1D-res} we compare the time evolution of the shell radius for grid resolutions of 0.5--4\,pc at ambient density $1\cmcube$. There is convergence towards the analytic result with increasing resolution. In comparing these results we apply resolution-dependent optimal radius $R_0\simeq7.9$, 8.5, 9.2 and 17 pc, respectively with increasing zone size, and with $f_{\rm kin}=0$. Alternatively, fixing $R_0=17$ pc, the minimum size required to resolve a sphere at 4~pc resolution, yields solutions that are consistently slower with increasing resolution. This is because, the remnant shell reaches higher densities, permitting more efficient cooling. Although a radius $R_0=17$\,pc sphere covers 4.25 grid zones at 4~pc resolution, this is not ideal to represent a sphere on a Cartesian grid. The evolving spherical shell is adequate with a radius of $R_0=24$\,pc, but with purely thermal energy injection is subject to excessive energy losses. Including the kinetic energy adjustment with $f_{\rm kin} > 0$ produces  excellent agreement at all resolutions, as intended. Such a solution at 4~pc resolution is included in figure\,\ref{fig:1D-res} to illustrate the effectiveness of the adjustment.

\begin{figure}
\begin{center}
\begin{minipage}{150mm}
\begin{center}
\subfigure[Resolution $\delta x=4.0$~pc.]{\resizebox*{15cm}{!}{\includegraphics[trim=0.75cm 0.45cm 0.5cm 0.3cm,clip=true]{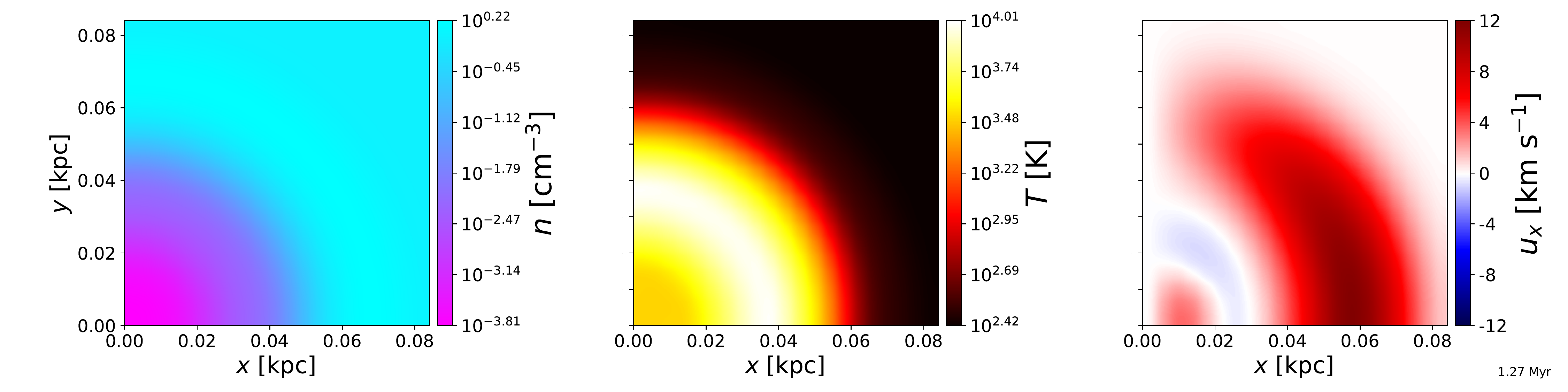}}}\\%
\subfigure[Resolution $\delta x=2.0$~pc.]{\resizebox*{15cm}{!}{\includegraphics[trim=0.75cm 0.45cm 0.5cm 0.3cm,clip=true]{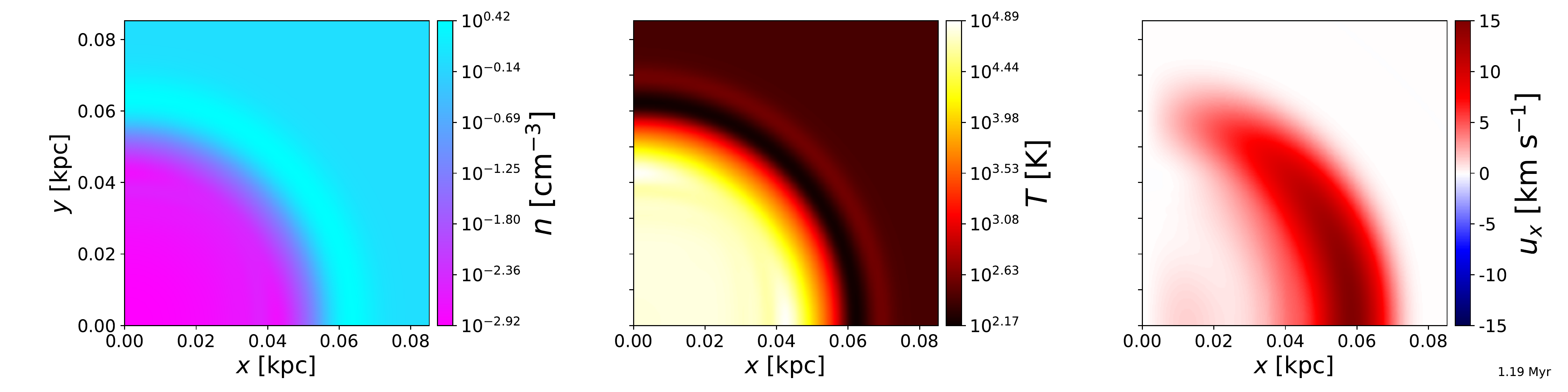}}}\\%
\subfigure[Resolution $\delta x=1.0$~pc.]{\resizebox*{15cm}{!}{\includegraphics[trim=0.75cm 0.45cm 0.5cm 0.3cm,clip=true]{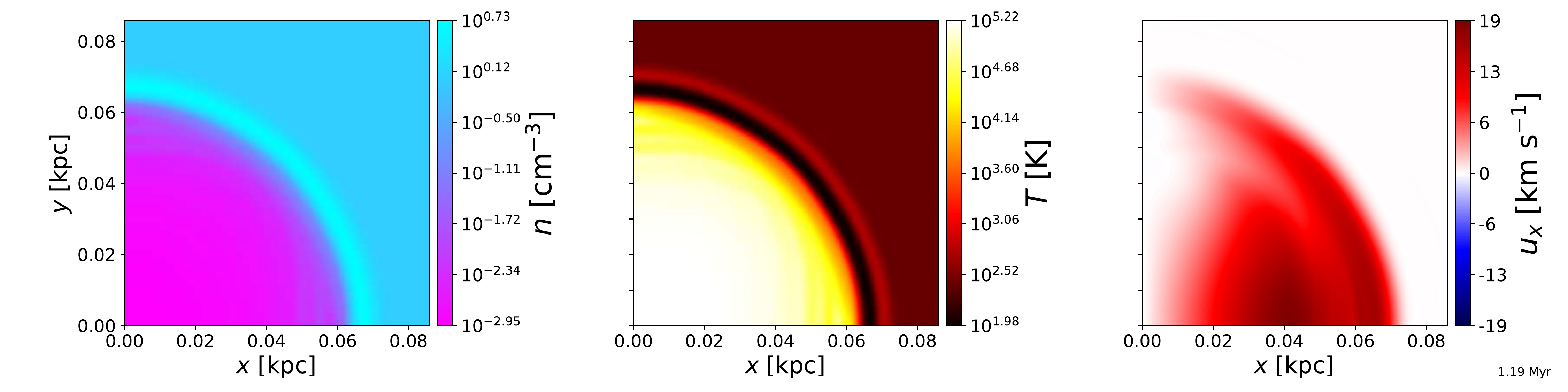}}}\\%
\caption{
Varying resolution models, with (a) $\delta x = 4$~pc, (b) 2~pc and (c) 1~pc. Two-dimensional quarter-slices from the $z=0$ plane depicting gas number density, temperature and $x$-component of velocity are shown. Ambient gas number density is 1\,cm$^{-3}$. The diffusivity coefficients used are ${c_{\rm shock}}=[1,{6,2}]$ and $\nu=c_s\,\delta x$. (Colour online)
}%
\label{fig:res-shells}
\end{center}
\end{minipage}
\end{center}
\end{figure}

Slices of snapshots from these runs while the shell is expanding at Mach\,{2} into the ambient ISM are displayed in figure\,\ref{fig:res-shells}. At 4\,pc resolution we see that the shell has not yet started to cool below the ambient ISM temperature (figure\,\ref{fig:res-shells} a, middle panel), while at 2\,pc and below, cooling occurs. In previous Pencil Code SN driven turbulence at 4\,pc resolution cold gas below 100\,K was present, but this result suggests they arose primarily from compression fronts at remnant interactions and general turbulent shocks.

The reverse shocks seen in the remnant interior at moderate resolution in figures\,\ref{fig:4pc-shells} and \ref{fig:res-shells} occur because the ambient pressure is non-zero, and the energy input is not an idealized point source. In the diffuse interior with low momentum these velocities can be higher than the remnant shell speed. The effect reduces with increasing resolution, since the coarse Cartesian grid defining the injection sphere exacerbates this effect.

\begin{figure}
\begin{center}
\begin{minipage}{150mm}
\begin{center}
\subfigure[$\nu=c_s \delta x$\,kpc km s$^{-1}$
]{\resizebox*{15cm}{!}{\includegraphics[trim=0.75cm 0.45cm 0.5cm 0.3cm,clip=true]{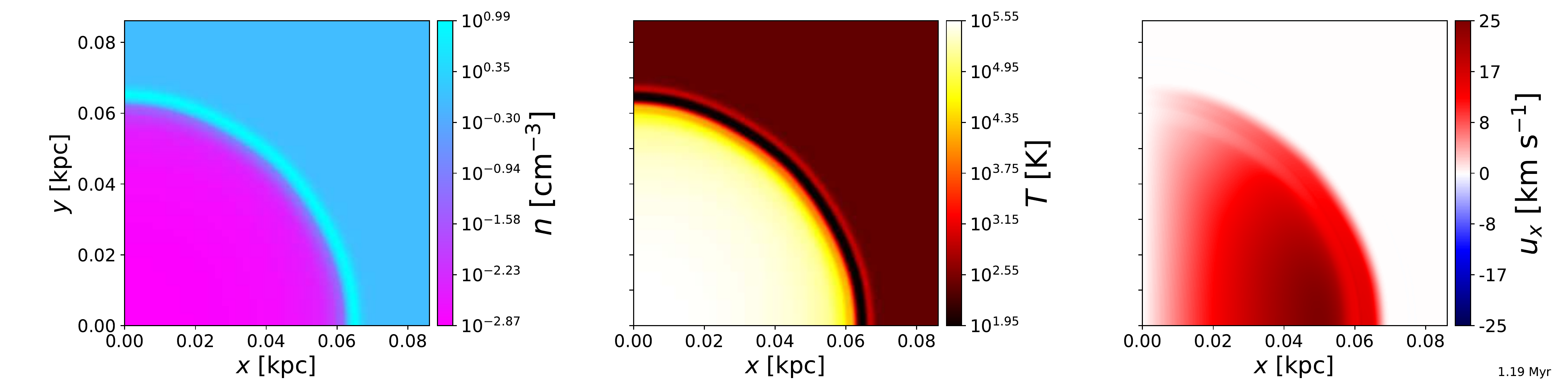}}}\\%
\subfigure[$\nu=0$
]{\resizebox*{15cm}{!}{\includegraphics[trim=0.75cm 0.45cm 0.5cm 0.3cm,clip=true]{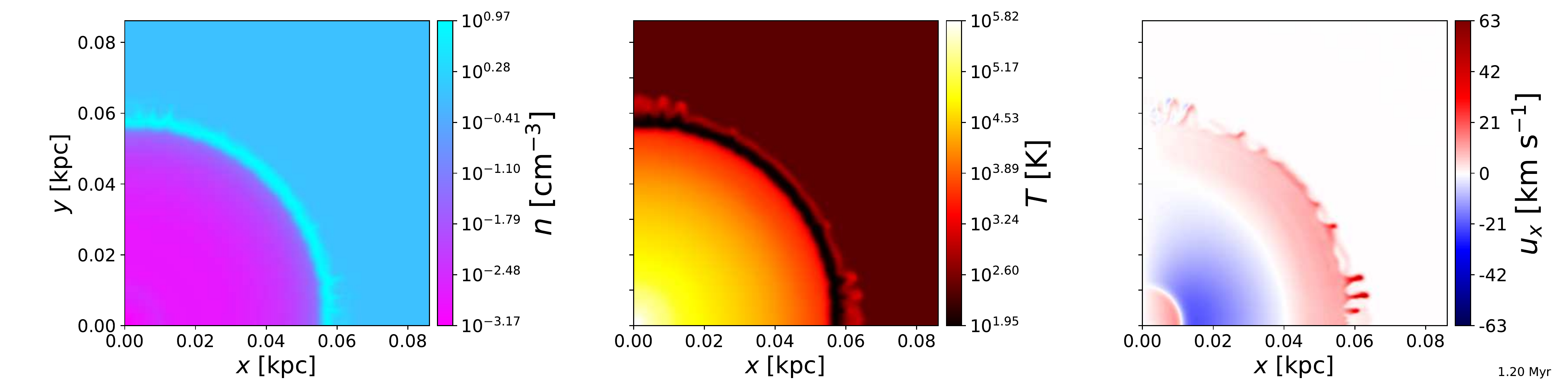}}}%
\caption{
Varying diffusivity coefficients ${c_{\rm shock}}=[1,{6,2}]$ and $\nu=5\cdot10^{-4}c_s$\,kpc\,km\,s$^{-1}$ (upper) and $\nu=0$ (lower). Two-dimensional quarter-slices from the $z=0$ plane depicting gas number density, temperature and $x$-component of velocity are shown. Ambient gas number density is 1\,cm$^{-3}$ and the grid resolution is 0.5~pc. The bottom panel shows Vishniac-Ostriker-Bertschinger  overstability, which is absent in the top panel where the shear viscosity suppresses the effect. (Colour online)
}%
\label{fig:hires-shells}
\end{center}
\end{minipage}
\end{center}
\end{figure}

Slices of the highest resolution model of 0.5\,pc grid spacing are shown in figure\,\ref{fig:hires-shells} for snaphots at the time when the shell reaches Mach\,2 for models including shear viscosity $\nu$ and without. With $\nu=0$ we see the emergence of the Vishniac-Ostriker-Bertschinger \citep{Vishniac83,VOB85} overstability arising from the cooling reducing the thickness of the shell. The overstability can be seen entering the nonlinear phase in the thin, dense, cooled shell \citep[see][for a detailed analysis]{MN93}, but this is suppressed with $\nu=0.0005c_s$.
 
\subsection{Timestep dependence}
\label{sect:timestep}

\begin{figure}
\begin{center}
\begin{minipage}{150mm}
\begin{center}
{\resizebox*{7.25cm}{!}{\includegraphics[trim=0.35cm 0.45cm 0.2cm 0.3cm,clip=true]{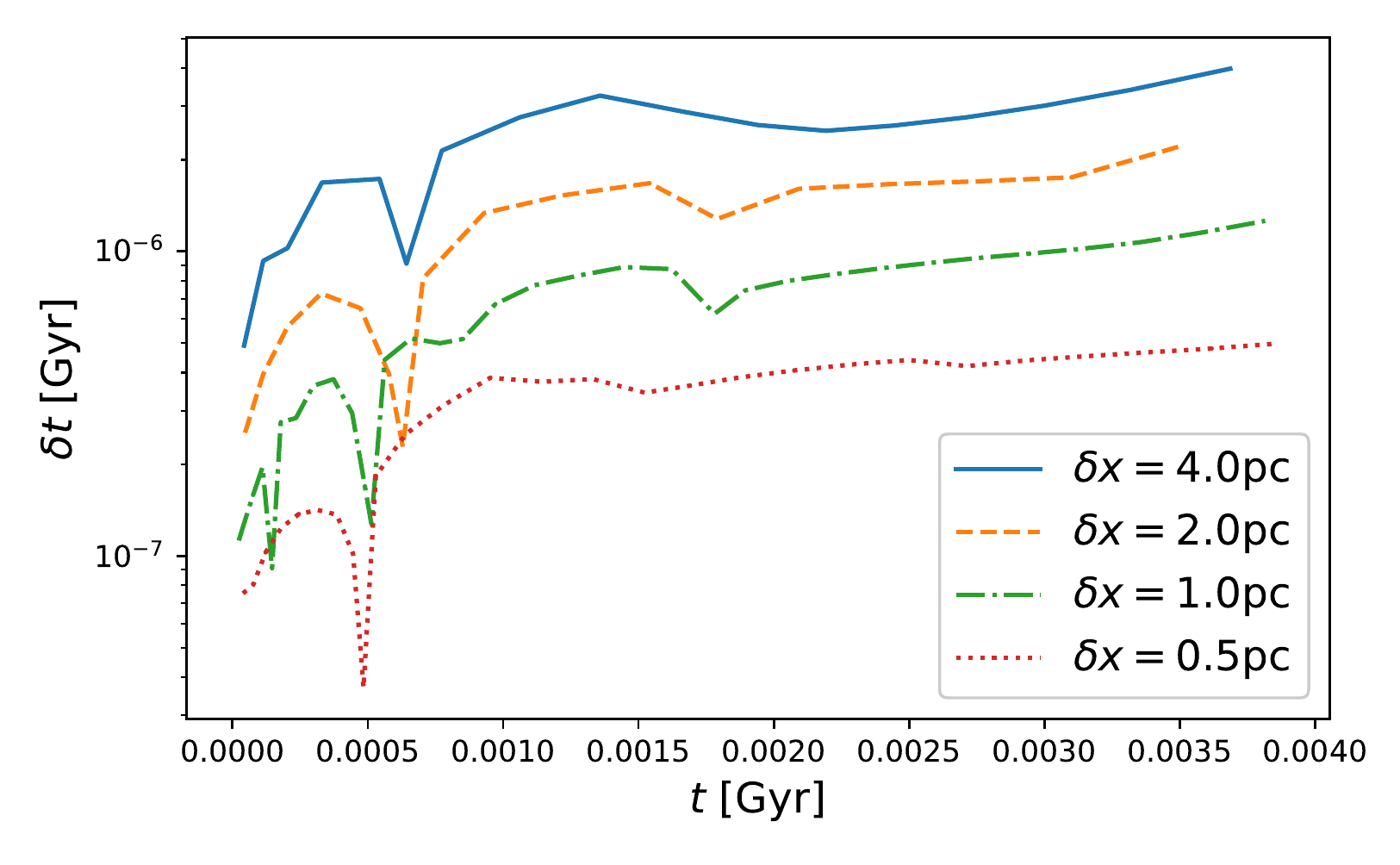}}}%
{\resizebox*{7.25cm}{!}{\includegraphics[trim=0.35cm 0.45cm 0.2cm 0.3cm,clip=true]{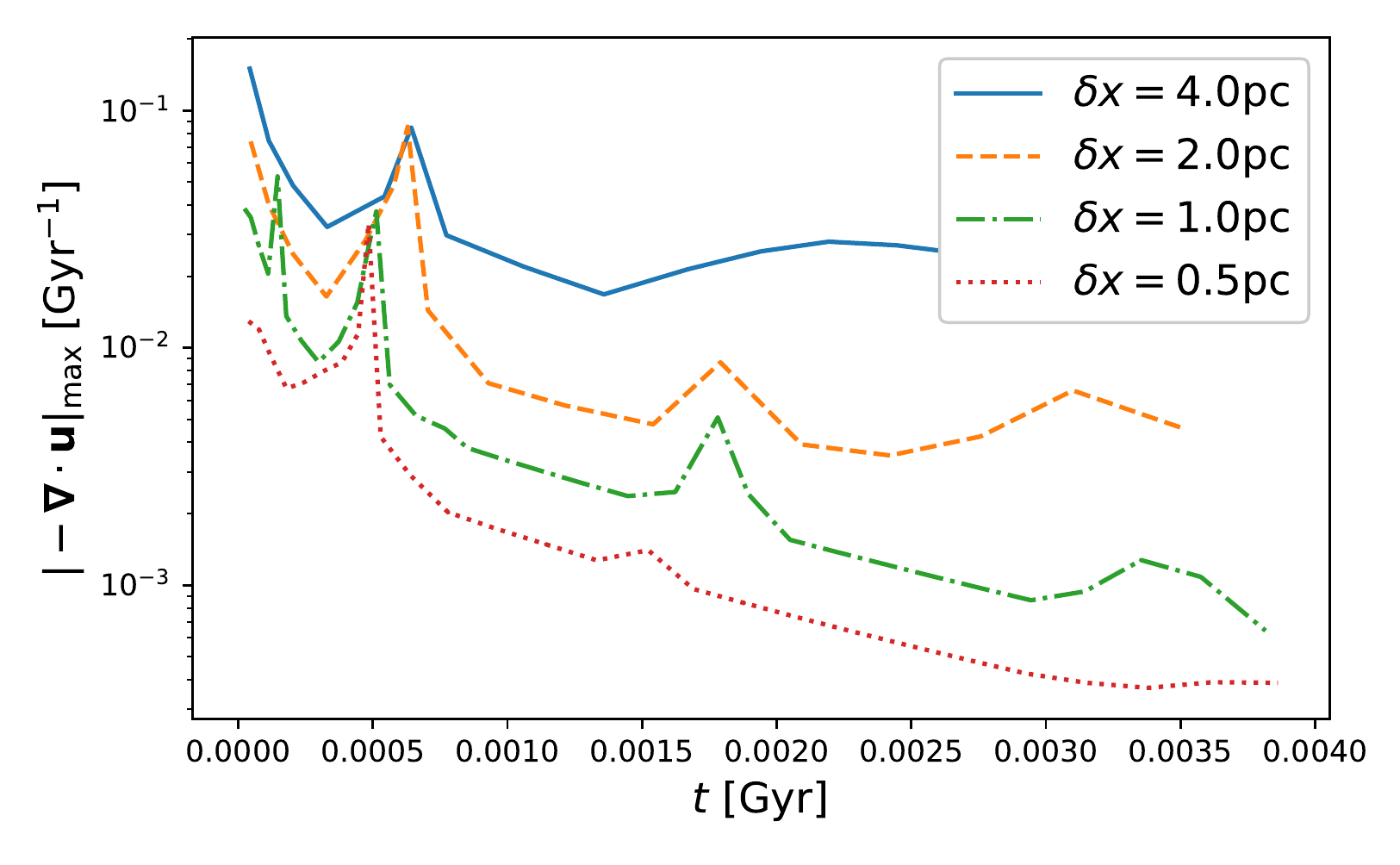}}}%
\caption{First panel: timestep evolution in unit ambient density  single SN explosion runs with grid resolution of 4, 2, 1 and 0.5\,pc. Second panel: evolution of the maximal value of convergence included in the artificial diffusion coefficients. (Colour online)
}%
\label{fig:dt-res}
\end{center}
\end{minipage}
\end{center}
\end{figure}

The timestep for snowplough tests (see section\,\ref{sect:pars}) with grid resolution 0.5--4\,pc, is plotted in the first panel of figure\,\ref{fig:dt-res}. Based on the diffusive timestep control, the dependence on resolution would be expected to be quadratic in the change in $\delta x$, such that the timestep for the $\delta x=0.5$\,pc run would be four times smaller than that of the $\delta x=1$\,pc run. In these relatively simple single SN expanding shocks, the magnitude of the viscous forces does not much exceed $10^6$\,km\,s$^{-1}$\,Gyr$^{-1}$ and the timestep control is dominated by the artificial thermal diffusivity $\zeta_\chi$. From the second panel of figure\,\ref{fig:dt-res} it is evident that the maximal convergence, and therefore also $\zeta_\chi$, approximately halves for each doubling in resolution. This indicates that the explicit artificial diffusivities are inversely proportional to the grid resolution. Thus, the timestep appears to drop only linearly with grid cell size, rather than with the quadratic dependence expected for constant diffusivity.

\begin{figure}
\begin{center}
\begin{minipage}{150mm}
\begin{center}
{\resizebox*{7.25cm}{!}{\includegraphics[trim=0.35cm 0.45cm 0.2cm 0.3cm,clip=true]{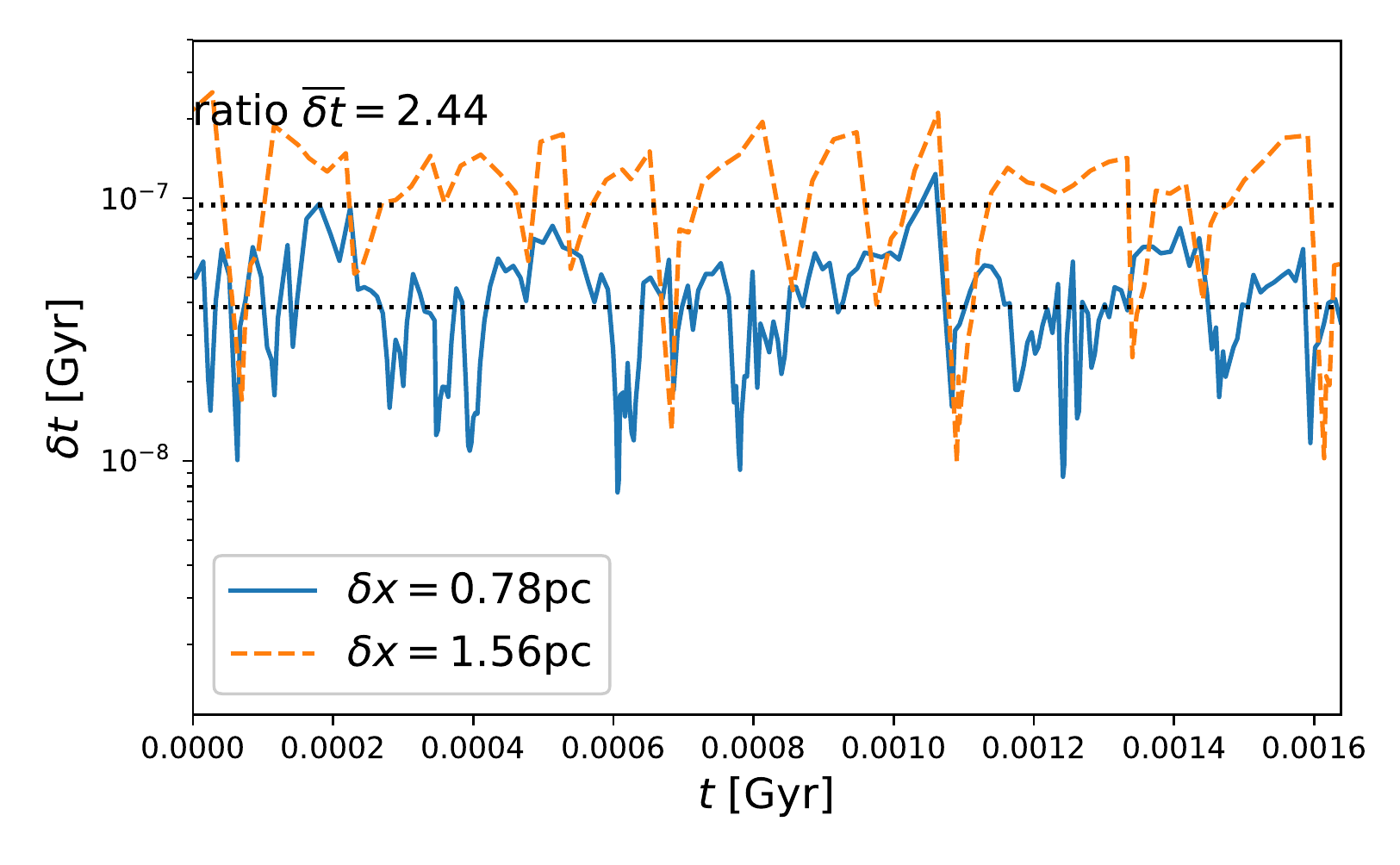}}}%
{\resizebox*{7.25cm}{!}{\includegraphics[trim=0.35cm 0.45cm 0.2cm 0.3cm,clip=true]{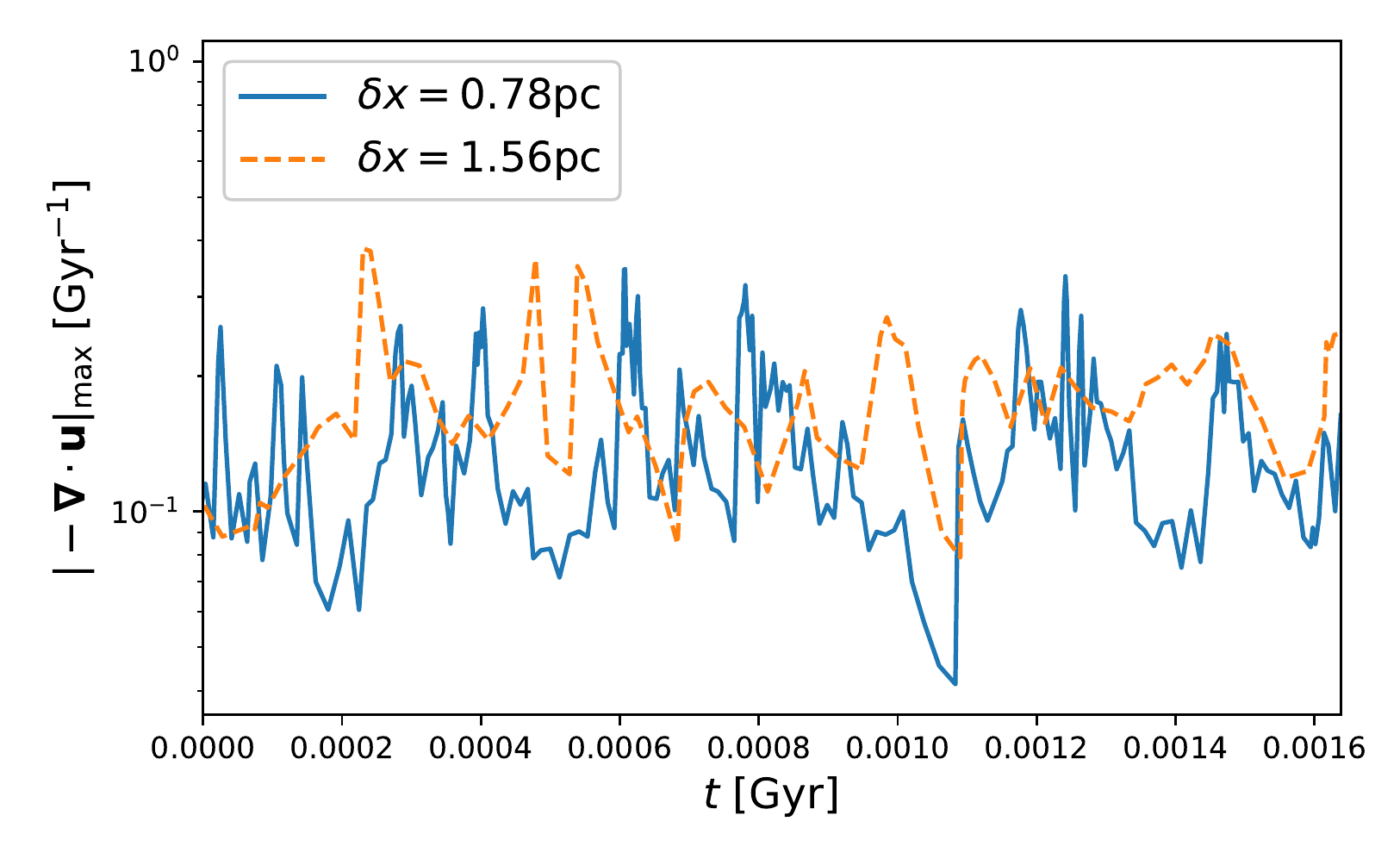}}}%
\caption{First panel: timestep evolution in SN driven turbulent simulations in unstratified periodic boxes with grid resolution of 1.56 and 0.78\,pc. Dotted lines indicate the average timestep for each simulation. $D_{\rm shock},\nu_{\rm shock},\chi_{\rm shock}=1,4,4$. Second panel: evolution of the maximal value of convergence included in the artificial diffusion coefficients. (Colour online)
}%
\label{fig:dt2-res}
\end{center}
\end{minipage}
\end{center}
\end{figure}

To demonstrate the effect on timestep in evolved turbulent systems, where the viscous forces and temperature gradients are large enough to impact the stability, the timestep and maximum convergence are plotted in figure\,\ref{fig:dt2-res} for simulations with grid spacing of 1.56\,pc and 0.78\,pc. The models apply random SN forcing to an unstratified  magnetized ambient ISM with gas number density $1\cmcube$ in a periodic slab.

The time series are extracted from a  statistical steady turbulent system, in which the magnetic field has saturated at a strength of a few $\mu$G. Temporal mean rms velocity, maximum velocity, and gas density and temperature extrema have similar values in each simulation. The timestep with the increased resolution is on average 2.44 times smaller. The magnitude of the maximal convergence displayed in the second panel of figure\,\ref{fig:dt2-res} does not differ much due to resolution, as in the case of the single SN simulations. The drop in timestep remains approximately linear with grid cell size, but not due to the lower strength of convergence. The timestep is most often controlled by the gradients in the flow and the temperature. Both are typically higher with increased resolution, but the cost is again lower than the quadratic dependence associated with constant diffusivity.

\section{Summary of results\label{sect:res}}

In order to stabilize shocks in the high-order Pencil code, we have demonstrated the use of a von Neumann artificial viscosity, as implemented for example by \citet{SN92}, in combination with the application at the shock front of artificial thermal diffusivity and mass diffusion to the energy and continuity equations. This combination reduces the vulnerability of the Pencil Code to numerical instability while significantly reducing the overall diffusivity of the model previously applied by \citet{Gent:2012} to SN driven turbulence in the ISM. The inclusion of artificial mass diffusion alters the determination of momentum and energy, and we implement a correction term to each equation to consistently conserve their properties. Similar artifical mass diffusion has been applied by, for example,  \citet[][]{RCCHS05} and \citet{JKM06,JYK09}, but without explicit corrections to the momentum and energy equations.

We have also introduced novel additional tools for controlling the timestep depending on the sum of all terms on the right hand side of the momentum equation and the energy equation. These stabilize the code by constraining the time step based on the maximal change in time of the force and heating at each iteration. Empirically, we find the viscous forces, viscous heating and temperature gradients tend to be extremely high in the SN turbulence, and can introduce numerical instabilities if the timestep is not sufficiently small to resolve the time evolution. The alternative of just reducing the Courant number for the diffusive timestep may work, but as it is based on the diffusion coefficients alone tend to reduce the timestep more universally. The adaption to the maximal forces or heating acts only when they are largest, often in the aftermath of an SN, so the timestep can recover subsequently. Another advantage is that the timestep can be less sensitive to decreases in grid spacing compared to the diffusive timestep, which scales inversely with the square of the grid spacing. This is because the unified time steps tend to be proportional to the size of the cell-to-cell velocity gradients, which decrease with decreasing cell size. Hence, increases in resolution need not be so comparatively numerically expensive. 

The code reaches reasonable agreement with the Riemann shock tube test for shocks exceeding Mach\,100 for a range of resolutions. A minimum value of $\nu_{\rm shock}={4}$ is required for the artificial viscosity coefficient for such high Mach numbers and some artificial thermal  diffusivity is required to dampen instabilities in the wake of the shock front.  The replication of the strong shock profile is relatively insensitive to changes in the coefficients $\nu_{\rm shock}$ and $\chi_{\rm shock}$, in the range {4}--8. Grid resolution, rather than the size of the artificial diffusivity coefficients, is the primary determinant of the level of smoothing present in the numerical shock profiles. The divergence of the flow is inversely proportional to the grid size, so the effective artificial diffusivity as function of resolution scales approximately as  $\nu_{\rm shock}\,\delta x^{-1}$.

The capacity of the code to model SN blast waves was tested against the Sedov-Taylor analytic and \citet{Cioffi88} semi-analytic solutions. Simulations were evolved until the blast wave becomes subsonic with respect to the ambient ISM. The models give good agreement with the Sedov-Taylor $t^{2/5}$ power law for the SN remnant shell radius evolution over a range of ambient ISM gas number density 0.1--$5\cmcube$ even for a coarse grid resolution of 4\,pc.

When cooling is included the numerical models agree reasonably with the\citet{Cioffi88} offset power law for the shell radial expansion rate, for ambient gas number densities $1\cmcube$ and below. Making allowances for preceding cooling losses by applying a fraction of the energy as kinetic \citep{KO15,SBHO15}, our models with 4\,pc grid resolution and more dense ambient ISM also yield good agreement with the analytic solutions. This is for both cooling functions tested, and notwithstanding that the semi-analytic solution was derived from simulations with a different cooling prescription, and were only performed for ambient density $0.1\cmcube$. We demonstrate that differences between the numerical and analytical solutions are well explained by the differences in the efficiency of the cooling function models.

The effect of resolution was examined, and convergence to the semi-analytic solution was evident for increases in grid resolution from 4\,pc to 0.5\,pc in the fiducial $1 \cmcube$ model. Using less than five grid cells to resolve the initial remnant radial profile tends to be insufficient to approximate a spherical energy source on the Cartesian domain; including the sphere origin explicitly as a gridpoint yields the optimal spherical evolution of the remnant. At high resolution a smaller injection radius improves convergence, but a larger number of grid zones than five to model the injection radial profile is required to retain a stable solution. For grid resolution below 1.0\,pc Vishniac-Ostriker-Bertschinger thin shell overstability begins to appear if no shear viscosity is implemented.

To directly induce cooling below the ambient temperature in the remnant shell, as opposed to relying on shock interactions, a grid resolution of 2\,pc or better is required. However, for the purposes of modelling turbulence and the dynamo, where it is adequate to capture the appropriate forcing energy, velocities and length scales to drive the turbulence, even the lowest grid resolution considered here provides good agreement with the analytic solutions.

This detailed analysis of the treatment of strong shocks and minimal diffusivity enables the code to now combine the large scale dynamo processes already present in \citet{Gent:2013b} with a prescription capable of supporting the small scale turbulent dynamo as present, for example, in \citet{Balsara04}. Further, investigation of the dependence and effect of individual SN explosions in idealised uniform and stably stratified ISM with magnetic fields and cosmic rays is a natural extension of the present study.

\section*{Acknowledgement}
The authors wish to acknowledge CSC -- IT Center for Science, Finland, for computational resources and the financial support by the Academy of Finland to the ReSoLVE Centre of Excellence (project no. 307411). M-MML was partly supported by US NSF grant AST18-15461. We thank the anonymous referees for their constructive comments.

\bibliographystyle{gGAF}
\bibliography{GGAF-2018-0035-Gent-refs}
\vspace{12pt}

\end{document}